\documentclass[fleqn,10pt]{wlscirep}
\usepackage[utf8]{inputenc}
\usepackage{array}
\usepackage[T1]{fontenc}
\usepackage{tabu}
\usepackage{mathtools}
\usepackage{bm}

\usepackage{calrsfs}
\DeclareMathAlphabet{\pazocal}{OMS}{zplm}{m}{n}
\DeclareMathAlphabet{\mathpzc}{OT1}{pzc}{m}{it}

\usepackage{float}

\newcommand{\thickhline}{\noalign{\hrule height 2pt}}
\newcolumntype{+}{!{\vrule width 2pt}}
\definecolor{lightgray}{gray}{0.9}

\title{Model-independent methods for embedding directed networks into Euclidean and hyperbolic spaces}

\author[1,*]{Bianka Kovács}
\author[1,2,3]{Gergely Palla}
\affil[1]{Dept.\ of Biological Physics, Eötvös Lor{\'a}nd University, H-1117 Budapest, P{\'a}zm{\'a}ny P.\ stny.\ 1/A, Hungary}
\affil[2]{ MTA-ELTE Statistical and Biological Physics Research Group, H-1117 Budapest, P{\'a}zm{\'a}ny P.\ stny.\ 1/A, Hungary}
\affil[3]{Health Services Management Training Centre, Semmelweis University, H-1125 Budapest, K{\'u}tv{\"o}lgyi {\'u}t 2, Hungary}

\affil[*]{bianka.kovacs@ttk.elte.hu}

\begin{abstract}

The arrangement of network nodes in hyperbolic spaces has become a widely studied problem, motivated by numerous results suggesting the existence of hidden metric spaces behind the structure of complex networks. Although several methods have already been developed for the hyperbolic embedding of undirected networks, approaches able to deal with directed networks are still in their infancy. Here, we propose a framework based on the dimension reduction of proximity matrices reflecting the network topology, coupled with a general conversion method transforming Euclidean node coordinates into hyperbolic ones even for directed networks. While proposing a new measure of proximity, we also incorporate an earlier Euclidean embedding method in our pipeline, demonstrating the widespread applicability of our Euclidean-hyperbolic conversion. Besides, we introduce a dimension reduction technique that maps the nodes directly into the hyperbolic space with the aim of reproducing a distance matrix measured on the given (un)directed network. According to mapping accuracy, graph reconstruction performance and greedy routing score, our methods are capable of producing high-quality embeddings for several real networks.

\end{abstract}
\begin{document}

\flushbottom
\maketitle

\thispagestyle{empty}



\section*{Introduction}

Networks offer an intuitive and general approach to the study of complex systems that has become extremely widespread in the recent decades~\cite{Laci_revmod,Dorog_book,Newman_Barabasi_Watts}. The staggering amount of research in this direction has shown that the statistics of the underlying graph structure can highlight previously unseen properties in systems ranging from interactions within the cell up to the level of the entire human society~\cite{Laci_revmod,Dorog_book,Newman_Barabasi_Watts,Jari_Holme_book,Vespignani_book}. The most well-known features that seem to be more or less universal across the majority of the complex networks are the small-world property~\cite{Milgram_small_world,Kochen_book}, the high clustering coefficient~\cite{Watts-Strogatz}, the scale-free degree distribution~\cite{Faloutsos,Laci_science} and a well-pronounced community structure~\cite{Fortunato_coms,Fortunato_Hric_coms,Cherifi_coms}. 

Grasping the above properties all at once with a simple network model is a challenging task for which hyperbolic approaches offer an intuitive framework. The basic idea of hyperbolic network models is to place the nodes in some representation of the hyperbolic space (in most cases the two-dimensional native disk~\cite{hyperGeomBasics}), and connect them with a probability decaying as the function of the hyperbolic distance~\cite{hyperGeomBasics,PSO,HyperMap,GPA_PSOsoftComms,nPSO,S1,S1softComms,ourEmbedding}. Remarkably, the networks generated in this way are usually small-world, highly clustered and scale-free~\cite{hyperGeomBasics,PSO}, and according to recent results they can easily display a strong community structure as well~\cite{commSector_commDetMethod,commSector_hypEmbBasedOnComms_2016,our_hyp_coms,GPA_PSOsoftComms,nPSO_New_J_Phys,S1softComms,our_PSO_Modularity}. In parallel with revealing the notable properties of hyperbolic models, several studies suggested the existence of hidden geometric spaces behind the structure of real networks as well, ranging from  protein interaction networks~\cite{Higham_geom_protein_2008,Kuchaiev_geom_protein_2009} through brain networks \cite{Cannistraci_brain_2013,Tadic_brain_2018} to the Internet~\cite{Boguna_2009_nat_phys, Boguna_Krioukov_Internet_2010,Jonckhere_Internet_2011,Bianconi_internet_2015,Chepoi_Internet_2017} or the world trade network~\cite{Boguna_trade_net_2016}, leading to important discoveries about the self-similarity~\cite{S1} and the navigability of networks~\cite{Boguna_2009_nat_phys,Gulyas_natcoms,Cannistraci_geom_congruency}.

These advancements opened a further frontier in the research focusing on the relationship between hyperbolic spaces and complex networks centred on the problem of hyperbolic embedding, where the task is to find an optimal arrangement of the network nodes in the hyperbolic space for a given network structure that we inputted~\cite{Boguna_Krioukov_Internet_2010}. A natural idea in this respect is likelihood optimisation, where a loss function is formulated (and minimised) based on the assumption that the input network was generated by a given hyperbolic network model. A prominent method following this idea is HyperMap~\cite{HyperMap}, working with a generalised version of the popularity-similarity optimisation (PSO) model~\cite{PSO} called the E-PSO model. Another possibility is the application of dimension reduction techniques to matrices that represent the network topology, such as in the Laplacian-based Network Embedding (LaBNE) technique~\cite{Alanis-Lobato_LE_embedding} (relying on the Laplacian matrix of the graph to be embedded) and the family of coalescent embeddings~\cite{coalescentEmbedding} (building on different matrices of distances measured along the graph after pre-weighting), where the dimension reduction yields a Euclidean embedding, the radial coordinates of which are converted then to hyperbolic ones in accordance with the PSO model, or the hydra (hyperbolic distance recovery and approximation) method~\cite{Hydra}, where the dimension reduction yields node positions in the hyperboloid model of the hyperbolic space that are finally converted to an embedding in the Poincaré ball representation. Dimension reduction and the optimisation of the angular node coordinates with respect to a given hyperbolic network model can be also combined as proposed in Ref.~\cite{Alanis-Lobat_liekly_LE_emb} for the Laplacian-based embedding and the E-PSO model, or as in the case of the Mercator method~\cite{Mercator} relying on the Laplacian embedding as well but optimising with respect to the so-called $\mathbb{S}^1/\mathbb{H}^2$ model~\cite{S1}, or as in Ref.~\cite{ourEmbedding}, where a coalescent embedding was combined with a local likelihood optimisation according to the E-PSO model. 

Although the aforementioned methods achieved notable success and have been shown to provide high-quality embeddings for a number of different networks, they all lack a very important capability: to take into account the link directions when dealing with directed network input. In general, directed connections 
can indicate asymmetric relations between the nodes (e.g., the dominant-subordinate relations in hierarchical networks~\cite{hierarchy_of_journals,hier_methyl}, the consumer-producer relations in food webs, etc.) or may signal some sort of flow over the links. Consequently, nodes with mainly incoming links may have a very different function in the system compared to nodes with mainly outgoing links or nodes having a balanced amount of in- and out-neighbours, and the directionality may play an important role also on the level of communities~\cite{Directed_CPM}. In this light, it seems that ignoring link directions during the preparation of an embedding can lead to a considerable amount of information loss. 



Motivated by the above, here we propose a general framework for embedding directed networks into hyperbolic spaces. Due to the possibly different functions of the sources and the targets in directed systems, our approach assigns separate source and target positions to each one of the network nodes, allowing large flexibility in how the directed nature of the input may affect the obtained embedding. This means that in the two-dimensional case, the output of the method can be visualised on a pair of native disks (one of which contains the nodes at their source coordinates and the other at their target positions), where the links always point from the "source disk" to the "target disk". In order to keep the approach model-independent, the calculation of the node positions is based on a dimension reduction of a proximity matrix encapsulating the distance relations in the network. The result of the dimension reduction can be already treated as a Euclidean embedding of the network. To obtain the hyperbolic coordinates from the Euclidean node arrangement, we introduce a transformation designed to preserve the attractiveness of a given radial position from the point of view of link creation. With the help of this transformation, we can incorporate the output of several directed Euclidean embedding methods for gaining a hyperbolic layout of the studied network. Along this line, in the present work we also apply the Euclidean HOPE algorithm~\cite{HOPE}, and transform its output in the same manner as the results of the newly proposed Euclidean embeddings. 

Finally, inspired by the undirected hyperbolic embedding method~\cite{Hydra}, we also introduce a directed embedding approach that yields hyperbolic coordinates based on the dimension reduction of a Lorentz product matrix calculated from node-node distances measured along the inputted network, providing a hyperbolic layout without embedding the network first into the Euclidean space. We test all the proposed methods both on synthetic and real networks, examining the mapping accuracy~\cite{mappingAccuracyAsSPLcorr}, and the performance in graph reconstruction and greedy routing problems. 


\section*{Results}

In this section, we first outline the studied embedding framework and describe the quality functions used for characterising the performance of the different methods. This is followed by the results obtained for a couple of directed real networks.



\subsection*{The studied embedding algorithms}

\begin{figure}[!h]
    \centering
    \includegraphics[width=0.84\textwidth]{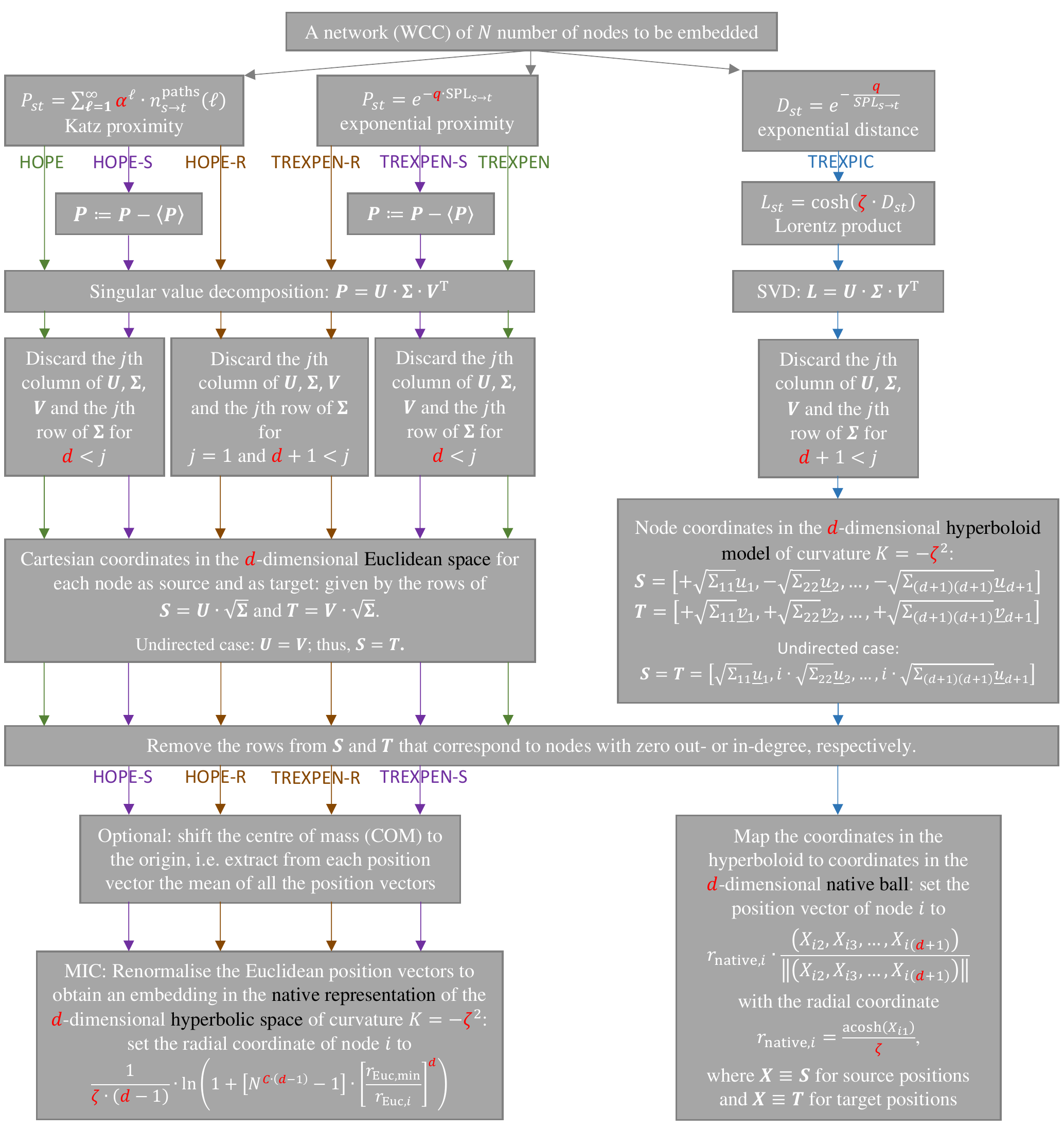}
    \caption{ {\bf Flowchart of the studied embedding algorithms.} The embedding parameters are written in red.
    }
    \label{fig:flowchart}
\end{figure}

In this paper, we consider embeddings of directed networks, which -- due to the possible different roles of the same node as a source or as a target of links -- result in two distinct sets of coordinates (i.e., source and target coordinates). In Fig.~\ref{fig:flowchart}, we provide a concise flowchart of the considered embedding methods, the full detailed description of which is given in Sect.~\ref{sect:embeddingMethodsInDetail} of the Supplementary Information. Note that all the studied methods are deterministic, yielding always the same node arrangement for a given network.

\subsubsection*{Embedding into the hyperbolic space through the conversion of a Euclidean node arrangement}

The three main steps of the algorithms described by the left side of the flowchart in Fig.~\ref{fig:flowchart} can be summarised in the following way:
\begin{enumerate}
    \item Preparation of a proximity matrix $\mathbf{P}$ based on the network topology. 
    \item Decomposition of this matrix for performing dimension reduction and obtaining a Euclidean embedding, i.e. a lower-dimensional representation in the Euclidean space.
    \item Model-independent conversion (MIC) of the Euclidean source and target coordinates into position vectors in the native representation of the hyperbolic space. 
\end{enumerate}
The proximity matrix can be defined in multiple alternative ways, and since steps 2) and 3) are always the same, we name the different methods based on the choice of $\mathbf{P}$. In the High-Order Proximity preserved Embedding (HOPE)~\cite{HOPE}, the applied proximity matrix is the Katz matrix, where the intuitive meaning of a matrix element is that it corresponds to the weighted sum of the paths between the corresponding pair of nodes, where longer paths are more or less suppressed with the help of the adjustable parameter $\alpha$. As an alternative for embedding via the Katz matrix, we introduce the method TRansformation of EXponential shortest Path lengths to EuclideaN measures, abbreviated as TREXPEN, where the proximity matrix $\bm{P}$ is composed of exponential shortest path lengths in the form of 
\begin{equation}
    P_{st}=e^{-q\cdot\mathrm{SPL}_{s\rightarrow t}},
    \label{eq:TREXPEN_Prox_matrix}
\end{equation}
where $\mathrm{SPL}_{s\rightarrow t}$ denotes the shortest path length from node $s$ to node $t$, and $0<q$ is a decay parameter similar in nature to the $\alpha$ parameter of the Katz matrix. Note that for node pairs $s$ and $t$ where $t$ is unreachable from $s$, the above matrix element $P_{st}$ and also the element of the Katz matrix becomes zero, which enable us to embed weakly connected components, not only strongly connected parts of directed networks. 

The usage of a proximity matrix (where large values indicate small distances or large similarities) has the advantage compared to distance matrices that it yields such Euclidean embeddings in which smaller topological distances can be associated primarily with larger inner products of the position vectors instead of smaller Euclidean distances, providing the possibility to effortlessly separate the contribution of the radial and the angular node coordinates in the geometric relations. However, when equating only non-negative proximity values with Euclidean inner products, the angular range of the node coordinates becomes restricted. Therefore, we also consider a centred version of the proximity matrices by shifting the mean of the matrix elements to zero, which is expected to broaden the angular range of the node coordinates. We shall refer to the embeddings where the mean of the proximities is set to zero before the matrix decomposition as HOPE-S and as TREXPEN-S (where the suffix "-S" refers to the shifting of the elements of $P$). Another alternative considered here is that we return to the original (non-shifted) proximity matrices, but discard the first and use from the second to the $d+1$th dimension for creating a $d$-dimensional embedding. The rationale behind this approach is that when embedding the network, we are interested in the positions of the nodes relative to each other, whereas the first component in the dimension reduction usually contains information mainly about the point cloud as a whole, relative to the origin. We shall refer to the embedding methods relying on the second to $d+1$th dimensions as HOPE-R and as TREXPEN-R (where the suffix "-R" refers to the removal of the first dimension). These circular Euclidean node arrangements in which the high connection probabilities are represented with high inner products can serve as a good candidate for a Euclidean-hyperbolic conversion that maps the high Euclidean inner products to small hyperbolic distances.

In our hyperbolic embedding methods, we used the native representation of the hyperbolic space~\cite{hyperGeomBasics}, which is commonly used both in hyperbolic network models~\cite{PSO,nPSO_New_J_Phys,RHG_d_dim_mathematics,dPSO} and hyperbolic embeddings~\cite{HyperMap,coalescentEmbedding,Mercator,ourEmbedding}. This representation visualises the $d$-dimensional hyperbolic space of curvature $K=-\zeta^2<0$ in the Euclidean space as a $d$-dimensional ball of infinite radius, in which the radial coordinate of a point (i.e., its Euclidean distance measured from the centre of the ball) is equal to the hyperbolic distance between the point and the centre of the ball, and the Euclidean angle formed by two hyperbolic lines is equal to its hyperbolic value. The hyperbolic distance is measured along a hyperbolic line, which is either an arc going through the points in question and intersecting the ball's boundary perpendicularly or -- if the ball centre falls on the Euclidean line that connects the examined points -- the corresponding diameter of the ball. According to the commonly applied approximating form of the hyperbolic distance~\cite{hyperGeomBasics} given by $x_{s\rightarrow t} \,\,\approx r_s^{\mathrm{source}}+r_t^{\mathrm{target}}+\frac{2}{\zeta}\cdot\ln\left(\frac{\theta_{s\rightarrow t}}{2}\right)$, a smaller hyperbolic distance $x_{s\rightarrow t}$ between the source position of node $s$ and the target position of node $t$ -- the indicator of a higher connection probability -- can originate from small radial coordinates $r_s^{\mathrm{source}}$ and $r_t^{\mathrm{target}}$ and/or a small angular distance $\theta_{s\rightarrow t}$. Another intuitive consequence of the above distance formula is that nodes with low radial coordinates are more attractive since their hyperbolic distance can become small in a larger angular region compared to nodes with large radial coordinates.

On the other hand, the Euclidean embedding methods we consider provide layouts where node pairs with high proximity values (and presumably, also high connection probabilities) obtain position vectors yielding a high inner product value. Since the inner product between the source position of node $s$ and the target position of node $t$ is simply $r_s^{\mathrm{source}}\cdot r_t^{\mathrm{target}}\cdot\cos(\theta_{s\rightarrow t})$, high connection probability in the Euclidean space can originate from large angular coordinates and/or small angular distances. Furthermore, in this case the attractive nodes correspond to the ones with the largest radial coordinates.

Since small angular distance is favourable from the point of view of both a large Euclidean inner product and a small hyperbolic distance, we transfer the angular coordinates from the Euclidean space without modification to the hyperbolic ball, similarly to the practice in several previous embedding algorithms from the literature~\cite{Alanis-Lobato_LE_embedding,coalescentEmbedding,Mercator}. However, the situation is more complex in terms of the radial coordinates, since a high inner product requires large radial coordinates in the Euclidean space, whereas a low hyperbolic distance favours small radial coordinates in the hyperbolic ball. Nevertheless, relying on the expectation that Euclidean and the hyperbolic radial arrangements of the same network should represent the same attractivity relations, we can presume that if the radial positions of the embedding from both geometries are converted to the same space, then the node arrangements that are formed in the common space must be consistent with each other. More precisely, we assume that the node arrangements obtained in the common space from the Euclidean and the hyperbolic radial coordinates reflect the same radial attractivity of any node compared to the highest one.

We use the linearly expanding half-line as the pass-through between the polynomially expanding Euclidean and the exponentially expanding hyperbolic spaces. For this, we take the well-known formulas for the spherical volume, and define the coordinate on the half-line of Euclidean and hyperbolic radial values to be equal to the volume of a sphere with the radius equal to the original radial coordinate in the given metric space, resulting in
\begin{align}
    r_{\mathrm{line}}(r_{\mathrm{Euc}}) &=V_d^{\rm Euc}(r_{\mathrm Euc})= \frac{\pi^{\frac{d}{2}}}{\Gamma(\frac{d}{2}+1)}\cdot r_{\mathrm{Euc}}^d, \label{eq:r_line_Euc}\\
     r_{\mathrm{line}}(r_{\mathrm{hyp}}) &= V_d^{\rm hyp}(r_{\mathrm hyp})= \frac{e^{\zeta\cdot(d-1)\cdot r_{\mathrm{hyp}}}-1}{\zeta\cdot(d-1)\cdot2^{d-1}}.
     \label{eq:r_line_hyp}
\end{align}
Then, our assumption about the reconcilability of the node coordinates calculated on the half-line from the Euclidean and the hyperbolic radial coordinates can be formalised for any node $i$ as
\begin{equation}
    \frac{r_{\mathrm{line}}(r_{\mathrm{Euc,max}})}{r_{\mathrm{line}}(r_{\mathrm{Euc,}i})} = \frac{r_{\mathrm{line}}(r_{\mathrm{hyp,}i})}{r_{\mathrm{line}}(r_{\mathrm{hyp,min}})},
    \label{eq:radial_attr_equiv}
\end{equation}
where we have also taken into account that the attractivity of the nodes increases in the Euclidean and decreases in the hyperbolic space with the radial coordinate (and that the most attractive node is at the maximal radial coordinate $r_{\mathrm{Euc,max}}$ in the Euclidean space, and at the minimal radial coordinate $r_{\mathrm{hyp,min}}$ in the hyperbolic space). 

By fixing the maximal radius in the hyperbolic space, we can use Eqs.~(\ref{eq:r_line_Euc})--(\ref{eq:radial_attr_equiv}) for calculating the hyperbolic radial coordinate of the nodes based on their Euclidean radial coordinate. Our suggestion for the largest possible radial coordinate in the hyperbolic ball is $r_{\mathrm{hyp,max}} = \frac{C}{\zeta}\cdot\ln(N)$, where $C$ is a constant. With this choice, the hyperbolic volume scales as $V_d^{\mathrm{hyp}}\sim N^{C\cdot(d-1)}$ with the number of nodes $N$, and at $C=2$ we obtain the same volume as we would have in a network generated by the PSO model~\cite{PSO,dPSO}. Based on that, the radial coordinate in the hyperbolic ball can be expressed as
\begin{equation}
    r_{\mathrm{hyp,}i}(r_{\mathrm{Euc,}i})=\frac{1}{\zeta\cdot(d-1)}\cdot\ln\left(1+[N^{C\cdot(d-1)}-1]\cdot\left[\frac{r_{\mathrm{Euc,min}}}{r_{\mathrm{Euc,}i}}\right]^d\right),
    \label{eq:rHyp_final}
\end{equation}
where further details of the calculation are given in Sect.~\ref{sect:EucHypConv} of the Supplementary Information. Besides, Sect.~\ref{sect:PSObasedVsOurConversion} of the Supplementary Information demonstrates that MIC, our new, model-independent Euclidean-hyperbolic conversion of the radial coordinates can outperform the widely used~\cite{Alanis-Lobato_LE_embedding,Alanis-Lobat_liekly_LE_emb,coalescentEmbedding,ourEmbedding} PSO-based transformation even on such hyperbolic networks that were generated by the PSO model.

\begin{figure}[!h]
    \centering
    \includegraphics[width=1.0\textwidth]{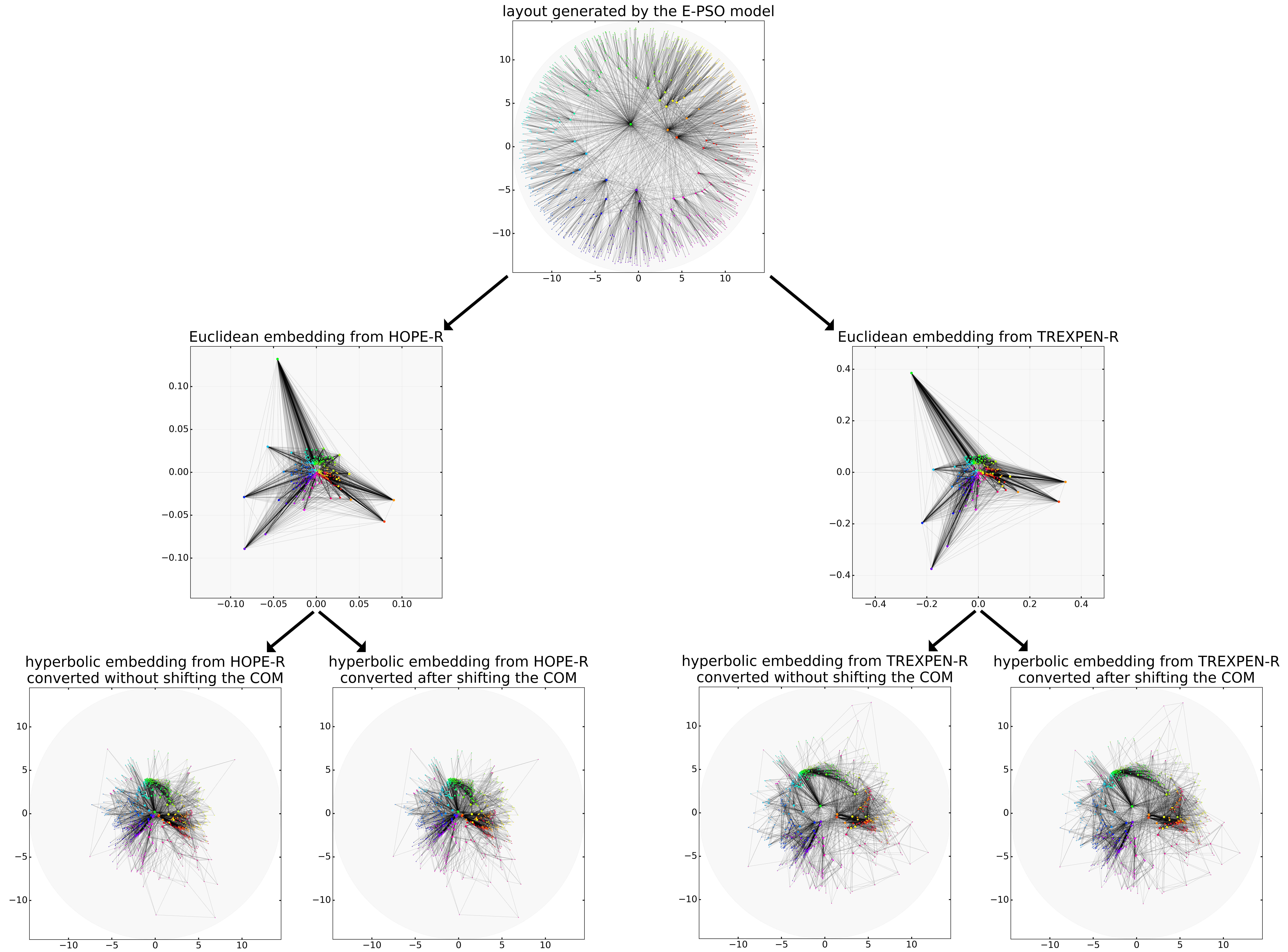}
    \caption{ {\bf Demonstrating the operation of the Euclidean-hyperbolic conversion MIC via examples of embeddings of an undirected network that was generated in the hyperbolic plane of curvature $K=-\zeta^2=-1$ by a generalised version of the popularity-similarity optimisation model.} The node degrees are indicated by the node sizes: nodes with more connections are depicted by larger markers. The nodes are coloured in each layout according to the angular coordinates originally assigned to the network nodes by the E-PSO model. The depicted HOPE-R embeddings were created using $\alpha=5.97\cdot10^{-3}$, while the TREXPEN-R layouts were obtained at $q=3.89$. We used $C=2$ and $\zeta=1$ for all the hyperbolic embeddings.} 
    \label{fig:EHconvOnPSO}
\end{figure}

As an illustration of our Euclidean-hyperbolic conversion method MIC, in Fig.~\ref{fig:EHconvOnPSO} we show two-dimensional embeddings of an undirected E-PSO network~\cite{HyperMap,ourEmbedding} that was generated from $N=1000$ number of nodes, setting the average degree to $\bar{k}\approx2\cdot(m+L)=2\cdot(3+2)=10$ (where one can interpret $m$ as the number of external links that emerge in each time step and $L$ as the net number of added and removed internal links per time step), the popularity fading parameter to $\beta=0.8$ (corresponding to the decay exponent $\gamma=1+1/\beta=2.25$ of the degree distribution $\pazocal{P}(k)\sim k^{-\gamma}$), and the temperature $T=0$ (resulting in an average clustering coefficient of $\bar{c}=0.806$). During the network generation, the nodes appeared one by one with increasing radial coordinate and connected to a given number of hyperbolically closest ones of the previously appeared nodes. Aiming at connections of small hyperbolic distances basically means that the new nodes tended to connect to nodes of small radial coordinates and/or small angular distance from them. In our Euclidean embeddings that represent small topological distances as large inner products, the early-appearing 
nodes that collected the highest number of links during the network formation become placed in the outermost positions, as the radial attractivity of the nodes increases outwards in this case. However, when transforming these layouts into hyperbolic ones, the largest hubs are transferred back to the innermost positions that possess the highest radial attractivity from the point of view of the minimisation of the hyperbolic distances. Besides, both our Euclidean and hyperbolic embeddings seem to preserve the angular arrangement of the nodes, reflecting the common preference of both geometries towards the relatively small angular distances of the connected pairs.

\subsubsection*{Embedding directly into the hyperbolic space with TREXPIC}

The above-discussed hyperbolic embedding methods rely on the implicit assumption that the Euclidean embedding obtained in the first stages of the algorithms is able to capture the most important features of the network structure. This dependence on the Euclidean methods can be avoided by embedding directly into the hyperbolic space, as it was done e.g. in the hydra approach~\cite{Hydra} on undirected networks. In order to provide also such an algorithm that follows this alternative path, we propose the method TRansformation of EXponential shortest Path lengths to hyperbolIC measures, abbreviated as TREXPIC in the following. 

According to Ref.~\cite{Hydra}, the Lorentz product defined between two position vectors as ${\underline{y}\circ\underline{z} = y_1z_1-(y_2z_2+y_3z_3+...+y_{d+1}z_{d+1})}$ enables the calculation of the hyperbolic distance in the hyperboloid representation of the $d$-dimensional hyperbolic space via the formula $x(\underline{y},\underline{z})=\frac{1}{\zeta}\cdot\mathrm{acosh}(\underline{y}\circ\underline{z})$. Thus, if we construct a distance matrix $\bm{D}$ between the nodes where the matrix element $D_{st}$ estimates the hyperbolic distance from node $s$ to node $t$, then using the formula $L_{st}=\cosh(\zeta\cdot D_{st})$ we obtain a matrix containing the estimated pairwise Lorentz products. 
Here, we suggest using 
\begin{equation}
    D_{st}=e^{-\frac{q}{\mathrm{SPL}_{s\rightarrow t}}},
    \label{eq:TREXPIC_D_matrix}
\end{equation}
where $q>0$ is an adjustable parameter that controls how fast our distance measure increases towards the larger shortest path lengths. The advantage of this choice compared to using simply the shortest paths themselves as in Ref.~\cite{Hydra} is that it makes all the matrix elements finite even in weakly connected components.

Based on the matrix of Lorentz products, we created low-dimensional hyperbolic embeddings in the hyperboloid model with the help of dimension reduction. For this, we used singular value decomposition (SVD) as opposed to hydra, which performs eigendecomposition. Then, using a mapping between the hyperboloid model and the native representation of the hyperbolic space, we obtained a layout in the native ball that is comparable with the output of the previous embedding methods.

\subsection*{Directed embedding into two-dimensional spaces}

\begin{figure}[!h]
    \centering
    \includegraphics[width=0.86\textwidth]{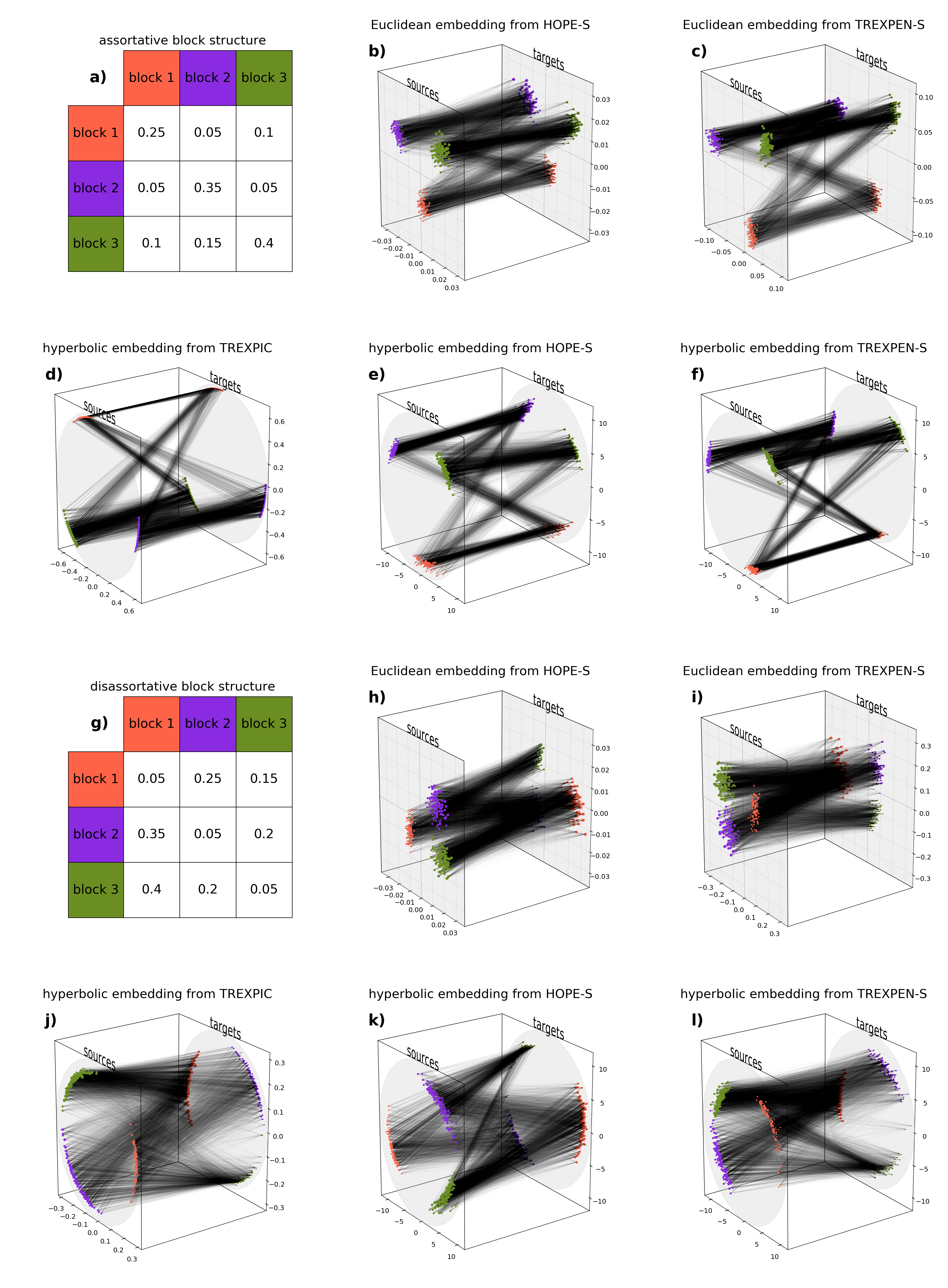}
    \caption{ {\bf Examples for the two-dimensional embeddings of SBM networks having an assortative and a disassortative block structure.} a) The assortative block matrix used for generating the input for the embeddings shown in panels b)--f), where the embedding method is named in the panel title. g) The disassortative block matrix used for generating the input for the embeddings shown in panels h)--l).
    }
    \label{fig:SBMLayout}
\end{figure}

As a first illustration of the results that can be obtained from our framework, in Fig.~\ref{fig:SBMLayout} we show the embeddings of synthetic directed networks generated by the stochastic block model (SBM)~\cite{simplestSBMarticle,dirSBMgeneration,SBMcode} in both Euclidean and hyperbolic spaces in the case of setting the number of dimensions to $d=2$, allowing the display of the achieved layouts in a simple manner. In the top half of the figure (Fig.~\ref{fig:SBMLayout}a--f) we show the results for a graph with an apparent community structure (where the diagonal elements of the connection probability matrix of the blocks are larger), while in the bottom half of the figure (Fig.~\ref{fig:SBMLayout}g--l) the embedded network has an "anti-community" structure (where the off-diagonal connection probabilities are larger). According to these layouts, the considered embedding methods were able to correctly separate the different blocks and provide an angular arrangement that reflects the most important features of the network structure in an easy-to-observe manner. Further layouts of the SBM networks are displayed in Sect.~\ref{sect:SBMlayouts} of the Supplementary Information.

Next, in Fig.~\ref{fig:polBlogsLayout} we present
embeddings of the network of political weblogs~\cite{polBlogsRef} (for which several quantitative results are provided in the next section) in both the Euclidean and the hyperbolic plane. As it can be seen here, the nodes of different attributes tend to become grouped into different angular regions in the embeddings. More examples of the automatic separation of the ground-truth communities of real networks are provided in Sect.~\ref{sect:realLayouts} of the Supplementary Information.

\begin{figure}[!h]
    \centering
    \includegraphics[width=1.0\textwidth]{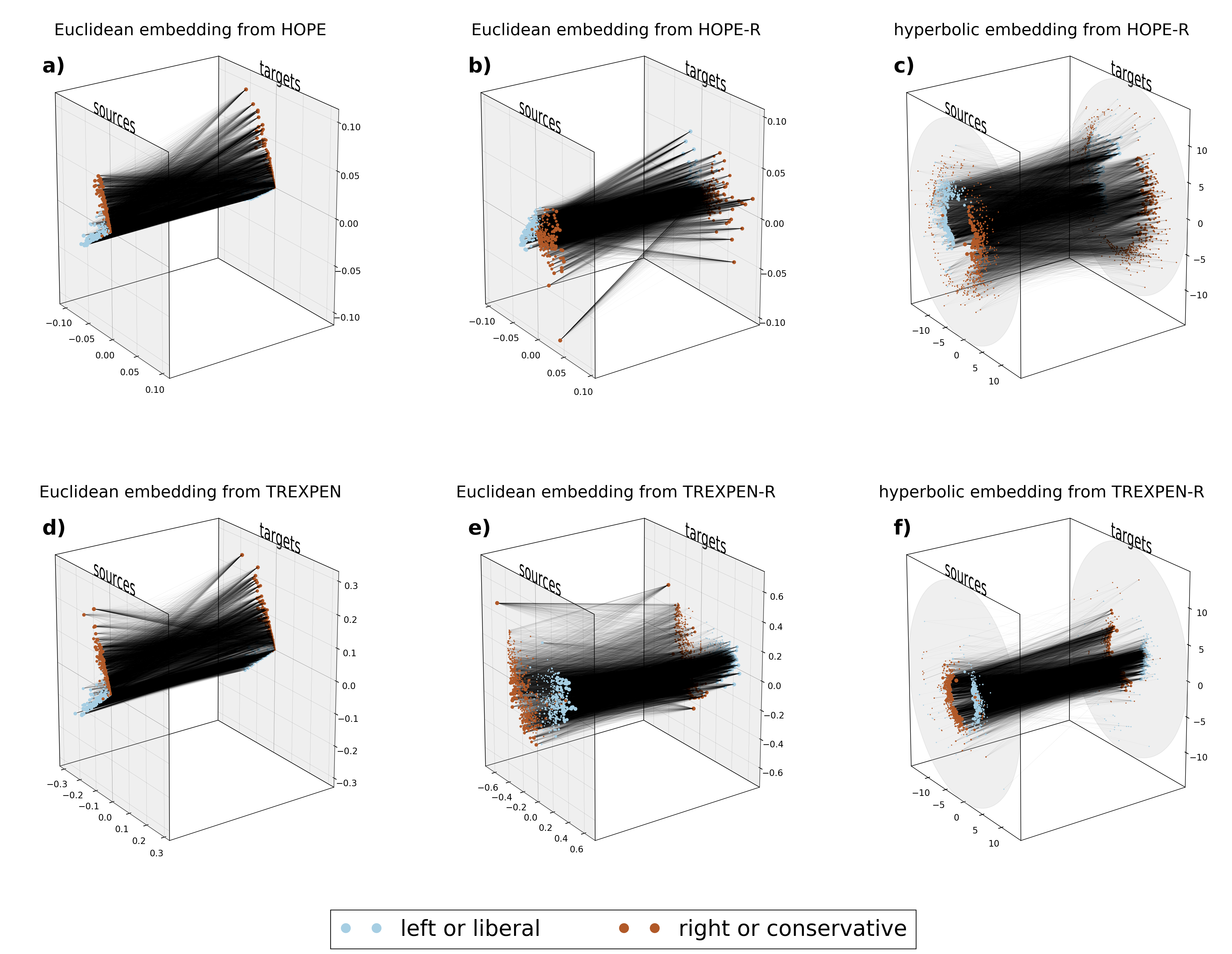}
    \caption{ {\bf Examples for the two-dimensional embeddings of the network of political blogs.} The colour of each node indicates the political leaning of the corresponding weblog. Larger node sizes on the source and the target plane correspond to larger out- and in-degrees, respectively.} 
    \label{fig:polBlogsLayout}
\end{figure}

\subsection*{Performance of HOPE, TREXPEN, their several variants and TREXPIC on real directed networks}
We tested our new embedding methods on the following directed real networks:
\begin{itemize}
    \item A subnetwork of $N=505$ number of nodes and $E=2081$ number of edges extracted from Wikipedia's norm network of 2015~\cite{wikipediaRef}, where Wikipedia pages are connected to each other with directed edges that correspond to hyperlinks. We created the subgraph by omitting all nodes for which the highest value of the topic distribution does not reach $80\%$, i.e. we kept only 
    the pages for which the topic was not too uncertain. 
    \item The transcriptional regulation network~\cite{yeastTransRef} of the yeast \textit{Saccharomyces cerevisiae}, describing $E=1063$ number of 
    interactions between $N=662$ number of regulatory proteins and genes. The links point from the regulating objects toward the regulated ones. The mode of regulation was considered to be the same in each case, i.e. we did not differentiate between activators and repressors.
    \item A network~\cite{polBlogsRef} of $E=19021$ hyperlinks among $N=1222$ number of U.S. political weblogs from before the 2004 presidential election. The blogs are characterised by their political leaning, forming 2 groups: left/liberal and right/conservative.
    \item A word association network~\cite{wordAssocRef} of $N=4865$ number of nodes and $E=41964$ number of links that point from the cue words toward the associated words. 
\end{itemize}
Note that we carried out the same analysis as below for four additional directed real networks in Sect.~\ref{sect:extraRealDirEmbeddings} of the Supplementary Information, and in Sect.~\ref{sect:undirEmb} of the Supplementary Information we also show some results regarding the embeddings of two undirected real networks, confirming that our new methods are able to compete with previous, well-known dimension reduction techniques. In addition, in Sect.~\ref{sect:directednessOfLinks} of the Supplementary Information, we show the significance of the directedness of the links in the examined directed real networks by comparing their directed embeddings to the embeddings of their undirected counterpart.



Since a node with a zero out-degree cannot have any role as a source, it can not be represented in the source layout (will not have a source coordinate), and similarly, a node with zero in-degree will not have a target coordinate. Therefore, we only embedded the largest weakly connected component (WCC) of each graph -- the above-listed $N$ and $E$ values refer to these. Throughout this section, we discarded the link weights given in some of the datasets and assigned the weight $1$ to each edge. To learn about how our embeddings treat real link weights, see Sect.~\ref{sect:realLinkWeights} of the Supplementary Information. 

In the following subsections, we evaluate the embedding performance on the above-listed four directed networks in three aspects: we examine mapping accuracy, graph reconstruction and greedy routing. The detailed description of the applied measures is provided in the Methods section. During the measurements, we took into consideration all the possible node pairs in each task for the two smaller graphs (namely the network of Wikipedia pages and the yeast transcription network), but -- because of the high computational intensity -- accomplished the evaluation of the embedding performance only on sampled sets of node pairs in the case of the two larger graphs (i.e. the network of political blogs and the word association network). The details of the applied sampling procedures are given in the Methods section.

We always tested HOPE-S, HOPE-R, TREXPEN-S and TREXPEN-R both with and without shifting the centre of mass (COM) of the node positions to the origin, but depicted here only the results of the better option. Note that shifting all the nodes by the same vector does not change the pairwise (Euclidean or hyperbolic) distances, but modifies the pairwise inner products of the nodes in a Euclidean embedding, and also changes the hyperbolic node arrangement that can be obtained from that via our Euclidean-hyperbolic conversion MIC. The difference between the quality scores achieved with or without shifting the COM is demonstrated by Sects.~\ref{sect:embParams} and \ref{sect:undirEmb} of the Supplementary Information: usually the Euclidean embeddings are hindered by the displacement of the COM, 
whereas MIC -- and the hyperbolic embeddings resulting from it -- can benefit from the balancing of the Euclidean node arrangement.

In every task, the tested number of dimensions were $d=2,3,4,8,...,2^n\leq\frac{N}{10}$, $n\in\mathbb{Z^+}$ for all the embedding methods, where the condition $d\leq N/10$ is intended to ensure a considerable dimension reduction. Note that while the embeddings obtained in high-dimensional spaces may be able to capture more information precisely, relatively high importance can be attributed also to the $d=2$ and the $d=3$ settings that are the only ones yielding directly visualisable node arrangements.

In HOPE and its variants, we tested 15 number of $\alpha$ values that we sampled from the interval $\left[\frac{1}{200\cdot\rho_{\mathrm{spectral}}(\bm{A})},\frac{1}{\rho_{\mathrm{spectral}}(\bm{A})}\right]$ for each network (see Sect.~\ref{sect:HOPE} of the Supplementary Information), where $\rho_{\mathrm{spectral}}(\bm{A})$ is the spectral radius of the adjacency matrix $\bm{A}$. In the case of TREXPEN and its variants, we always tested 15 number of $q$ values sampled from the interval $[-\ln(0.9)/\mathrm{SPL}_{\mathrm{max}},-\ln(10^{-50})/\mathrm{SPL}_{\mathrm{max}}]$ (see Sect.~\ref{sect:TREXPEN} of the Supplementary Information), where $\mathrm{SPL}_{\mathrm{max}}$ is the largest finite shortest path length occurring in the given network. For TREXPIC, we tested 15 number of $q$ values from the interval $[\ln(1.0/0.9999)\cdot\mathrm{SPL}_{\mathrm{max}},\ln(10)\cdot\mathrm{SPL}_{\mathrm{max}}]$ for each network (see Sect.~\ref{sect:TREXPIC} of the Supplementary Information). The suitability of these parameter intervals is demonstrated by Sect.~\ref{sect:embParams} of the Supplementary Information, where we show through the example of the Wikipedia network that the performance of the examined methods typically reaches a maximum within these ranges and declines at the boundaries. It is important to emphasize that we did not try to find the exact optimum of the embedding parameters, meaning that slight variances between the different embedding methods have to be treated with caution since these may simply be a consequence of the imperfection of the parameter settings and the method that seems to be worse may prevail over the other at a better parameter setting. 

The curvature $K=-\zeta^2$ of the hyperbolic space was set to $-1$ for all the hyperbolic embeddings -- the role of the curvature is discussed in Sect.~\ref{sect:embParams_curvature} of the Supplementary Information. And lastly, we always used $C=2$ in MIC, which choice is supported by Sect.~\ref{sect:embParams} of the Supplementary Information.

\subsubsection*{Mapping accuracy}

A simple measure of the embedding quality is provided by the mapping accuracy~\cite{mappingAccuracyAsSPLcorr}, defined as the Spearman's correlation coefficient~\cite{SpearmanCorrCode} between the shortest path lengths and given geometric measures of the node pairs in an embedded network. In this study, the examined geometric measures were the Euclidean distance and the additive inverse of the inner product in the case of the Euclidean embeddings, and the hyperbolic distance for the hyperbolic node arrangements. In all cases, we considered the quality of 
the embeddings to be better, 
that yielded higher positive values of the correlation coefficient, meaning that we expected all the investigated methods to minimise the distances and/or maximise the inner products between the positions of 
the nodes that are close to each other according to the network topology. 

\begin{figure}[!h]
    \centering
    \includegraphics[width=1.0\textwidth]{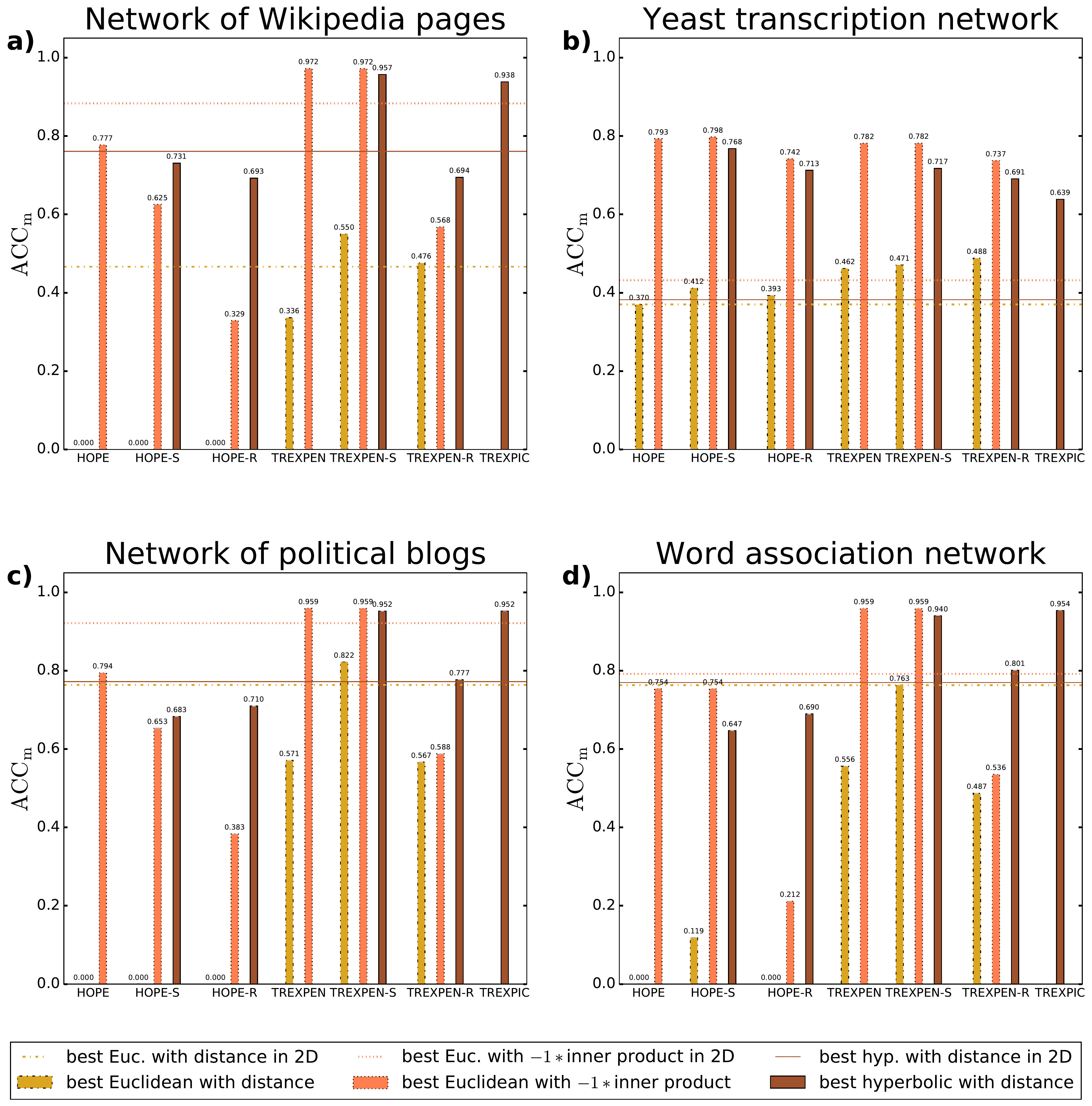}
    \caption{ {\bf Mapping accuracy on directed real networks.} Each panel refers to a real network named in the title of the panel. For the networks in panels a) and b), we measured the mapping accuracy examining each node pair connected by at least one directed path, whereas for the larger networks in panels c) and d), the mapping accuracy was measured on 3 samples of $500000$ node pairs connected by at least one directed path. In the case of the larger networks, we always considered the average of the performances over the 3 samples and depicted the corresponding standard deviations with error bars. The colours indicate the used geometric measure, as listed in the common legend at the bottom of the figure. We plotted only the best results in each panel, obtained with the parameter setting that yielded the highest values of the mapping accuracy. Note that the 0 values denote that the given methods have not achieved any positive value. The bars were created considering all the tested number of dimensions, whereas the horizontal lines show the best two-dimensional performances achieved among all the embedding methods.
    }
    \label{fig:mapAccMain}
\end{figure}

In Fig.~\ref{fig:mapAccMain} we show the mapping accuracy on the four test networks, i.e. the network of Wikipedia pages, the transcription network, the network of political blogs, and the word association network. As expected, TREXPEN, its variants and TREXPIC yield higher correlations between the shortest path lengths and the geometric measures compared to HOPE and its variants in most of the cases since HOPE considers all the paths between the nodes to a certain extent, not only the shortest ones. The best overall results were produced by Euclidean embeddings, but the hyperbolic methods do not fall behind much and, in the meantime, typically prevail over the Euclidean node arrangements when considering the distances between the nodes instead of the inner products.

\subsubsection*{Graph reconstruction}

To quantify the ability of the node arrangements provided by our embedding methods to reflect the topology of the inputted networks, we accomplished graph reconstruction trials aiming at the differentiation between the connected and the unconnected node pairs of the examined networks based on pairwise geometric measures. For this, we embedded the whole largest WCC for each one of the studied networks, and ranked the source-target node pairs according to the Euclidean distance, the inner product or the hyperbolic distance between them, assuming simply that smaller distances and/or higher inner products refer to higher proximities along the graph, and thus, larger connection probabilities. 

As a baseline, we measured the graph reconstruction performance of some local methods that, contrary to the embeddings, do not use the whole graph to give an estimation of the connection probability of a given node pair. We associated higher connection probabilities with higher numbers of common neighbours~\cite{commonNeighbors}, higher node degrees (preferential attachment~\cite{prefAttachmentAsProximity}) and higher values of 3 directed variations of the resource allocation index created from the undirected version described in Ref.~\cite{resourceAllocationIndex} -- for details, see the Methods section. In our figures, we always indicate for each quality measure only the best result obtained among these (altogether 5) tested local methods.

We evaluated the graph reconstruction performance with 3 measures: $\mathrm{Prec}\in[0,1]$ denotes the precision obtained when treating the number of links $\mathpzc{E}$ that have to be reconstructed as a known input (i.e. the proportion of the actual links among the first $\mathpzc{E}$ node pairs in the order assigned by the given connection probability measure), the area under the precision-recall (PR) curve $\mathrm{AUPR}\in(0,1]$ 
and the area under the receiver operating characteristic (ROC) curve $\mathrm{AUROC}\in[0,1]$. All of these are increasing functions of the graph reconstruction performance. For more details, see the Methods section.

\begin{figure}[!h]
    \centering
    \includegraphics[width=0.75\textwidth]{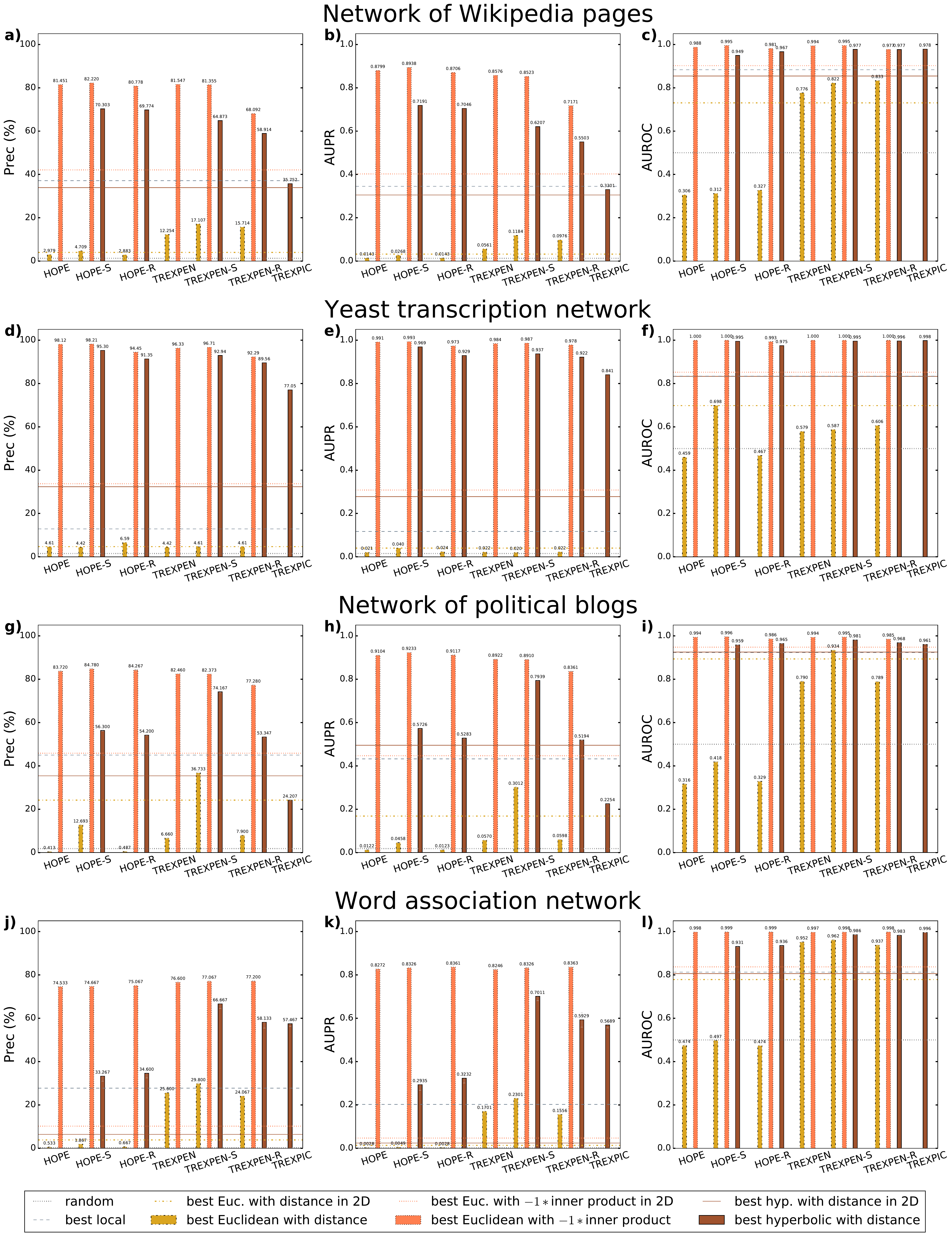}
    \caption{ {\bf Graph reconstruction performance on directed real networks.} For the networks in panels a)--f), the task was to reconstruct all the links ($E_{\mathrm{sampled}}=E$), whereas for the network of political blogs in panels g)--j) and for the word association network in panels j)--l), due to the large network size, the task was to reconstruct $3$ samples of $E_{\mathrm{sampled}}=5000$ and $E_{\mathrm{sampled}}=500$ number of links, respectively. In the case of the larger networks, we always considered the average of the quality scores over the 3 samples and depicted the corresponding standard deviations with error bars. Each row of panels refers to a real network indicated in the row title, while the different columns show the different quality measures that we studied, given by the precision obtained when reconstructing the first $E_{\mathrm{sampled}}$ most probable links (1$^{\rm st}$ column), the area under the precision-recall (PR) curve (2$^{\rm nd}$ column), and the area under the ROC curve (3$^{\rm rd}$ column). The colours indicate the applied geometric measure, as listed in the common legend at the bottom of the figure. Using the bars, we plotted only the best results regarding all the performance measures, considering all the tested number of dimensions. The horizontal lines in colour show the best two-dimensional performances achieved among all the embedding methods, whereas the grey horizontal lines correspond to the baselines provided by the random predictor and the best local method.
    }
    \label{fig:graRecMain}
\end{figure}

Figure~\ref{fig:graRecMain} presents the embedding quality with respect to the graph reconstruction task of the examined four networks. The usage of Katz proximities (in HOPE and its variants) and the exponential proximities (in TREXPEN and its variants) or distances (in TREXPIC) both seem to be expedient in this task. While generally the inner product in the Euclidean embeddings seems to be the best proxy for the connection probability, in the network of political blogs, with regard to the area under the PR curve (Fig.~\ref{fig:graRecMain}h) the best method in the two-dimensional case is a hyperbolic one. Furthermore, when focusing on the distance-based representations of the network topology, the hyperbolic embeddings clearly outperform the Euclidean ones that often even struggle to surpass the performance of the local methods.

\subsubsection*{Greedy routing}

The navigability of an embedded network can be measured via the greedy routing~\cite{Kleinberg_greedy_routing,Boguna_2009_nat_phys,Muscoloni_Cannistraci_navigability}, corresponding to the process when a walker tries to reach a given destination node from a starting node, always knowing only the position of the end of 
the links that spring from the current node compared to the position of the destination node. In our hyperbolic embeddings, we minimised in each step among the current neighbours their hyperbolic distance from the position of the destination node occupied as a target of links, while in Euclidean embeddings we tested both the minimisation of the Euclidean distance and the maximisation of the inner product. An embedded network is considered to be more navigable if its greedy routing score~\cite{coalescentEmbedding} $\mathrm{GR\text{-}score}\in[0,1]$ is higher, expressing a larger success rate in reaching the destination node and/or a smaller hop-length of the successful greedy routes. 

\begin{figure}[!h]
    \centering
    \includegraphics[width=0.73\textwidth]{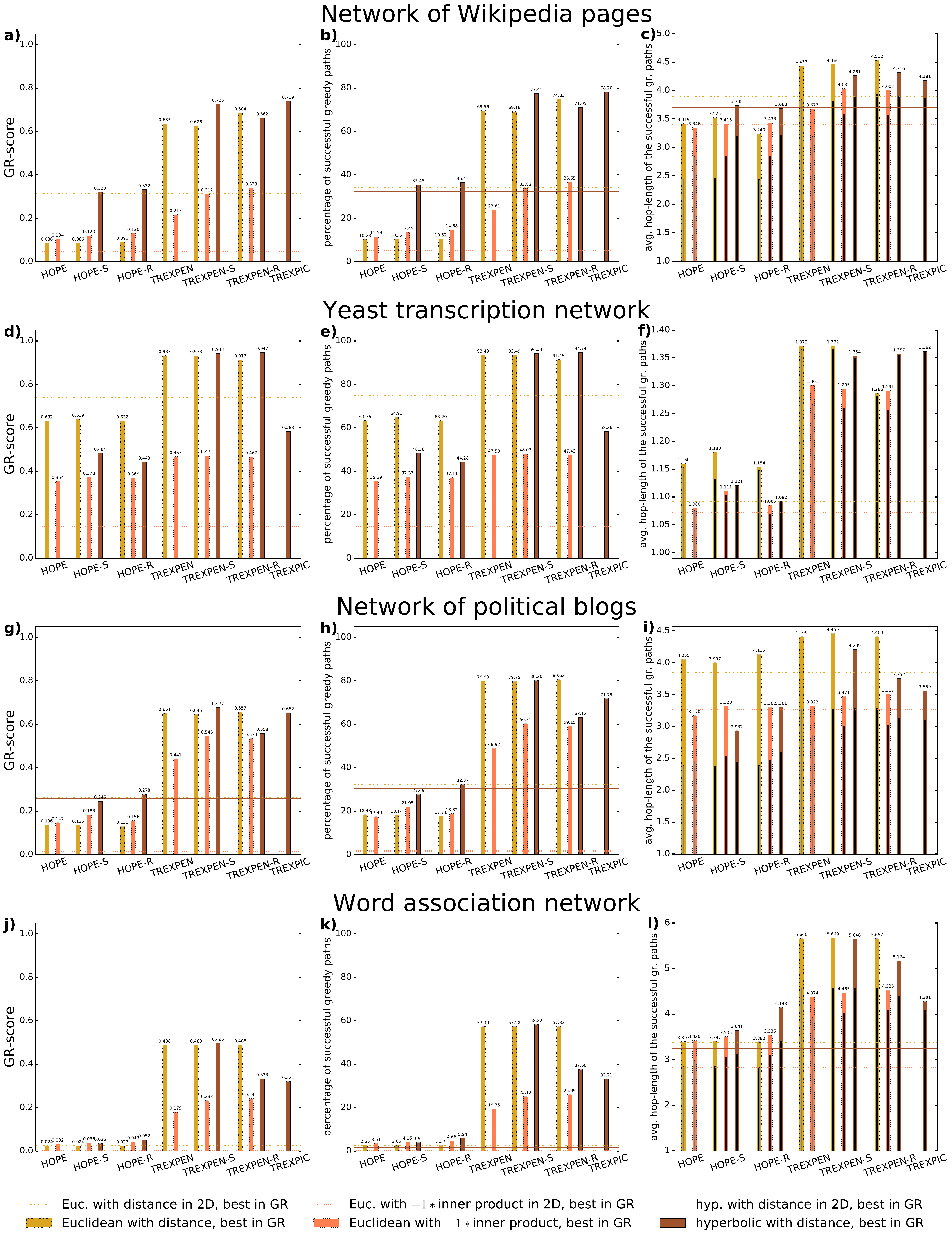}
    \caption{ {\bf Greedy routing performance on directed real networks.} For the networks in panels a)--f), the task was to perform greedy routing between each node pair connected by at least one directed path, whereas for the larger networks in panels g)--l), 
    the task was to perform greedy routing in 3 samples of $500000$ node pairs connected by at least one directed path. In the case of the larger networks, we always considered the average of the quality scores over the 3 samples and depicted the corresponding standard deviations with error bars. The colours indicate the used geometric measure as listed in the common legend at the bottom of the figure. We plotted in each panel for each method only the result of the parameter setting that turned out to be the best according to the $\mathrm{GR\text{-}score}$. 
    The bars were created considering all the tested number of dimensions, whereas the horizontal lines show the best two-dimensional average performances achieved among all the embedding methods. 
Each row of panels refers to a real network named in the row title, and the different columns correspond to different quality measures: the 1$^{\rm st}$ column shows the greedy routing score (the higher the better), the 2$^{\rm nd}$ column corresponds to
the success rate of greedy routing (the higher the better), and 
the 3$^{\rm rd}$ column depicts the average hop-length of the successful greedy paths (the smaller the better), where the grey bars indicate the average of the hop-length of the shortest paths connecting the node pairs for which the greedy routing was successful.
}
    \label{fig:GRmain}
\end{figure}

In Fig.~\ref{fig:GRmain}, we depict the achieved greedy routing scores with the corresponding success rates and average hop-lengths for the examined starting node-destination node pairs in the studied four real networks. For all of these networks, the best GR-scores are achieved in the hyperbolic space; however, the distance-based routing performed in the Euclidean space is usually also effective. The inner product generally does not seem to be well usable for navigating on networks in the Euclidean space. Besides, in this task HOPE and its variants clearly fall behind our new methods that build on exponential proximities or distances instead of Katz proximities.

\section*{Discussion}



The High-Order Proximity preserved embedding (HOPE)~\cite{HOPE} method place the nodes of undirected and directed networks in the Euclidean space of any number of dimensions in a relatively simple and fast way -- using dimension reduction -- and represent the connection probabilities with the inner products of the position vectors of the nodes. In the present work, we introduced a new Euclidean embedding method TREXPEN (TRansformation of EXponential shortest Path lengths to EuclideaN measures) that, similarly to HOPE, not only deals with undirected graphs but can also capture the asymmetries of the connection probability emerging in the presence of directed links by assigning both a source and a target position vector to each network node. Considering solely the length of the shortest paths instead of all the path lengths, the proximity matrix used in TREXPEN was shown to be similarly or even more suitable to obtain higher embedding qualities. This was especially striking in the case of the greedy routing score, where the usage of our new exponential proximities instead of Katz proximities~\cite{HOPE} was proven to be strongly advantageous. In addition, our new proximity measure can be applied without any difficulty also on weighted networks, as it is described in Sect.~\ref{sect:realLinkWeights} of the Supplementary Information.

Motivated by the emergence of several hyperbolic embedding methods besides the Euclidean algorithms, we also proposed a model-independent conversion method MIC, using which such circular Euclidean node arrangements that represent high connection probabilities with large inner products can be transformed into hyperbolic ones without assigning any specific hyperbolic network model as the origin of the network to be embedded. We showed that with some modification not only our new method TREXPEN but also the well-known HOPE is capable of generating such Euclidean embeddings that can be converted to hyperbolic node arrangements of high quality. Besides, inspired by the hydra method~\cite{Hydra} that embeds undirected graphs directly into the hyperbolic space, we developed a further method named TREXPIC (TRansformation of EXponential shortest Path lengths to hyperbolIC measures) that creates hyperbolic node arrangements for both directed and undirected networks without the need of also creating a Euclidean embedding as an intermediate step. As far as we know, the thus obtained hyperbolic embedding methods are the first model-independent embedding methods developed for arranging even directed networks in the hyperbolic space. Treating the number of dimensions of the embedding space as a free parameter, all of our methods can utilize the benefits of the increased number of dimensions, even in the hyperbolic space. We demonstrated the excellent usability of HOPE, TREXPEN, their variants and TREXPIC for different tasks via experiments carried out on real networks of several disciplines, including e.g. networks between webpages, word associations, and a transcriptional regulation network.

In general, it can be concluded that the two-dimensional hyperbolic layouts obtained with MIC are more pleasant to the human eye compared to their Euclidean counterpart as in the hyperbolic disk the large number of radially unattractive nodes become collected in the outer regions instead of gathering them around the origin. Meanwhile, due to the relatively small differences in the hyperbolic radial coordinates, the radial arrangement provided by TREXPIC on the hyperbolic plane seems to be not so informative visually, although the measured quality scores sustain the applicability of the TREXPIC method too besides the proposed conversion-based hyperbolic algorithms.

It is worth emphasizing that in our measurements regarding the mapping accuracy, the graph reconstruction performance and the navigability, the hyperbolic distance was the only geometric measure using which relatively good quality scores have been achieved in all of the different tasks: among the examined three measures, the Euclidean distance performed the worst in mapping accuracy and especially in graph reconstruction, where it was often outperformed even by the simple local methods that we tested, while the results obtained using the Euclidean inner product lagged behind both that of the Euclidean and the hyperbolic distances in greedy routing. These findings clearly justify the competitiveness of the hyperbolic embeddings. In recent years, several studies examined the emergent properties of random networks of different geometries~\cite{hyperGeomBasics,generateNetworkOnLine,EucInnProd_networkGeneration} and the indicators of different hidden geometries behind networks~\cite{measuringHyperbolicity,detectingHyperbolicGeometry_triangles}. In this work, we did not pursue to reveal how certain network properties are connected to the type and the dimension of the geometrical space underlying the networks; however, our embedding framework may contribute to further investigations on this topic by enabling the placement of real networks in different geometrical spaces of any number of dimensions.

\section*{Methods} 
This section provides the exact definition of the measures and methods used for evaluating the embedding performance. Note that none of the examined quality indicators assumes any specific model as the generator of the embedded network, i.e., all the applied evaluation processes are model independent, just like our embedding methods. For the details and the explanations regarding the studied embedding algorithms, see Section S1 in the Supplementary Information.

\subsection*{Mapping accuracy}
To evaluate the performance of the embedding methods in expressing the distance relations measured along the graph by means of geometric measures, we calculated a mapping accuracy measure $\mathrm{ACC}_{\mathrm{m}}\in[-1,+1]$ also used for undirected networks in Ref.~\cite{mappingAccuracyAsSPLcorr}. It was defined as the Spearman's correlation coefficient~\cite{SpearmanCorrCode} between the shortest path lengths of a network and the pairwise distances between the network nodes in the embedding space -- either Euclidean or hyperbolic. However, in the case of the Euclidean embeddings, the Euclidean distance was not the only geometric measure that was examined, but the correlation of the shortest path lengths with the inner products was also calculated. 

Naturally, in directed networks we took into account the directedness of the paths and compared the hop-length of the shortest path from node $s$ to node $t$ to the distance or the inner product measured between the source position vector of node $s$ and the target position vector of node $t$. 
We always discarded those $s-t$ node pairs in our calculations, for which the examined graph does not contain any connecting paths, i.e. between which the shortest path length is infinity, and also disregarded the pairing of each node with itself (characterised by a shortest path length of $0$) since the location of the target representation of a node compared to its own source position does not influence the quality of the embedding in itself, but only via the relations of the node's two representations with the other nodes. Besides, to reduce the computational cost, in networks having more than $500000$ number of start-destination node pairs that could be used for the evaluation of the mapping accuracy, we estimated this quality measure based on $3$ random samples of $500000$ proper node pairs. Note that when all the proper node pairs of a network are considered, then the calculation of the mapping accuracy is deterministic, and thus, there is no need for the repetition of its computation.

\subsection*{Evaluation of the embedding performance in graph reconstruction}

We examined how precisely the embedding methods can represent the presence and the absence of the pairwise connections of an inputted network via the graph reconstruction task (also considered e.g. in Refs.~\cite{HOPE,comparisonOfSeveralEmbeddingsInSeveralTasks}), where the question is whether the connected and the unconnected node pairs can be distinguished based on pairwise measures that are derived with full knowledge of the network topology and can be interpreted as a proxy of the connection probability. Regarding the embedding techniques, this means that we embedded the whole largest WCC of a network in the Euclidean or the hyperbolic space, arranged the node pairs in the increasing order of the Euclidean distance, the additive inverse of the inner product or the hyperbolic distance and compared the set of node pairs appearing at the beginning of the order (i.e. below a given threshold of the applied geometric measure) to the list of links in the network. Besides the embeddings, we also tested \textit{local} methods in graph reconstruction, where the decreasing order of the connection probability is estimated by the decreasing order of such measures that depend solely on the immediate neighbourhood of the two nodes in question. The assumptions of the applied local methods were the following: 
\begin{itemize}
    \item Common neighbours: In undirected networks, the larger number of common neighbours of two nodes are often associated with a larger connection probability~\cite{commonNeighbors}. In directed networks, we assumed that the larger the number of paths of hop-length 2 from node $s$ to node $t$, the higher the probability of the link from node $s$ to node $t$.
    \item Preferential attachment: In undirected networks, a simple proximity measure is given by the product of the node degrees in the examined node pair~\cite{prefAttachmentAsProximity}. In the directed case, we applied this concept as the following: the larger the product of the out-degree of node $s$ and the in-degree of node $t$ (considering also the link $s\rightarrow t$ since we deal with graph reconstruction and not link prediction), the higher the probability of the link from node $s$ to node $t$.
    \item Resource allocation index: The resource allocation index $\mathrm{RAI}$ applies one of the simplest ways for reducing the contribution of the common neighbours of high degrees to the connection probability and assigning more weight to the common neighbours of low degrees, which provide more specific connections between the examined two nodes. 
    For undirected networks, the resource allocation index is defined~\cite{resourceAllocationIndex} as
    \begin{equation}
        \mathrm{RAI}(i,j) = \sum_{c\in\mathrm{CN}(i,j)} \frac{1}{k_c},
    \label{eq:ResArcIndDef_undir}
    \end{equation}
    where $\mathrm{CN}(i,j)$ denotes the set of the common neighbours of the examined two nodes $i$ and $j$, and $k_c$ stands for the degree of the common neighbour $c$. Larger values of $\mathrm{RAI}$ are presumed to indicate larger connection probabilities. For directed networks, we identified the set of common neighbours $\mathrm{CN}(s,t)$ for the ordered node pair $s,\,t$ as the nodes that are reachable from node $s$ in one step and from which node $t$ is reachable in one step, and tested 3 versions of $\mathrm{RAI}(s,t)$, in which we substituted $k_c$ in Eq.~(\ref{eq:ResArcIndDef_undir}) with either the out-degree, the in-degree, or the total degree of the common neighbour $c$. 
\end{itemize}
In every case, the order between node pairs that have the same value of the given measure of connection probability was set randomly.

In the smaller networks, we considered all the possible node pairs in the graph reconstruction task with the exception of the pairing of each node with itself (since self-loops are disregarded by the embeddings) and those node pairs in which the out-degree of the source node or the in-degree of the target node is 0 (since to a node with 0 out- or in-degree no position is assigned by the embedding methods as source or target, respectively). In those larger graphs where the total number of the proper source-target pairs exceeds $500000$, we applied a random sampling of the connected and the unconnected node pairs. To obtain such samples that well represent the total dataset, it is important to set the ratio between the number of sampled links and the total number of sampled node pairs equal to the ratio between the total number of links and the total number of proper node pairs in the network~\cite{linkPredAspects,samplingInGraRecAndLinkPred}. In order to keep the computational cost within reasonable limits, we set the number of links $E_{\mathrm{sampled}}$ in each sample low enough to ensure that the total size of the sample (i.e. the sum of the number of links and the corresponding number of unconnected node pairs) remains under $500000$. When measuring the embedding quality on such samples, we always repeated the sampling and the reconstruction of the given links $3$ times. However, since -- at proper settings of the embedding parameters -- it is very rare that the same value of the given geometric measure (i.e. the same connection probability) becomes assigned to more than one node pair yielding an indefinite ordering between them, and therefore, the graph reconstruction itself is rather deterministic, we did not repeat the evaluation of the graph reconstruction performance in those cases where all the proper node pairs were considered. 

We characterised the embedding performance in graph reconstruction with the following 3 measures (also used e.g. in Ref.~\cite{descriptionOfMeasuresOfGraRecAndLinkPred} for evaluating link prediction accuracies), each of which is an increasing function of the embedding quality:
\begin{itemize}
    \item The precision at $\mathpzc{E}$ number of node pairs labelled as connected, i.e. $\mathrm{Prec}\in[0,1]$ is defined as the proportion of the actual links among the $\mathpzc{E}$ number of guesses corresponding to the first $\mathpzc{E}$ node pairs in the decreasing order of the given measure of the connection probability. In our measurements, we always set $\mathpzc{E}$ to the number of links that have to be reconstructed, that is, to the total number of links in the smaller WCCs and to the number of sampled links $E_{\mathrm{sampled}}$ in the case of the larger networks. For a random predictor, $\mathrm{Prec}$ was calculated for each network as the ratio between the number of actual links and all the node pairs in the examined set.
    \item The precision-recall (PR) curve~\cite{precRecCurveCode} depicts the proportion of the actual links among all the node pairs that are labelled as connected (i.e. the precision) as a function of the proportion of the links that are successfully identified among all the links that have to be restored (i.e. the recall or true positive rate), where moving between the different points of the curve corresponds to changing the threshold value of the given connection probability measure or, in other words, shifting the point in the node pair order that separates the node pairs that we label as connected from those that we label as unconnected. To give an overall description of the performances obtained at the different thresholds, we calculated $\mathrm{AUPR}\in(0,1]$ that is the area~\cite{areaUnderCurveCode} under the PR curve~\cite{areaEstimationsForPRcurve}. In the case of a random predictor, the precision-recall curve is a horizontal line at the precision value given by the ratio between the number of actual links and all the node pairs in the examined set, yielding an $\mathrm{AUPR}$ equal to this constant precision value.
    \item The receiver operating characteristic (ROC) curve~\cite{ROCcurveCode} presents the proportion of the links that are successfully identified among all the links that have to be restored (i.e. the recall or true positive rate) as a function of the proportion of the actually unconnected node pairs that are labelled as connected (i.e. the false positive rate) obtained using different threshold values of the given measure associated with the connection probability. To summarize this curve in a single number, we calculated $\mathrm{AUROC}\in[0,1]$ that is the area~\cite{areaUnderCurveCode} under the ROC curve, corresponding to the probability that a randomly chosen connected node pair gets ranked over a randomly chosen unconnected node pair in the order of the examined connection probability measure~\cite{AUROCmeaningInGeneral,AUROCmeaningInLinkPred}. For a random predictor, the ROC curve is a straight line between the points $(0,0)$ and $(1,1)$ with $\mathrm{AUROC}=0.5$.
\end{itemize}

\subsection*{Evaluation of the embedding performance in greedy routing}
To characterise the navigability of the embedded networks, we measured the efficiency of the greedy routing~\cite{Kleinberg_greedy_routing,Boguna_2009_nat_phys,Muscoloni_Cannistraci_navigability} on them, similarly to Refs.~\cite{HyperMap,coalescentEmbedding,ourEmbedding,mappingAccuracyAsSPLcorr}. The aim of greedy routing is to walk along the network's edges from a starting node $s$ to a destination node $t$ using the possible least number of steps, leaning solely on local information, namely the geometric distance of the current neighbours from the destination. 

In our measurements, we adopted a rather general stepping rule, where the greedy router being at node $i$ always moves along that outgoing link of node $i$ that points toward the neighbour having a target position of the smallest geometric measure in relation to the target position of the destination node among all the current neighbours. The examined geometric measures for which the local minimisation was performed were the Euclidean distance or the additive inverse of the inner product in the Euclidean embeddings, and the hyperbolic distance in the hyperbolic cases. Returning to a node that has already been visited in the current walk indicates that the walk between the given pair of starting and destination nodes can not be accomplished in a greedy way. Thus, two simple measures of the greedy routing's quality are the average hop-length of the successful greedy routes (that reached the destination and have not stopped any other node) and the fraction of successful greedy walks. Besides, we also measured the greedy routing score~\cite{coalescentEmbedding} ($\mathrm{GR\text{-}score}\in[0,1]$, the higher the better), which we define for directed networks as 
\begin{equation}
    \mathrm{GR\text{-}score} = \frac{1}{N_{\mathrm{paths}}}\cdot\sum\limits_{s\in S}\,\,\sum\limits_{t\in T_s}\frac{\ell_{s\rightarrow t}^{\mathrm{(SP)}}}{\ell_{s\rightarrow t}^{\mathrm{(GR)}}},
\label{eq:GRscoreDef}
\end{equation}
where $\ell_{s\rightarrow t}^{\mathrm{(SP)}}$ stands for the shortest path length from node $s$ to another node $t$ -- which is infinity if there is no path in the graph leading from $s$ to $t$ --, and $\ell_{s\rightarrow t}^{\mathrm{(GR)}}$ denotes the greedy routing hop-length between the same pair of starting and destination nodes -- which is set to infinity if the routing fails to reach node $t$ from node $s$. To allow the investigation of weakly connected networks where not all the nodes are reachable from every node, we always took into account only those starting node-destination node pairs that are connected by at least one path in the graph, i.e., for which the greedy routing is at least theoretically possible. Therefore, the total number $N_{\mathrm{paths}}$ of the examined start-destination pairs can be smaller than $N\cdot(N-1)$, and the summations in Eq.~(\ref{eq:GRscoreDef}) go over only the nodes that function as a source of links in the network, i.e. the nodes of non-zero out-degree (contained by the set $S$) and the destinations to which leads at least one directed path from node $s$ (contained by the set $T_s$ for a given starting node $s$, not including node $s$).

For large networks, it is not feasible to take into consideration each possible node pair, but using a large enough random sample of the node pairs, the performance of an embedding in greedy routing can still be well estimated. In this study, we maximised the number of start-destination node pairs for which the greedy routing was attempted at $500000$ for each network, meaning that in those networks where the total number of node pairs connected by at least one path of finite length was larger than this limit, we randomly sampled $500000$ number of such node pairs and performed the greedy routing only between the selected starting and destination nodes. For those networks where thus not all the possible node pairs were examined, we repeated the node pair sampling and the greedy routing $3$ times. Otherwise, since -- at proper settings of the embedding parameters -- it is very rare that two or more neighbouring nodes have the exact same geometric relation with the destination and the greedy router has to choose randomly between them, and therefore, the greedy routing itself is rather deterministic, we carried out greedy routing only once for all the proper node pairs of a network.


\section*{Data availability}
All data generated during the current study are available from the corresponding author upon request.

\section*{Code availability}
\begin{sloppypar}
The code used for embedding undirected/directed, unweighted/weighted networks using HOPE, TREXPEN, their variants and TREXPIC will be available at https://github.com/BianKov/TREXPEN\_TREXPIC upon publication.
\end{sloppypar}

\section*{Acknowledgements}

B.K. thanks Dániel Molnár for the several useful discussions throughout the research. 
The project was partially supported by the Hungarian National Research, Development and Innovation Office (grant no. K 128780, NVKP\_16-1-2016-0004), by the European Union’s Horizon 2020 research and innovation programme, VEO (grant agreement No. 874735) and the Thematic Excellence Programme (Tématerületi Kiválósági Program, 2020-4.1.1.-TKP2020) of the Ministry for Innovation and Technology in Hungary, within the framework of the DigitalBiomarker thematic programme of the Semmelweis University.

\section*{Additional information}

\subsection*{Author contributions statement}
B.K. and G.P. developed the concept of the study, B.K. implemented and tested the embedding methods, B.K. pre-processed the network data, performed the analyses and prepared the figures, B.K. and G.P. wrote the paper. All authors reviewed the manuscript. 

\subsection*{Competing Interests}
The authors declare no competing interests.

\clearpage
\setcounter{section}{0}
\renewcommand{\thefigure}{S\arabic{figure}}
\renewcommand{\thetable}{S\arabic{table}}
\renewcommand{\theequation}{S\arabic{equation}}
\renewcommand{\thesection}{S\arabic{section}}

\section*{\huge{Model-independent methods for embedding directed networks into Euclidean and hyperbolic spaces -- Supplementary Information}}
\bigskip

\section{The embedding methods in detail}
\label{sect:embeddingMethodsInDetail}
\setcounter{figure}{0}
\setcounter{table}{0}
\setcounter{equation}{0}
\renewcommand{\thefigure}{S1.\arabic{figure}}
\renewcommand{\thetable}{S1.\arabic{table}}
\renewcommand{\theequation}{S1.\arabic{equation}}

In this section, we detail the examined embedding methods that determine two positions for each node in a directed network: one representing its behaviour as a source of links (source position) and one that characterises it as a target of links (target position). First, we expound on how singular value decomposition (SVD) of a proximity matrix can be used for arranging nodes in a $d$-dimensional Euclidean space, representing the network topology by the inner products of the nodes' position vectors. Second, we specify how the proximity matrix can be derived for a network in the case of the well-known method HOPE~\cite{HOPE} and in our newly-introduced technique named TREXPEN, abbreviating {\it TRansformation of EXponential shortest Path lengths to EuclideaN measures}. Then, we describe how HOPE and TREXPEN can be used to create such Euclidean embeddings that are spread out on the whole angular range and how these circular node arrangements can be transformed with a model-independent conversion MIC into hyperbolic embeddings, where higher connection probabilities are reflected by smaller hyperbolic distances. Finally, we present TREXPIC ({\it TRansformation of EXponential shortest Path lengths to hyperbolIC measures}), a new embedding method that is built on the singular value decomposition of a matrix obtained from distances measured along the graph and maps the network nodes directly into the $d$-dimensional hyperbolic space.

\subsection{Euclidean embedding based on the singular value decomposition of a proximity matrix}
\label{sect:SVDofProxMatrix}
\setcounter{figure}{0}
\setcounter{table}{0}
\setcounter{equation}{0}
\renewcommand{\thefigure}{S1.1.\arabic{figure}}
\renewcommand{\thetable}{S1.1.\arabic{table}}
\renewcommand{\theequation}{S1.1.\arabic{equation}}

Singular value decomposition (SVD) have been successfully used in several different embedding methods in the past few years~\cite{ncMCE,HOPE,coalescentEmbedding,ourEmbedding}. Given the singular value decomposition $\bm{P}=\bm{U}\cdot\bm{\Sigma}\cdot\bm{V}^T$ of a proximity matrix $\bm{P}$ derived from the network topology, for the Euclidean coordinate matrices written in the form of 
\begin{equation}
    \bm{S} = \bm{U}\cdot\sqrt{\bm{\Sigma}}
    \label{eq:sourceCoordMatrix}
\end{equation}
and
\begin{equation}
    \bm{T} = \left(\sqrt{\bm{\Sigma}}\cdot\bm{V}^T\right)^T = \bm{V}\cdot\sqrt{\bm{\Sigma}},
    \label{eq:targetCoordMatrix}
\end{equation}
it is obvious that $\bm{S}\cdot\bm{T}^T=\bm{U}\cdot\sqrt{\bm{\Sigma}}\cdot\sqrt{\bm{\Sigma}}\cdot\bm{V}^T=\bm{U}\cdot\bm{\Sigma}\cdot\bm{V}^T=\bm{P}$, meaning that the elements of the proximity matrix $\bm{P}$ correspond to the inner products between the Cartesian position vectors represented by the rows of the above-defined $\bm{S}$ and $\bm{T}$ matrices. Accordingly, for a proximity matrix consisting of only non-negative elements, the inner products of the source and target position vectors given by $\bm{S}$ and $\bm{T}$ can not become negative, yielding angular distances not larger than $\pi/2$.

As it was described in Ref.~\cite{HOPE}, if $d$-dimensional position vectors are needed (i.e., the number of columns of $\bm{S}$ and $\bm{T}$ has to be $d$), then an optimal solution for minimizing the L2-norm of $\bm{P}-\bm{S}\cdot\bm{T}^T$ is to perform the SVD of the proximity matrix $\bm{P}$, keep only the first (largest) $d$ number of singular values from the diagonal matrix $\bm{\Sigma}$, and calculate the $N\times d$-sized position matrices $\bm{S}$ and $\bm{T}$ based on Eqs.~(\ref{eq:sourceCoordMatrix}) and (\ref{eq:targetCoordMatrix}) using only the first $d$ number of columns (singular vectors) in the matrices $\bm{U}$ and $\bm{V}$, i.e., use the formulas
\begin{equation}
    \bm{S} = [\sqrt{\sigma_1}\cdot\underline{u}_1\,,\,\sqrt{\sigma_2}\cdot\underline{u}_2\,,\,\sqrt{\sigma_3}\cdot\underline{u}_3\,,\,...\,,\,\sqrt{\sigma_d}\cdot\underline{u}_d]
    \label{eq:sourceCoordMatrix_reduced_innProd}
\end{equation}
and
\begin{equation}
    \bm{T} = [\sqrt{\sigma_1}\cdot\underline{v}_1\,,\,\sqrt{\sigma_2}\cdot\underline{v}_2\,,\,\sqrt{\sigma_3}\cdot\underline{v}_3\,,\,...\,,\,\sqrt{\sigma_d}\cdot\underline{v}_d].
    \label{eq:targetCoordMatrix_reduced_innProd}
\end{equation}
In this study, for the embedding of a network that consists of $N$ number of nodes, the tested number of dimensions were $d=2,3,4,8,16,32,...,2^n\leq N/2, n\in\mathbb{Z^+}$.

For undirected networks, the proximity matrix $\bm{P}$ is symmetric; thus, $\bm{U}=\bm{V}$ and $\bm{S}=\bm{T}$. Otherwise, $\bm{S}\neq\bm{T}$, meaning that each node is characterised by two positions that can differ from each other. In order to open up the possibility to deal also with weakly connected directed networks and not only strongly connected ones, it is expedient to decompose such matrixes with SVD that represent infinite distances -- measured in the absence of any paths from a given node to an other -- as finite matrix elements. 
However, although the number of rows in the matrices $\bm{S}$ and $\bm{T}$ -- created from a proximity matrix $\bm{P}$ of size $N\times N$ -- is always equal to the number of nodes $N$ of the embedded graph, meaning that $\bm{S}$ and $\bm{T}$ assign a coordinate array for each one of the network nodes, not all of these positions are meaningful for weakly connected networks. Since a node with $0$ out-degree does not function as a source of links in the network, its source position given by the corresponding row of the matrix $\bm{S}$ is meaningless. The same can be said regarding the target positions assigned by the rows of the matrix $\bm{T}$ to nodes of $0$ in-degree. Consequently, in our investigations we considered only those nodes to be placed in the embedding space as a source of links that have non-zero out-degree and assigned a target position to only those nodes of which the in-degree is larger than $0$.

\subsection{HOPE: High-Order Proximity preserved Embedding}
\label{sect:HOPE}
\setcounter{figure}{0}
\setcounter{table}{0}
\setcounter{equation}{0}
\renewcommand{\thefigure}{S1.2.\arabic{figure}}
\renewcommand{\thetable}{S1.2.\arabic{table}}
\renewcommand{\theequation}{S1.2.\arabic{equation}}

In the High-Order Proximity preserved Embedding (HOPE) method~\cite{HOPE}, the proximity matrix $\bm{P}$ to be decomposed contains the pairwise Katz indexes of the nodes~\cite{KatzIntroduction} defined as
\begin{equation}
    P_{st}=\sum_{\ell=1}^{\infty} \alpha^{\ell}\cdot n_{s\rightarrow t}^{\mathrm{paths}}(\ell),
    \label{eq:KatzDefinition}
\end{equation}
where $n_{s\rightarrow t}^{\mathrm{paths}}(\ell)$ denotes the number of paths of hop-length $\ell>0$ connecting node $s$ to node $t$ and $0<\alpha<1$ is a decay parameter controlling how fast the contribution of the paths decreases with their length $\ell$ (the smaller the value of $\alpha$, the more important the shorter paths compared to the longer ones). When a node can not be reached from an other, then the corresponding matrix element is $0$, because the number of paths is $0$ for any finite path length $\ell$ and paths of infinite length do not have a contribution to the sum in Eq.~(\ref{eq:KatzDefinition}) since $\alpha<1$. Note that the Katz index $P_{ss}$ of a node $s$ with itself is determined solely by the number of cycles that include the given node and the paths of length $0$ are not considered in Eq.~(\ref{eq:KatzDefinition}), meaning that the highest Katz indexes do not necessarily fall in the diagonal of the matrix $\bm{P}$.

To enable a faster computation, the matrix of Katz indexes is usually calculated for unweighted networks from the adjacency matrix $\bm{A}$ of size $N\times N$ ($A_{ij}=0$ if there is no link pointing from node $i$ to $j$, otherwise $A_{ij}=1$; in the absence of self-loops, $A_{ii}=0\,\,\,\forall i$) and the identity matrix $\bm{I}$ of size $N\times N$ as~\cite{KatzDescription}
\begin{equation}
    \bm{P}=\sum_{\ell=1}^{\infty} \alpha^{\ell}\cdot\bm{A}^{\ell}=(\bm{I}-\alpha\cdot\bm{A})^{-1}-\bm{I},
    \label{eq:KatzMatrixElements}
\end{equation}
where the second step uses the series expansion $(\bm{I}-\alpha\cdot\bm{A})^{-1}=\bm{I}+\alpha\cdot\bm{A}+\alpha^2\cdot\bm{A}^2+\mathcal{O}
(\bm{A}^3)$ and assumes that the decay parameter $\alpha$ is lower than the reciprocal of the spectral radius of the adjacency matrix $\bm{A}$, i.e. $\alpha<1/\rho_{\mathrm{spectral}}(\bm{A})$, where the spectral radius is the largest absolute value of the eigenvalues of $\bm{A}$. In our measurements, we always tested 15 number of $\alpha$ values, which we sampled between $\alpha_{\mathrm{min}}=\frac{1}{200\cdot\rho_{\mathrm{spectral}}(\bm{A})}$ and $\alpha_{\mathrm{max}}=\frac{1}{\rho_{\mathrm{spectral}}(\bm{A})}$ equidistantly on a logarithmic scale.

Note that in Ref.~\cite{HOPE}, instead of the usual singular value decomposition of the above-described Katz proximity matrix, a method named JDGSVD was used, which does not require the calculation of the proximity values $P_{st}$ but decomposes directly two composing matrices of $\bm{P}$, and thereby speeds up the embedding procedure. However, for the sake of simplicity, in our study we followed the instructions of the previous section, i.e. we used the usual singular value decomposition of the proximity matrix given by Eq.~(\ref{eq:KatzMatrixElements}).

\subsection{TREXPEN: TRansformation of EXponential shortest Path lengths to EuclideaN measures}
\label{sect:TREXPEN}
\setcounter{figure}{0}
\setcounter{table}{0}
\setcounter{equation}{0}
\renewcommand{\thefigure}{S1.3.\arabic{figure}}
\renewcommand{\thetable}{S1.3.\arabic{table}}
\renewcommand{\theequation}{S1.3.\arabic{equation}}

In our new Euclidean embedding method TREXPEN, the proximity matrix $\bm{P}$ subjected to dimension reduction is derived from the pairwise shortest path lengths measured on the network to be embedded. Namely, the matrix element assigned to the node pair $s-t$ is calculated from the shortest path length $\mathrm{SPL}_{s\rightarrow t}$ measured from node $s$ to node $t$ as 
\begin{equation}
    P_{st}=e^{-q\cdot\mathrm{SPL}_{s\rightarrow t}},
    \label{eq:ourProxMatrixElements}
\end{equation}
yielding larger values for smaller distances measured on the graph. The possible largest value of the here-defined proximity is $1$, yielded by $0$ shortest path lengths in the diagonal. If there is no path connecting node $s$ to node $t$, then $\mathrm{SPL}_{s\rightarrow t}=\infty$, which would be hard to handle in itself; however, Eq.~(\ref{eq:ourProxMatrixElements}) converts infinite distances to matrix elements of $0$, and thereby, it provides the opportunity to easily embed also those networks that are not strongly, just weakly connected. The multiplying factor $q>0$ controls the speed of the decay of the proximity with the increase in the shortest path length. For extremely small values of $q$, according to the Taylor series expansion $e^{-q\cdot\mathrm{SPL}_{s\rightarrow t}}\approx 1-q\cdot\mathrm{SPL}_{s\rightarrow t}$, the decay from $1$ is approximately linear and all the non-zero proximity values remain close to $1$, while for large $q$ factors, all the off-diagonal elements of the proximity matrix $\bm{P}$ will fall very close to $0$.

It is important to notice that although it is possible to consider mainly just the shortest ones of the path lengths in the Katz index by setting the decay parameter $\alpha$ to extremely small values, but, at the same time, these small $\alpha$ values will also reduce the importance of the longer shortest paths compared to the lowest path lengths of the graph. Our newly introduced proximity given by Eq.~(\ref{eq:ourProxMatrixElements}) provides, however, a solution for this problem, as it enables various weighting of the shortest paths of different lengths. More precisely, while our proximity at large values of $q$ is similar to a Katz index of small $\alpha$, the decrease in $q$ has different effects compared to the increase in $\alpha$: the former increases the relative importance of the longer shortest paths, whilst the latter increases the importance of all the longer paths, not just the ones that connect two nodes via the least hops.

In our experiments regarding TREXPEN, we varied the occurring smallest non-zero proximity (obtained from Eq.~(\ref{eq:ourProxMatrixElements}) at the largest finite shortest path length $\mathrm{SPL}_{\mathrm{max}}$ in the given network to be embedded) between $0.9$ and $10^{-50}$ by testing $q$ values between ${q_{\mathrm{min}}=-\ln(0.9)/\mathrm{SPL}_{\mathrm{max}}}$ and $q_{\mathrm{max}}=-\ln(10^{-50})/\mathrm{SPL}_{\mathrm{max}}$. To accomplish an exhaustive exploration of the parameter space and ensure that both the small and the large $q$ values are sufficiently represented in our parameter sampling, we tested in each task 15 number of $q$ values, from which 7 were sampled from the interval $[q_{\mathrm{min}},q_{\mathrm{mid}})$ equidistantly on a logarithmic scale and 8 were sampled from $[q_{\mathrm{mid}},q_{\mathrm{max}}]$ equidistantly on a linear scale, where $q_{\mathrm{mid}}=e^{\frac{\ln(q_{\mathrm{min}})+\ln(q_{\mathrm{max}})}{2}}$ is the logarithmic midpoint of the examined interval $[q_{\mathrm{min}},q_{\mathrm{max}}]$, i.e., the geometric mean of $q_{\mathrm{min}}$ and $q_{\mathrm{max}}$.

\subsection{Creation of circular patterns}
\label{sect:circPatterns}
\setcounter{figure}{0}
\setcounter{table}{0}
\setcounter{equation}{0}
\renewcommand{\thefigure}{S1.4.\arabic{figure}}
\renewcommand{\thetable}{S1.4.\arabic{table}}
\renewcommand{\theequation}{S1.4.\arabic{equation}}

Coalescent embedding techniques yielding a circular arrangement of the network nodes were introduced in Ref.~\cite{coalescentEmbedding} with two possible realisations: either the distance matrix was centred before the dimension reduction, or it was not centred but the first dimension of the embedding was discarded. We found that both HOPE and TREXPEN are able to produce layouts that are not restricted to a narrow angular range if we transfer the above-mentioned procedures from the coalescent embedding techniques. Relying on this finding, we propose the methods HOPE-S and TREXPEN-S, denoting those versions of the above-described Euclidean embeddings, in which the mean of the proximity matrix elements given by Eqs.~(\ref{eq:KatzMatrixElements}) and (\ref{eq:ourProxMatrixElements}) is \textit{shifted} to $0$ before carrying out the singular value decomposition, by subtracting the mean of all the proximity values from each element of~$\bm{P}$. Furthermore, we introduce HOPE-R and TREXPEN-R, where we treat the first dimension of the embedding to be \textit{redundant} and use the singular values from the second to the $d+1$th one to create a $d$-dimensional embedding of a network. Contrarily, just like in the case of the original, non-circular HOPE and TREXPEN algorithms, in HOPE-S and TREXPEN-S the first $d$ singular values are retained for creating a $d$-dimensional embedding and none of them is neglected.

According to our experience, the centre of mass (COM) of the resulted circular patterns may not coincide with the origin of the coordinate system of the embedding. In our experiments, we examined two settings regarding the location of the centre of mass in the methods HOPE-S, TREXPEN-S, HOPE-R and TREXPEN-R: besides the original embeddings, we always tested the more balanced version of these node arrangements too, which we obtained by shifting their centre of mass to the origin. Note that in order to not distort the topology-reflecting relations between the source and the target positions, the location of the centre of mass was calculated for the source and target node positions jointly.

\subsection{Transformation of circular Euclidean embeddings into hyperbolic node arrangements with MIC}
\label{sect:EucHypConv}
\setcounter{figure}{0}
\setcounter{table}{0}
\setcounter{equation}{0}
\renewcommand{\thefigure}{S1.5.\arabic{figure}}
\renewcommand{\thetable}{S1.5.\arabic{table}}
\renewcommand{\theequation}{S1.5.\arabic{equation}}

In Ref.~\cite{coalescentEmbedding} introducing coalescent embedding, several circular Euclidean embeddings were transformed into hyperbolic ones by replacing the Euclidean radial coordinates with new ones having the highest likelihood according to the popularity-similarity optimisation (PSO) model~\cite{PSO} of growing hyperbolic networks. The embedding method Mercator~\cite{Mercator} also mixes a Euclidean, machine learning technique (namely the Laplacian eigenmaps~\cite{LE}) with a maximum likelihood estimation process, but it optimizes the node arrangement in accordance with the hyperbolic $\mathbb{S}^1/\mathbb{H}^2$ (or RHG) model of static networks~\cite{hyperGeomBasics,S1} instead of the PSO model. These approaches start from such dimension reduction techniques that represent high connection probabilities as small Euclidean distances and utilize only the angular node arrangement obtained in the Euclidean space to create hyperbolic embeddings. Here, based on our Euclidean embedding methods HOPE-S, HOPE-R, TREXPEN-S and TREXPEN-R that
map high proximities measured along the graph to high inner products in the Euclidean space and do not restrict the range of
angular distances (and the occupied volumes in general) between the nodes like HOPE or TREXPEN, we propose MIC, a Euclidean-hyperbolic conversion of the radial coordinates that does not assume any specific hyperbolic network model as the generator of the network to be embedded. Moreover, MIC is capable of taking into account the possible interdependencies between the radial and the angular coordinates in the Euclidean embedding by not disregarding the Euclidean radial coordinates that emerged from the dimension reduction. Thereby, especially when the angular node distribution yielded by the Euclidean embedding is strongly inhomogeneous, our transformation eases the concept that in the hyperbolic space, the radial position of a node determines its overall attractivity -- i.e., its popularity or degree -- in itself.

In MIC, we used the native ball representation~\cite{hyperGeomBasics} of the $d$-dimensional hyperbolic space of curvature $K<0$, in which the hyperbolic distance $x_{s\rightarrow t}$ between the source position of node $s$ and the target position of node $t$ fulfills the hyperbolic law of cosines written as
\begin{equation}
    \mathrm{cosh}(\zeta x_{s\rightarrow t})=\mathrm{cosh}(\zeta r_s^{\mathrm{source}})\,\mathrm{cosh}(\zeta r_t^{\mathrm{target}})-\mathrm{sinh}(\zeta r_s^{\mathrm{source}})\,\mathrm{sinh}(\zeta r_t^{\mathrm{target}})\,\mathrm{cos}(\theta_{s\rightarrow t}),
    \label{eq:hypDist}
\end{equation}
where $\zeta=\sqrt{-K}$, $r_s^{\mathrm{source}}\equiv\|\underline{s}_s\|$ and $r_t^{\mathrm{target}}\equiv\|\underline{t}_t\|$ denote the Euclidean norms of the position vectors $\underline{s}_s$ and $\underline{t}_t$, while ${\theta_{s\rightarrow t}=\mathrm{arccos}(\frac{\underline{s}_s\cdot\underline{t}_t}{\|\underline{s}_s\|\,\|\underline{t}_t\|})}$ is the angular distance between the examined points. Using the Cartesian coordinate matrices $\bm{S}$ and $\bm{T}$ given by Eqs.~(\ref{eq:sourceCoordMatrix}) and (\ref{eq:targetCoordMatrix}), the source position vector $\underline{s}_s$ and the target position vector $\underline{t}_t$ can be written as $\mathrm{row}_s\bm{S}=(S_{s1},S_{s2},...,S_{sd})$ and
$\mathrm{row}_t\bm{T}=(T_{t1},T_{t2},...,T_{td})$, respectively, and thus, the radial coordinates can be expressed as $r_s^{\mathrm{source}}=\sqrt{\sum_{j=1}^d (S_{sj})^2}$ and $r_t^{\mathrm{target}}=\sqrt{\sum_{j=1}^d (T_{tj})^2}$, while the scalar product $\underline{s}_s\cdot\underline{t}_t$ can be calculated as $\sum_{j=1}^d S_{sj}\cdot T_{tj}$.
For $r_s^{\mathrm{source}}=0$ simply $x_{s\rightarrow t}=r_t^{\mathrm{target}}$, and if $r_t^{\mathrm{target}}=0$ then $x_{s\rightarrow t}=r_s^{\mathrm{source}}$. In the case of $\theta_{s\rightarrow t}=0$, $x_{s\rightarrow t}=|r_s^{\mathrm{source}}-r_t^{\mathrm{target}}|$, while for $\theta_{s\rightarrow t}=\pi$, $x_{s\rightarrow t}=r_s^{\mathrm{source}}+r_t^{\mathrm{target}}$.

According to Ref.~\cite{hyperGeomBasics}, for sufficiently large values of $\zeta r_s^{\mathrm{source}}$ and $\zeta r_t^{\mathrm{target}}$ with an angular distance $\theta_{s\rightarrow t}$ larger than ${2\cdot\sqrt{e^{-2\zeta r_s^{\mathrm{source}}}+e^{-2\zeta r_t^{\mathrm{target}}}}}$ but small enough to replace $\sin(\theta_{s\rightarrow t}/2)$ with $\theta_{s\rightarrow t}/2$, the hyperbolic distance between node $s$ as source and node $t$ as target can be approximated as
\begin{equation}
    x_{s\rightarrow t} {\color{black}\,\,\approx r_s^{\mathrm{source}}+r_t^{\mathrm{target}}+\frac{2}{\zeta}\cdot\ln\left(\sin{\left(\frac{\theta_{s\rightarrow t}}{2}\right)}\right)}\approx r_s^{\mathrm{source}}+r_t^{\mathrm{target}}+\frac{2}{\zeta}\cdot\ln\left(\frac{\theta_{s\rightarrow t}}{2}\right),
    \label{eq:hypDistApprox}
\end{equation}
showing that smaller hyperbolic distances -- the indicators of higher connection probabilities -- basically originate from small radial coordinates and/or small angular distances. Note that in the measurements presented in this study, we always calculated the hyperbolic distances according to the exact formula given by Eq.~(\ref{eq:hypDist}) and did not rely on the approximations of Eq.~(\ref{eq:hypDistApprox}).

As described in the previous sections, HOPE and TREXPEN create such Euclidean node arrangements, in which the proximities measured along the graph to be embedded are reflected by the inner products between the nodes' position vectors, meaning that larger inner products can be interpreted as higher connection probabilities. Considering the formula
\begin{equation}
    \underline{s}_s \cdot \underline{t}_t = r_s^{\mathrm{source}}\cdot r_t^{\mathrm{target}}\cdot\cos(\theta_{s\rightarrow t})
    \label{eq:innProdVSangDist&r}
\end{equation}
of the scalar product between the Euclidean source position vector $\underline{s}_s$ of node $s$ and the target position vector $\underline{t}_t$ of node $t$, expressed with the radial coordinates (Euclidean norms) $r_s^{\mathrm{source}}$, $r_t^{\mathrm{target}}$ and the angular distance $\theta_{s\rightarrow t}$, it is easy to see that from the point of view of higher inner products, larger radial coordinates and smaller angular distances are preferable. Since the pursuit for a large inner product in a Euclidean embedding and a small hyperbolic distance in a hyperbolic embedding implies the same type of preference regarding the angular distances (namely, the smaller the more attractive), we transfer the angular node arrangement of circular Euclidean embeddings to the corresponding hyperbolic embeddings without modification. However, while in the inner product-based Euclidean embeddings the larger radial coordinates are more attractive than the smaller ones, in the hyperbolic case -- according to Eq.~(\ref{eq:hypDistApprox}) -- the smaller radial coordinates are those that have higher attractivity. Therefore, the radial coordinates obviously have to be subjected to a transformation in order to turn a Euclidean embedding into a hyperbolic one. Note that in such Euclidean embeddings in which the small topological distances of a network are mapped (primarily) to small Euclidean distances instead of large inner products, the contribution of the radial and the angular node positions to the geometric relations could not be well separated, meaning that in those cases it would be less straightforward to handle the radial and the angular coordinates in a disjoint manner during the Euclidean-hyperbolic conversion. 

The main idea behind our Euclidean-hyperbolic conversion MIC is that since the Euclidean and the hyperbolic radial arrangements of the same network aim at representing the same attractivity relations, thus, by converting these radial arrangements to the same space, we must arrive from both a Euclidean and a hyperbolic embedding to such node arrangements that are easily reconcilable to each other. Similarly to the principle known as \textit{universality of the uniform} stating that data from any continuous probability distribution can be converted to a sample that follows the simplest, uniform probability distribution, we assume that the radial positions in an embedding of any geometry can be converted to a node arrangement on a half-line (\textit{universality of the half-line}). I.e., as the pass-through between the polynomially expanding Euclidean and the exponentially expanding hyperbolic spaces, we used the linearly expanding half-line. We calculated the radial coordinates on this half-line as
\begin{equation}
    r_{\mathrm{line}}(r_{\mathrm{Euc}}) = \frac{\pi^{\frac{d}{2}}}{\Gamma(\frac{d}{2}+1)}\cdot r_{\mathrm{Euc}}^d
    \label{eq:convertEucRtoLine}
\end{equation}
from radial coordinates given in the $d$-dimensional Euclidean space and as
\begin{equation}
    r_{\mathrm{line}}(r_{\mathrm{hyp}}) = \frac{e^{\zeta\cdot(d-1)\cdot r_{\mathrm{hyp}}}-1}{\zeta\cdot(d-1)\cdot2^{d-1}}
    \label{eq:convertHypRtoLine}
\end{equation}
from radial coordinates given in the native representation of the $d$-dimensional hyperbolic space of curvature $K=-\zeta^2$, since using these formulas, the volume $V_d^{\mathrm{Euc}}(r_{\mathrm{Euc}})=\frac{\pi^{\frac{d}{2}}}{\Gamma(\frac{d}{2}+1)}\cdot r_{\mathrm{Euc}}^d$ and (for not too small values of the radius $r_{\mathrm{hyp}}$~\cite{RHG_d_dim_mathematics}) ${V_d^{\mathrm{hyp}}=\frac{e^{\zeta\cdot(d-1)\cdot r_{\mathrm{hyp}}}-1}{\zeta\cdot(d-1)\cdot2^{d-1}}}$ of $d$-dimensional Euclidean and hyperbolic balls become linear functions $V_d^{\mathrm{E}}=r_{\mathrm{line}}(r_{\mathrm{Euc}})$ and $V_d^{\mathrm{h}}=r_{\mathrm{line}}(r_{\mathrm{hyp}})$ of the transformed radial coordinates. 
Note that when considering these volume formulas for the radial conversion, it is implicitly assumed that we deal with such embeddings that utilize relatively well how the given geometrical space expands by occupying a relatively large portion of a Euclidean or hyperbolic sphere.


In the case of a Euclidean embedding, the most attractive radial coordinate is the largest one, $r_{\mathrm{Euc,max}}$, while the most attractive radial coordinate in a hyperbolic embedding is the smallest one, $r_{\mathrm{hyp,min}}$. The equivalent of these coordinates are the positions $r_{\mathrm{line}}(r_{\mathrm{Euc,max}})$ and $r_{\mathrm{line}}(r_{\mathrm{hyp,min}})$ on the half-line. Since the positions on the half-line obtained from the two types of embeddings grasp the same attractivity relations as both the Euclidean and the hyperbolic embeddings of the given network, we expect 
that both node arrangements obtained on the half-line reflect the same radial attractivity of any node $i$ compared to the highest one. 
Thus, we can write 
\begin{equation}
    \frac{r_{\mathrm{line}}(r_{\mathrm{Euc,max}})}{r_{\mathrm{line}}(r_{\mathrm{Euc,}i})} = \frac{r_{\mathrm{line}}(r_{\mathrm{hyp,}i})}{r_{\mathrm{line}}(r_{\mathrm{hyp,min}})},
    \label{eq:radAttrRatiosEquivalence}
\end{equation}
i.e., according to Eqs.~(\ref{eq:convertEucRtoLine}) and (\ref{eq:convertHypRtoLine}),
\begin{equation}
    \frac{r_{\mathrm{Euc,max}}^d}{r_{\mathrm{Euc,}i}^d} = \frac{e^{\zeta\cdot(d-1)\cdot r_{\mathrm{hyp},i}}-1}{e^{\zeta\cdot(d-1)\cdot r_{\mathrm{hyp,min}}}-1},
    \label{eq:radAttrRatiosEquivalence_final}
\end{equation}
where both sides are larger than or equal to $1$. Note that it is assumed here that the occurring smallest radial coordinates $r_{\mathrm{Euc,min}}$ and $r_{\mathrm{hyp,min}}$ are larger than $0$, providing that each node has in both the Euclidean and the hyperbolic embedding a definite angular position that can be retained when switching between the different geometries. If, however, due to numerical errors we obtain $r_{\mathrm{Euc,min}}=0$ from a Euclidean embedding, we first execute the Euclidean-hyperbolic conversion only for the nodes of nonzero $r_{\mathrm{Euc,}i}$ values, and then we place the nodes that fell in the origin in the Euclidean embedding to an extremely large radial coordinate (namely, 10 times larger than the largest hyperbolic radial coordinate achieved among the nodes that were not in the origin in the Euclidean case) and a random angular position in the hyperbolic space.

To enable the calculation of hyperbolic radial coordinates $r_{\mathrm{hyp}}$ from the radial coordinates $r_{\mathrm{Euc}}$ of a Euclidean embedding using the equivalence of the ratios written in Eq.~(\ref{eq:radAttrRatiosEquivalence_final}), we have to choose one node to which we assign a hyperbolic radial position. Since this choice inherently determines the extent of the hyperbolic layout, it is reasonable to treat the outermost position as an adjustable parameter. In order to make the majority of the hyperbolic radial coordinates large enough to ensure the validity of the approximating formula of the hyperbolic distance given by Eq.~(\ref{eq:hypDistApprox}) -- based on which we assumed that in the case of the hyperbolic distance, the contribution of the radial and the angular node positions can be similarly well separated from each other as in the case of the inner product in the Euclidean embeddings, meaning that the radial coordinates can be transformed independently of the angular coordinates during the Euclidean-hyperbolic conversion~--, 
we used the formula
\begin{equation}
    r_{\mathrm{hyp,max}} = \frac{C}{\zeta}\cdot\ln(N)
    \label{eq:fixedHypMaxR}
\end{equation}
with $C=2$ for calculating the largest hyperbolic radial coordinate of both the source and the target representation of the nodes. 
With this choice, the volume of the hyperbolic ball within which we distribute the nodes will scale as $V_d^{\mathrm{hyp}}=\frac{e^{\zeta\cdot(d-1)\cdot r_{\mathrm{hyp,max}}}-1}{\zeta\cdot(d-1)\cdot2^{d-1}}=\\=\frac{N^{C\cdot(d-1)}-1}{\zeta\cdot(d-1)\cdot2^{d-1}}\sim N^{C\cdot(d-1)}$ with the number of nodes $N$ of the embedded graph.

Once $r_{\mathrm{hyp,max}}$ is fixed, the most attractive hyperbolic radial coordinate $r_{\mathrm{hyp,min}}$ can be determined by applying Eq.~(\ref{eq:radAttrRatiosEquivalence_final}) to the least attractive radial positions $r_{\mathrm{Euc,min}}$ and $r_{\mathrm{hyp,max}}$ as
\begin{equation}
    \frac{r_{\mathrm{Euc,max}}^d}{r_{\mathrm{Euc,min}}^d} = \frac{e^{\zeta\cdot(d-1)\cdot r_{\mathrm{hyp,max}}}-1}{e^{\zeta\cdot(d-1)\cdot r_{\mathrm{hyp,min}}}-1}.
    \label{eq:radAttrRatiosEquivalence_extremums}
\end{equation}
Using Eq.~(\ref{eq:fixedHypMaxR}), the smallest hyperbolic radial coordinate can be expressed from Eq.~(\ref{eq:radAttrRatiosEquivalence_extremums}) as
\begin{equation}
    r_{\mathrm{hyp,min}}=\frac{1}{\zeta\cdot(d-1)}\cdot\ln\left(1+[e^{\zeta\cdot(d-1)\cdot r_{\mathrm{hyp,max}}}-1]\cdot\left[\frac{r_{\mathrm{Euc,min}}}{r_{\mathrm{Euc,max}}}\right]^d\right)=\frac{1}{\zeta\cdot(d-1)}\cdot\ln\left(1+[N^{C\cdot(d-1)}-1]\cdot\left[\frac{r_{\mathrm{Euc,min}}}{r_{\mathrm{Euc,max}}}\right]^d\right).
    \label{eq:rHypMin}
\end{equation}
Finally, knowing the formula of $r_{\mathrm{hyp,min}}$, the radial coordinate of any node $i$ can be converted from its Euclidean value $r_{\mathrm{Euc,}i}$ to its hyperbolic equivalent $r_{\mathrm{hyp,}i}$ based on Eq.~(\ref{eq:radAttrRatiosEquivalence_final}). After some rearrangement and the substitution of Eq.~(\ref{eq:rHypMin}), we arrive from Eq.~(\ref{eq:radAttrRatiosEquivalence_final}) at the conversion formula
\begin{equation}
    r_{\mathrm{hyp,}i}(r_{\mathrm{Euc,}i})=\frac{1}{\zeta\cdot(d-1)}\cdot\ln\left(1+[e^{\zeta\cdot(d-1)\cdot r_{\mathrm{hyp,min}}}-1]\cdot\left[\frac{r_{\mathrm{Euc,max}}}{r_{\mathrm{Euc,}i}}\right]^d\right)=\frac{1}{\zeta\cdot(d-1)}\cdot\ln\left(1+[N^{C\cdot(d-1)}-1]\cdot\left[\frac{r_{\mathrm{Euc,min}}}{r_{\mathrm{Euc,}i}}\right]^d\right).
    \label{eq:rHyp(rEuc)}
\end{equation}
Naturally, these formulas fulfil the expectations that $r_{\mathrm{hyp}}(r_{\mathrm{Euc,max}})=r_{\mathrm{hyp,min}}$ and $r_{\mathrm{hyp}}(r_{\mathrm{Euc,min}})=r_{\mathrm{hyp,max}}$. It is important to notice that although we set the occurring largest source and target radial coordinate to the same value $r_{\mathrm{hyp,max}}$ in the hyperbolic embedding, because of the possible differences in the radial coordinates of the source and target radial coordinates yielded by the Euclidean embedding, the hyperbolic radial coordinates other than the largest one are not restricted to be the same for the nodes' source and target representation. Thus, the hyperbolic arrangements obtained using a single value of $r_{\mathrm{hyp,max}}$ can still have the ability to capture the differences between the radial attractivity relations of the nodes as sources and as targets.

\subsection{TREXPIC: TRansformation of EXponential shortest Path lengths to hyperbolIC measures}
\label{sect:TREXPIC}
\setcounter{figure}{0}
\setcounter{table}{0}
\setcounter{equation}{0}
\renewcommand{\thefigure}{S1.6.\arabic{figure}}
\renewcommand{\thetable}{S1.6.\arabic{table}}
\renewcommand{\theequation}{S1.6.\arabic{equation}}


When creating a hyperbolic embedding with the conversion of a Euclidean node arrangement, it is implicitly assumed that the given Euclidean embedding was already able to capture the network topology relatively well. This dependence on the Euclidean geometry can be eliminated from the hyperbolic embedding process if the network nodes are mapped directly into the hyperbolic space. A dimension reduction technique that realizes such a hyperbolic mapping of undirected networks was proposed recently under the name hydra (hyperbolic distance recovery and approximation)~\cite{Hydra}. Here, inspired by the algorithm of hydra, we introduce a new hyperbolic embedding method, TREXPIC, that can be used on directed networks too.

According to Ref.~\cite{Hydra}, in the hyperboloid model of the $d$-dimensional hyperbolic geometry, the hyperbolic distance $x(\underline{y},\underline{z})$ between the positions denoted by $\underline{y}\in\mathbb{R}^{d+1}$ and $\underline{z}\in\mathbb{R}^{d+1}$ can be calculated from the Lorentz product
\begin{equation}
    \underline{y}\circ\underline{z} = y_1z_1-(y_2z_2+y_3z_3+...+y_{d+1}z_{d+1})
    \label{eq:LorentzProductDefinition}
\end{equation}
of the two position vectors as
\begin{equation}
    x(\underline{y},\underline{z})=\frac{1}{\zeta}\cdot\mathrm{acosh}(\underline{y}\circ\underline{z}),
    \label{eq:hypDistOnHyperboloid}
\end{equation}
where $\zeta$ characterises the curvature $K<0$ of the hyperbolic space represented by the hyperboloid, namely $\zeta=\sqrt{-K}$. Therefore, given a hyperbolic distance $x$, $\cosh(\zeta\cdot x)$ is the corresponding Lorentz product measured in the hyperboloid model. Thus, given a distance (or dissimilarity) matrix $\bm{D}$ -- derived from the network topology -- consisting of elements that are interpreted as the estimations of the pairwise hyperbolic distances between the network nodes, the elements of the matrix
\begin{equation}
    \bm{L}=\cosh(\zeta\cdot\bm{D})
    \label{eq:LmatrixFromD}
\end{equation}
can be identified as Lorentz products (measured in the hyperboloid model) that are expected to be reproduced by the embedding. 

In hydra~\cite{Hydra}, the distance matrix $\bm{D}$ from which the matrix $\bm{L}$ of Lorentz products is created consists of the pairwise shortest path lengths of the graph to be embedded. However, in order to enable TREXPIC to embed also weakly connected networks besides strongly connected ones, infinite distances -- that indicate the absence of any paths from a given node to an other -- have to be converted to finite values. Therefore, we measured the distance of node $t$ from node $s$ along the graph as
\begin{equation}
    D_{st}=e^{-\frac{q}{\mathrm{SPL}_{s\rightarrow t}}},
    \label{eq:ourFiniteDistMatrixElements}
\end{equation}
where $\mathrm{SPL}_{s\rightarrow t}$ is the hop-length of the shortest path along which one can reach node $t$ from node $s$ and $q>0$ is an adjustable parameter that controls how fast our distance measure increases towards the larger shortest path lengths. The distances calculated from Eq.~(\ref{eq:ourFiniteDistMatrixElements}) fall in the interval $[0,1]$, where $0$ is the distance of each node from itself (occurring in the diagonal of $\bm{D}$ and the value $1$ corresponds to $\mathrm{SPL}_{s\rightarrow t}=\infty$. For small enough values of the multiplying factor $q$, $e^{-\frac{q}{\mathrm{SPL}_{s\rightarrow t}}}\approx 1-q/\mathrm{SPL}_{s\rightarrow t}$, and the increase in $q$ shifts the non-unit off-diagonal matrix elements towards $0$. In our investigations, we always tested $15$ settings of the $q$ parameter of TREXPIC, which we sampled equidistantly on a logarithmic scale between $q_{\mathrm{min}}=\ln(1.0/0.9999)\cdot\mathrm{SPL}_{\mathrm{max}}$ and $q_{\mathrm{max}}=\ln(10)\cdot\mathrm{SPL}_{\mathrm{max}}$, where $\mathrm{SPL}_{\mathrm{max}}$ denotes the largest finite shortest path length measured along the graph in question.

Based on Eq.~(\ref{eq:LorentzProductDefinition}), the matrix of the pairwise Lorentz products between the source and target position vectors given by the rows of the matrixes $\bm{S}$ and $\bm{T}$ can be written as
\begin{equation}
    \bm{L} = \bm{S}\cdot\bm{J}\cdot\bm{T}^T,
    \label{eq:LorentzProductOfCoordinateMatrixes}
\end{equation}
where $\bm{J}$ is the diagonal matrix of size $(d+1)\times(d+1)$ containing the values $+1,-1,-1,...,-1$ in the diagonal. Accordingly, given the singular value decomposition
\begin{equation}
    \bm{L} = \bm{U}\cdot\bm{\Sigma}\cdot\bm{V}^T,
    \label{eq:SVDofLorentzProductMatrix}
\end{equation}
one proper definition of such matrixes of position vectors in the $d$-dimensional hyperboloid that well approximate the Lorentz products in $\bm{L}$ is provided by the formulas
\begin{equation}
    \bm{S} = [+\sqrt{\sigma_1}\cdot\underline{u}_1\,,\,-\sqrt{\sigma_2}\cdot\underline{u}_2\,,\,-\sqrt{\sigma_3}\cdot\underline{u}_3\,,\,...\,,\,-\sqrt{\sigma_{d+1}}\cdot\underline{u}_{d+1}]
    \label{eq:sourceCoordMatrix_LorentzProduct}
\end{equation}
and
\begin{equation}
    \bm{T} = [+\sqrt{\sigma_1}\cdot\underline{v}_1\,,\,+\sqrt{\sigma_2}\cdot\underline{v}_2\,,\,+\sqrt{\sigma_3}\cdot\underline{v}_3\,,\,...\,,\,+\sqrt{\sigma_{d+1}}\cdot\underline{v}_{d+1}],
    \label{eq:targetCoordMatrix_LorentzProduct}
\end{equation}
with $\underline{u}_j$ and $\underline{v}_j$ denoting the $j$th column of $\bm{U}$ and $\bm{V}$, i.e., the $j$th one of the left and the right singular vectors, respectively, and $\sigma_i=\Sigma_{jj}$ is the $j$th value in the descending order of the singular values. As in the above-described methods, these coordinate matrixes can contain meaningless rows that correspond to nodes with $0$ out- or in-degree, but we remove these rows from $\bm{S}$ and $\bm{T}$ at this point. To select the upper sheet of the two-sheet hyperboloid, we always use the non-negative version of the leading singular vectors $\underline{u}_1$ and $\underline{v}_1$, the existence of which is ensured by the Perron–Frobenius theorem, stating for the non-negative matrixes $\bm{L}\cdot\bm{L}^T$ and $\bm{L}^T\cdot\bm{L}$ that their leading eigenvector -- corresponding to $\underline{u}_1$ and $\underline{v}_1$, respectively -- can be chosen to have only strictly positive components. Note that instead of the singular value decomposition, hydra calculates the eigendecomposition $\bm{L}=\bm{Q}\cdot\bm{\Lambda}\cdot\bm{Q}^{-1}$, where the diagonal matrix $\Lambda$ contains both positive and negative eigenvalues, while in SVD all the singular values in $\Sigma$ are always non-negative. The reason behind our preference toward the SVD contrary to the eigendecomposition is that in the SVD the connection between the first and the third terms of the product is not restricted as much as in the case of the eigendecomposition, where the diagonal matrix $\Lambda$ is multiplied by a matrix and its inverse, and thus, SVD gives more freedom for the emergence of differences between the source and the target position vectors.

In the case of undirected networks with a symmetric $\bm{L}$ yielding $\bm{U}=\bm{V}$, in order to obtain only one position for each node, we set $\bm{S}=\bm{T}$ by splitting the $-1$ multiplying factors -- that we introduced in Eq.~(\ref{eq:sourceCoordMatrix_LorentzProduct}) to fulfill Eq.~(\ref{eq:LorentzProductOfCoordinateMatrixes}) -- equally between the source and the target coordinate matrixes, redefining them as
\begin{equation}
    \bm{S} = [+\sqrt{\sigma_1}\cdot\underline{u}_1\,,\,\mathrm{i}\cdot\sqrt{\sigma_2}\cdot\underline{u}_2\,,\,\mathrm{i}\cdot\sqrt{\sigma_3}\cdot\underline{u}_3\,,\,...\,,\,\mathrm{i}\cdot\sqrt{\sigma_{d+1}}\cdot\underline{u}_{d+1}] = [+\sqrt{\sigma_1}\cdot\underline{v}_1\,,\,\mathrm{i}\cdot\sqrt{\sigma_2}\cdot\underline{v}_2\,,\,\mathrm{i}\cdot\sqrt{\sigma_3}\cdot\underline{v}_3\,,\,...\,,\,\mathrm{i}\cdot\sqrt{\sigma_{d+1}}\cdot\underline{v}_{d+1}] = \bm{T},
    \label{eq:symmCoordMatrix_LorentzProduct}
\end{equation}
where $\mathrm{i}=\sqrt{-1}$ stands for the imaginary unit. 

To obtain embeddings in the native representation of the $d$-dimensional hyperbolic space, we map the coordinates given in the hyperboloid by Eqs.~(\ref{eq:sourceCoordMatrix_LorentzProduct})--(\ref{eq:symmCoordMatrix_LorentzProduct}) to coordinates in the $d$-dimensional ball that represents the hyperbolic space. The position vector corresponding to the native ball's origin is $\underline{o}\equiv(o_1,o_2,o_3,...,o_{d+1})=(1,0,0,...,0)$ in the hyperboloid. According to the distance formula given by Eq.~(\ref{eq:hypDistOnHyperboloid}), the hyperbolic distance between any position $\underline{y}\in\mathbb{R}^{d+1}$ and $\underline{o}$ can be calculated as $\frac{1}{\zeta}\cdot\mathrm{acosh}(\underline{y}\circ\underline{o})=\frac{1}{\zeta}\cdot\mathrm{acosh}(y_1\cdot 1-(y_2\cdot 0+y_3\cdot 0+...+y_{d+1}\cdot 0))=\frac{1}{\zeta}\cdot\mathrm{acosh}(y_1)$ in the hyperboloid. On the other hand, the hyperbolic distance between any point and the origin in the native ball is equal to the radial coordinate of the given point (i.e., the Euclidean norm of its position vector) -- see Eq.~(\ref{eq:hypDist}) at $r_{\mathrm{origin}}=0$. Consequently, to preserve the hyperbolic distance of any node from the origin, the source and the target radial coordinates of node $i$ can be calculated in the native ball from its first coordinates $S_{i1}$ and $T_{i1}$ given in the hyperboloid as 
\begin{equation}
    r_{\mathrm{native,}i}^{\mathrm{source}}=\frac{1}{\zeta}\cdot\mathrm{acosh}(S_{i1})
    \label{eq:rNativeFromrHyperboloid_source}
\end{equation}
and
\begin{equation}
    r_{\mathrm{native,}i}^{\mathrm{target}}=\frac{1}{\zeta}\cdot\mathrm{acosh}(T_{i1}),
    \label{eq:rNativeFromrHyperboloid_target}
\end{equation}
respectively. The direction of the corresponding position vector in the native ball is specified by the hyperboloid coordinates from the second to the $(d+1)$th one. Thus, the source and target position vectors that are given in the hyperboloid by the $i$th row of the matrixes $\bm{S}$ and $\bm{T}$ can be written as 
\begin{equation}
    r_{\mathrm{native},i}^{\mathrm{source}}\cdot\frac{(S_{i2},S_{i3},...,S_{i(d+1)})}{\|(S_{i2},S_{i3},...,S_{i(d+1)})\|} = \frac{\mathrm{acosh}(S_{i1})}{\zeta}\cdot\frac{(S_{i2},S_{i3},...,S_{i(d+1)})}{\sqrt{S_{i2}^2+S_{i3}^2+...+S_{i(d+1)}^2}}
    \label{eq:sourcePosVectorInNative}
\end{equation}
and
\begin{equation}
    r_{\mathrm{native},i}^{\mathrm{target}}\cdot\frac{(T_{i2},T_{i3},...,T_{i(d+1)})}{\|(T_{i2},T_{i3},...,T_{i(d+1)})\|} = \frac{\mathrm{acosh}(T_{i1})}{\zeta}\cdot\frac{(T_{i2},T_{i3},...,T_{i(d+1)})}{\sqrt{T_{i2}^2+T_{i3}^2+...+T_{i(d+1)}^2}}
    \label{eq:targetPosVectorInNative}
\end{equation}
in the native ball, respectively. Note that it is assumed here that in the hyperboloid the occurring smallest first coordinates $\min\limits_{1\leq i\leq N}S_{i1}$ and $\min\limits_{1\leq i\leq N}T_{i1}$ are not smaller than $1$. If, however, due to numerical errors this condition gets violated, we simply set the problematic native radial coordinates to $\mathrm{acosh}(1)/\zeta=0$. It is also important to remark that if all the coordinates from the second to the $d+1$th one are purely imaginary in the hyperboloid, the direction described by them is the same as if they all would be real numbers, meaning that the imaginary multiplying factors in Eq.~(\ref{eq:symmCoordMatrix_LorentzProduct}) do not raise any issues.



\newpage
\section{Euclidean-hyperbolic conversion of embeddings of synthetic hyperbolic networks generated by the popularity-similarity optimisation model}
\label{sect:embOfPSO}
\setcounter{figure}{0}
\setcounter{table}{0}
\setcounter{equation}{0}
\renewcommand{\thefigure}{S2.\arabic{figure}}
\renewcommand{\thetable}{S2.\arabic{table}}
\renewcommand{\theequation}{S2.\arabic{equation}}

The popularity-similarity optimisation (PSO) model~\cite{PSO} generates growing undirected networks in the native representation~\cite{hyperGeomBasics} of the hyperbolic plane, associating the radial coordinate of a node with its popularity and interpreting the angular proximity of two nodes as their similarity. The model is able to create graphs that are small-world, scale-free (having a degree distribution that decays as $\pazocal{P}(k)\sim k^{-\gamma}$) and, in the meantime, highly clustered too, which are all known as fundamental common features of real networks. Besides, these properties can even be accompanied by an automatically emerging, strong community structure~\cite{our_hyp_coms}. In the basic definition of the PSO model, each node connects at its appearance to $m\in\mathbb{Z^+}$ number of previously appeared nodes, making the average internal degree of the subgraphs given by the nodes of larger degree than a certain threshold independent of the value of the threshold, which might be unrealistic~\cite{Mercator}. However, by generalising the model to also simulate the creation of so-called internal links that form between two nodes later than their appearance time~\cite{PSO,HyperMap} or the deletion of previously emerged links~\cite{ourEmbedding}, the above-mentioned average degrees can be turned into an increasing or a decreasing function of the degree threshold, respectively~\cite{ourEmbedding}. The model that simulates the change $L\in\mathbb{Z}$ in the number of connections between the older nodes per time step by defining the number of \textit{external} links (that connect a node at its appearance to the previously appeared ones) as a proper function of the appearance time is called E-PSO model~\cite{HyperMap,ourEmbedding}, which gives back the original PSO model~\cite{PSO} at $L=0$. The algorithm of the E-PSO model is described in detail in Sect.~\ref{sect:undir2dEPSOmodel}.

Due to its well-adjustable and realistic features, the E-PSO model of $L\geq 0$ is often used for creating hyperbolic embeddings of networks: in Ref.~\cite{HyperMap} the whole node arrangement is determined with the aim of maximizing the probability that it would emerge from the E-PSO model with the given adjacency matrix, while in Refs.~\cite{Alanis-Lobato_LE_embedding,coalescentEmbedding} a dimension reduction technique is applied, from which only the angular node arrangement is retained and the obtained Euclidean radial coordinates are simply replaced with those that are the most probable according to the E-PSO model. Based on the E-PSO model of $L\geq0$, the most probable radial coordinate in the native representation of the hyperbolic plane of curvature $K=-\zeta^2$ can be written for the node that is the $\ell$th in the decreasing order of node degrees in the simple form of $r_{\ell}=\beta\cdot\frac{2}{\zeta}\cdot\ln\ell+(1-\beta)\cdot\frac{2}{\zeta}\cdot\ln N$, where $N$ is the total number of nodes in the network and we only need to calculate $\beta$ from the degree decay exponent $\gamma$ (which can be determined by fitting a power-law $\pazocal{P}(k)\sim k^{-\gamma}$ to the tail of the degree distribution of the network to be embedded) as $\beta=1/(\gamma-1)$. Note that at $L<0$, the node degree in an E-PSO network can become an increasing function of the radial coordinate if $|L|$ is large enough compared to $m$ and $\beta$ is small enough, meaning that when a negative value of $L$ is assumed for a given network to be embedded, then the decreasing order of the node degrees does not necessarily correspond to the most probable radial order of the nodes, and thus, these cases should be taken with caution. However, decreasing the value of the popularity fading parameter $\beta$ not only enables the negative $L$ values to affect the relationship between the order of the radial coordinates and the node degrees but simultaneously decreases the differences between the radial coordinates, mitigating the impact of the choice of the radial node order. It is also important to notice that this method always assigns different radial attractivity to all the nodes, even those having the same degree, and the radial order between such nodes can be set arbitrarily, meaning that deciding it with a random choice, the PSO-based hyperbolic embedding methods are not completely deterministic. 

To enable the conversion of higher-dimensional Euclidean embeddings into hyperbolic ones, one can use the $d$PSO model~\cite{dPSO} that is the extension of the original, two-dimensional PSO model~\cite{PSO} to any number of dimensions $2\leq d\in\mathbb{Z}^+$. Regarding the radial arrangement of the nodes based on the $d$PSO model, two approaches were proposed in Ref.~\cite{dPSO} that are both rational: we can either connect the popularity fading parameter $\beta$ to the degree decay exponent $\gamma$ as
\begin{equation}
    \beta=\frac{1}{\gamma-1}
    \label{eq:betaInPSOembedding1}
\end{equation}
and calculate the radial coordinate of the node having the $\ell$th ($\ell=1,2,...,N$) largest degree (with ties in the order of node degrees broken arbitrarily) as
\begin{equation}
    r_{\ell}=\beta\cdot\frac{2}{\zeta\cdot(d-1)}\cdot\ln{\ell}+(1-\beta)\cdot\frac{2}{\zeta\cdot(d-1)}\cdot\ln{N},
    \label{eq:radCoordInPSOembedding1}
\end{equation}
or use the formulas
\begin{equation}
    \beta=\frac{1}{(d-1)\cdot(\gamma-1)}
    \label{eq:betaInPSOembedding2}
\end{equation}
and
\begin{equation}
    r_{\ell}=\beta\cdot\frac{2}{\zeta}\cdot\ln{\ell}+(1-\beta)\cdot\frac{2}{\zeta}\cdot\ln{N}
    \label{eq:radCoordInPSOembedding2}
\end{equation}
together. Note that in the 2-dimensional case, the two pairs of formulas given by Eqs.~(\ref{eq:betaInPSOembedding1})--(\ref{eq:radCoordInPSOembedding1}) and Eqs.~(\ref{eq:betaInPSOembedding2})--(\ref{eq:radCoordInPSOembedding2}) become the same.

Finally, the E-PSO model of undirected networks can be easily generalised to generate directed networks, as it is described in Sect.~\ref{sect:dir2dEPSOmodel}. The PSO-based Euclidean-hyperbolic radial conversion can be simply applied to Euclidean embeddings of directed networks where two position vectors belong to each node -- one that describes its behaviour as a source and one as a target of links -- by carrying out the procedure applied for the node degrees in undirected networks twice: once using the out-degrees and their distribution (i.e., $\gamma^{\mathrm{out}}$) to obtain the source radial coordinates and once using the in-degrees and their distribution (i.e., $\gamma^{\mathrm{in}}$) to determine the radial coordinates as targets. It is important to remark that in such directed networks where each link has its backwards-pointing counterpart, i.e. when the directedness of the links is actually not relevant, the out-degree of each node is equal to its in-degree, and thus, the source and target radial coordinates can become the same (if the ranking between the nodes of same degrees is set identical), as expected.

In Sect.~\ref{sect:PSObasedVsOurConversion}, we demonstrate the efficiency of our new, model-independent Euclidean-hyperbolic conversion MIC by comparing its performance in graph reconstruction to that of the PSO-based conversion on Euclidean embeddings created by HOPE-S, HOPE-R, TREXPEN-S and TREXPEN-R for synthetic networks generated by the two-dimensional undirected E-PSO model (described in Sect.~\ref{sect:undir2dEPSOmodel}). 
An important difference between MIC and the PSO-based conversion is that although both of them transform the radial coordinates independently of the angular coordinates (that are kept unaltered), i.e. both algorithms assume that the role of the radial and the angular positions can be separated, the PSO-based conversion simply discards the Euclidean radial coordinates obtained from the dimension reduction, while MIC builds on these, and thereby, can retain the interdependencies between the radial and the angular coordinates grasped by the Euclidean embedding. According to Sect.~\ref{sect:PSObasedVsOurConversion}, considering the original, Euclidean radial coordinates can be to the great advantage of the Euclidean-hyperbolic conversion even when embedding PSO networks.

\subsection{The undirected E-PSO model}
\label{sect:undir2dEPSOmodel}
\setcounter{figure}{0}
\setcounter{table}{0}
\setcounter{equation}{0}
\renewcommand{\thefigure}{S2.1.\arabic{figure}}
\renewcommand{\thetable}{S2.1.\arabic{table}}
\renewcommand{\theequation}{S2.1.\arabic{equation}}

The E-PSO model~\cite{HyperMap,ourEmbedding} simulates the network growth in the native representation of the two-dimensional hyperbolic space. The nodes appear one by one on the native disk that represents the hyperbolic plane and each node creates connections at its appearance with the previously arrived nodes, with probabilities depending on their hyperbolic distance. The model parameters are the following:
\begin{itemize}
	\item The curvature $K<0$ of the hyperbolic plane, controlled by $\zeta=\sqrt{-K}>0$. Changing $\zeta$ corresponds to a simple rescaling of all the hyperbolic distances. Usually, $\zeta$ is simply set to $1$, yielding a curvature $K=-1$.
	\item The number of nodes $N\in\mathbb{Z}^+$ at the end of the network growth.
	\item The expected number of links $m>0$ that are assumed to emerge in each time step (after the appearance of the first $m$ nodes) indeed as external links, connecting the newly appearing node to the older nodes.
	\item The parameter $L\in\mathbb{R}$ estimates the change in the number of internal links (connections between the previously appeared nodes) in each time step. A value of $L>0$ means that the model simulates the formation of (in expected value) $L$ number of such connections per time step that are assumed to emerge actually as an internal link, while in the case of $L<0$, the model simulates the deletion of (in expected value) $|L|$ number of internal links at each time step. The average degree of the network is $\bar{k}\approx2\cdot(m+L)$. The setting $L=0$ (simulating the case when the same number of internal links are added and removed at each time step) gives back the original PSO model~\cite{PSO}. For $L>0$ (i.e. at the dominance of the creation of internal links) or $L<0$ (i.e. at the dominance of the deletion of internal links), the average internal degree of the subgraph of nodes that have larger degrees than a certain threshold value becomes an increasing or a decreasing function of the threshold, respectively.
    \item The popularity fading parameter $\beta\in(0,1]$, controlling the outward drift of the nodes on the hyperbolic plane. 
    The exponent $\gamma$ of the power-law decaying tail of the degree distribution can be expressed with the popularity fading parameter as $\gamma=1+1/\beta$.
    \item The temperature $T\in[0,1)$, controlling the average clustering of the network, where lower temperature results in a higher average clustering coefficient.
\end{itemize}

At the starting point of the network growth, the network is empty. Then, at each time step $j=1,2,...,N$ the following steps are accomplished:
\begin{enumerate}
    \item Node $j$ appears on the native disk with the radial coordinate $r_{jj} = \frac{2}{\zeta}\ln j$ and a uniformly random angular coordinate $\theta_j\in[0,2\pi)$.
    \item The previously (i.e., at time $i<j$) appeared nodes are shifted outwards on the native disk to simulate popularity fading: the radial coordinate of node $i$ is updated at time $j$ according to $r_{ij} = \beta r_{ii}+(1-\beta)r_{jj}$.
    \item The new node $j$ connects, in expected value, with $\bar{m}_j$ number of previously appeared nodes, where
    \begin{equation}
        \bar{m}_j = \left\lbrace \begin{array}{ll} 
         m+L\cdot\frac{1}{\left[1-N^{-0.5}\right]^2}\cdot\ln\left(\frac{N}{j}\right)\cdot\left[1-j^{-0.5}\right] & \mathrm{if}\;\; \beta = 0.5, \\
         
         m+L\cdot\frac{2}{\ln^2(N)}\cdot\left[\frac{N}{j}-1\right]\cdot\ln(j) & \mathrm{if}\;\; \beta = 1.0, \\
         
         m+L\cdot\frac{2\cdot(1-\beta)}{\left[1-N^{-(1-\beta)}\right]^2\cdot(2\cdot\beta -1)}\left[\left(\frac{N}{j}\right)^{2\cdot\beta-1}-1\right]\left[1-j^{-(1-\beta)}\right] & \mathrm{otherwise}. \end{array} \right.
        \label{eq:expNumOfLinksATi_EPSO}
    \end{equation}
    To avoid the emergence of isolated nodes that do not have any connections, it is expedient to continuously repeat the connection trials until exactly $\lfloor \bar{m}_j \rfloor$ or $\lceil \bar{m}_j \rceil$ number of links are created instead of make only one connection attempt with each one of the older nodes. By choosing $m_j\equiv\lceil \bar{m}_j \rceil$ over $m_j\equiv\lfloor \bar{m}_j \rfloor$ with a probability corresponding to the fractional part of $\bar{m}_j$, the number of links $m_j$ of the $j$th time step averaged over different network realisations becomes $\bar{m}_j$ as expected. If the number of nodes that appeared before node $j$ is not larger than $m_j$, then node $j$ connects to all of them. Otherwise (i.e., for $m_j+1<j$), 
    \begin{itemize}
        \item[a)] if the temperature $T$ is $0$, then node $j$ connects to the $m_j$ hyperbolically closest nodes, while
        \item[b)] at temperatures $0<T$, any older node $i=1,2,...,j-1$ gets connected to node $j$ with probability
        \begin{equation}
            p(x_{ij})=\frac{1}{1+\mathrm{e}^{\frac{\zeta}{2T}(x_{ij}-R_j)}},
            \label{eq:PSO_link_prob}
        \end{equation}
        where the hyperbolic distance $x_{ij}$ between the two nodes can be expressed from Eq.~(\ref{eq:hypDist}) and the cutoff distance $R_j$ can be calculated as
        \begin{equation}
            R_j = \left\lbrace \begin{array}{ll} 
            r_{jj}-\frac{2}{\zeta}\ln\left(\frac{2T}{\sin(T\pi)}\cdot
             \frac{1-\mathrm{e}^{-\frac{\zeta}{2}(1-\beta)r_{jj}}}{m(1-\beta)}\right) & \mathrm{if}\;\; \beta < 1, \\
             r_{jj} - \frac{2}{\zeta}\ln\left(\frac{T}{\sin(T\pi)}\cdot \frac{\zeta r_{jj}}{m}\right) & \mathrm{if} \;\; \beta=1. \end{array} \right.
             \label{eq:cutoff_EPSO}
        \end{equation}
    \end{itemize}
\end{enumerate}

\subsection{The directed version of the E-PSO model}
\label{sect:dir2dEPSOmodel}
\setcounter{figure}{0}
\setcounter{table}{0}
\setcounter{equation}{0}
\renewcommand{\thefigure}{S2.2.\arabic{figure}}
\renewcommand{\thetable}{S2.2.\arabic{table}}
\renewcommand{\theequation}{S2.2.\arabic{equation}}

The above described E-PSO model can be easily generalised to generate directed networks by adding each node to the network twice: once as a source of links and once as a target of links, creating $m_j^{\mathrm{out}}$ number of outgoing and $m_j^{\mathrm{in}}$ number of incoming links for the $j$th source and the $j$th target node, respectively. Then, the relation between the two appearance (or radial) orders of the nodes and the two sets of angular coordinates become an adjustable feature in the model. For example, 3 simple, but fundamentally different choices for the rank of the source and the target radial coordinates can be given by the followings: the two ranks are the same (meaning that node $j$ appears at time $j$ both as source and target), they are the opposite of each other (i.e., at time $j=1,2,...,N$ the node that appears as a source of links is indexed by $j$, while the node that appears as a target is indexed by $N-j+1$ and node $j$ appears as target only at time $N-j+1$) or the two ranks are not correlated (meaning that the node that appears as the $j$th one among the source representations becomes added to the network as a target of links at a randomly chosen time step). Besides, not only the order of the out- and the in-degrees but their distributions can also be adjusted independently by defining different popularity fading parameters $\beta^{\mathrm{out}}$ and $\beta^{\mathrm{in}}$ that determine the speed of the radial shift for the source and the target node positions, respectively. Regarding the relation between the source and the target angular coordinates, when assuming that each node has similar attributes (or, in other words, finds roughly the same set of nodes similar to itself) as source and target, a relatively simple but rational solution is to sample the source coordinate $\theta_j^{\mathrm{source}}$ of each node $j$ uniformly at random from $[0,2\pi)$ and sample the corresponding target angular coordinate $\theta_j^{\mathrm{target}}$ from a normal distribution centred around the angle $\theta_j^{\mathrm{source}}$, having a relatively small standard deviation (taking the modulo with $2\pi$ if the sampled angle falls out of the range $[0,2\pi)$). Note, however, that several other treatments of both the radial orders and the angular coordinates are feasible. Naturally, using $\theta_j^{\mathrm{source}}\equiv\theta_j^{\mathrm{target}}$, $m_j^{\mathrm{out}}=m_j^{\mathrm{in}}$, $\beta^{\mathrm{out}}=\beta^{\mathrm{in}}$ and the same source and target radial order, at $T=0$ yields a network that can be easily interpreted as an undirected one since it contains the reversed counterpart for each one of the links.

Just like in the undirected model, the number of links of the newly appearing nodes is so adjusted in time that the model, despite the usage of external links only, eventually simulates such a network growth, where the expected number of links that point from the newly appearing source node to the older targets is $m^{\mathrm{out}}\in\mathbb{R}^+$, the expected number of links that point from the older source representations to the new target node is $m^{\mathrm{in}}\in\mathbb{R}^+$ and the expected change in the number of internal links that connect older pairs of nodes is $L\in\mathbb{R}$ in each time step. Assuming that node $s$ appears at time $j$ as a source of links, the expected number of the links that become attached to node $s$ at its arrival can be calculated in the directed E-PSO model through the following steps:
\begin{itemize}
    \item The probability that the $j$th source node (denoted here by $j$) becomes connected at the $t$th time step ($j<t$) to a given target node that appeared at time $i<j$ (denoted here by $i$) can be expressed with the radial positions $r_{jt}^{\mathrm{source}}$ and $r_{it}^{\mathrm{target}}$ occupied at time $t$ as $\Pi(j\rightarrow i,t)=L\cdot\frac{\mathrm{exp}\left(-\frac{\zeta}{2}\cdot\left(r_{jt}^{\mathrm{source}}+r_{it}^{\mathrm{target}}\right)\right)}{\int_1^t\int_1^t{\mathrm{exp}\left(-\frac{\zeta}{2}\cdot\left(r_{jt}^{\mathrm{source}}+r_{it}^{\mathrm{target}}\right)\right)\,\mathrm{d}i\,\mathrm{d}j}}$, where the factor of $2$ in the formula of the undirected case given in Section VIII of the Supplementary Information of Ref.~\cite{PSO} is omitted due to distinguishing the node pairs $i$--$j$ and $j$--$i$ from each other.
    \item The probability that the source node $j$ that appeared at time $j$ becomes connected to a given target node $i$ that appeared at time $i<j$ during the whole network growth, i.e. by time $N$ is $\tilde{\Pi}(j\rightarrow i)=\int_j^N{\Pi(j\rightarrow i,t)\,\mathrm{d}t}$.
    \item Thus, at the end of the network growth, the expected number of internal links pointing from the source node that appeared at time $j$ to all the older target nodes $i$ (appeared at time $i<j$) can be written as $\bar{L}_j^{\mathrm{out}}=\int_1^j{\tilde{\Pi}(j\rightarrow i)\,\mathrm{d}i}$.
\end{itemize}
The expected final number $\bar{L}_j^{\mathrm{in}}$ of internal links pointing toward the target node that appeared at time $j$ can also be derived similarly. For calculating the integrals, the approximations described in Ref.~\cite{HyperMap} can be used. The resulted formulas of the expected numbers of connections $\bar{m}_j^{\mathrm{out}}=m^{\mathrm{out}}+\bar{L}_j^{\mathrm{out}}$ and $\bar{m}_j^{\mathrm{in}}=m^{\mathrm{in}}+\bar{L}_j^{\mathrm{in}}$ emerging at time $j$ in the E-PSO model are given by Eqs.~(\ref{eq:expNumOfLinksATi_dirEPSO_out}) and (\ref{eq:expNumOfLinksATi_dirEPSO_in}), respectively.\bigskip


In general, the growth of a directed E-PSO network can be realised in the $j$th time step as follows:
\begin{enumerate}
    \item Node $s$ appears on the native disk as a source of links with the radial coordinate $r^{\mathrm{source}}_{sj} = \frac{2}{\zeta}\ln j$ and a uniformly random angular coordinate $\theta^{\mathrm{source}}_s\in[0,2\pi)$. Besides, node $t$ appears on the native disk as a target of links with the radial coordinate $r^{\mathrm{target}}_{tj} = \frac{2}{\zeta}\ln j$ and a uniformly random angular coordinate $\theta^{\mathrm{target}}_t\in[0,2\pi)$.
    \item The previously (i.e., at time $i<j$) appeared nodes are shifted outwards on the native disk to simulate popularity fading: the radial coordinate of the source representation of the node $p$ that appeared at time $i$ is updated at time $j$ according to $r_{pj}^{\mathrm{source}} = \beta^{\mathrm{out}} r_{pi}^{\mathrm{source}}+(1-\beta^{\mathrm{out}})r_{sj}^{\mathrm{source}}$, while the radial coordinate of the target representation of the node $q$ that appeared at time $i$ is updated at time $j$ according to $r_{qj}^{\mathrm{target}} = \beta^{\mathrm{in}} r_{qi}^{\mathrm{target}}+(1-\beta^{\mathrm{in}})r_{tj}^{\mathrm{target}}$.
    \item The new source node $s$ connects, in expected value, to $\bar{m}_j^{\mathrm{out}}$ number of previously appeared target nodes, where
    \begin{equation}
        \bar{m}_j^{\mathrm{out}} = \left\lbrace \begin{array}{ll} 
         m^{\mathrm{out}}+L\cdot\frac{1-\beta^{\mathrm{out}}}{\left[1-N^{-(1-\beta^{\mathrm{out}})}\right]\cdot\left[1-N^{-(1-\beta^{\mathrm{in}})}\right]}\cdot\ln\left(\frac{N}{j}\right)\cdot\left[1-j^{-\left(1-\beta^{\mathrm{in}}\right)}\right] & \mathrm{if}\;\; \beta^{\mathrm{in}}+\beta^{\mathrm{out}} = 1, \\
         
         m^{\mathrm{out}}+L\cdot\frac{1-\beta^{\mathrm{out}}}{\left[1-N^{-(1-\beta^{\mathrm{out}})}\right]\cdot\ln(N)\cdot\beta^{\mathrm{out}}}\cdot\left[\left(\frac{N}{j}\right)^{\beta^{\mathrm{out}}}-1\right]\cdot\ln(j) & \mathrm{if}\;\; \beta^{\mathrm{in}} = 1,\, \beta^{\mathrm{out}}<1, \\
         
         m^{\mathrm{out}}+L\cdot\frac{1}{\ln(N)\cdot\left[1-N^{-(1-\beta^{\mathrm{in}})}\right]\cdot\beta^{\mathrm{in}}}\cdot\left[\left(\frac{N}{j}\right)^{\beta^{\mathrm{in}}}-1\right]\cdot\left[1-j^{-\left(1-\beta^{\mathrm{in}}\right)}\right] & \mathrm{if}\;\; \beta^{\mathrm{in}} < 1,\, \beta^{\mathrm{out}}=1, \\
         
         m^{\mathrm{out}}+L\cdot\frac{1}{\ln^2(N)}\cdot\left[\frac{N}{j}-1\right]\cdot\ln(j) & \mathrm{if}\;\; \beta^{\mathrm{in}} = 1,\, \beta^{\mathrm{out}} = 1, \\
         
         m^{\mathrm{out}}+L\cdot\frac{1-\beta^{\mathrm{out}}}{\left[1-N^{-(1-\beta^{\mathrm{out}})}\right]\cdot\left[1-N^{-(1-\beta^{\mathrm{in}})}\right]\cdot\left(\beta^{\mathrm{in}}+\beta^{\mathrm{out}}-1\right)}\left[\left(\frac{N}{j}\right)^{\beta^{\mathrm{in}}+\beta^{\mathrm{out}}-1}-1\right]\cdot\left[1-j^{-(1-\beta^{\mathrm{in}})}\right] & \mathrm{otherwise}. \end{array} \right.
        \label{eq:expNumOfLinksATi_dirEPSO_out}
    \end{equation}
    
    In the meantime, in expected value $\bar{m}_j^{\mathrm{in}}$ number of previously appeared source nodes connect to the new target node $t$, where
    \begin{equation}
        \bar{m}_j^{\mathrm{in}} = \left\lbrace \begin{array}{ll} 
         m^{\mathrm{in}}+L\cdot\frac{1-\beta^{\mathrm{in}}}{\left[1-N^{-(1-\beta^{\mathrm{out}})}\right]\cdot\left[1-N^{-(1-\beta^{\mathrm{in}})}\right]}\cdot\ln\left(\frac{N}{j}\right)\cdot\left[1-j^{-\left(1-\beta^{\mathrm{out}}\right)}\right] & \mathrm{if}\;\; \beta^{\mathrm{in}}+\beta^{\mathrm{out}} = 1, \\

         m^{\mathrm{in}}+L\cdot\frac{1}{\left[1-N^{-(1-\beta^{\mathrm{out}})}\right]\cdot\ln(N)\cdot\beta^{\mathrm{out}}}\cdot\left[\left(\frac{N}{j}\right)^{\beta^{\mathrm{out}}}-1\right]\cdot\left[1-j^{-\left(1-\beta^{\mathrm{out}}\right)}\right] & \mathrm{if}\;\; \beta^{\mathrm{in}}=1, \, \beta^{\mathrm{out}} < 1, \\
         
         m^{\mathrm{in}}+L\cdot\frac{1-\beta^{\mathrm{in}}}{\ln(N)\cdot\left[1-N^{-(1-\beta^{\mathrm{in}})}\right]\cdot\beta^{\mathrm{in}}}\cdot\left[\left(\frac{N}{j}\right)^{\beta^{\mathrm{in}}}-1\right]\cdot\ln(j) & \mathrm{if}\;\; \beta^{\mathrm{in}}<1, \, \beta^{\mathrm{out}} = 1, \\
         
         m^{\mathrm{in}}+L\cdot\frac{1}{\ln^2(N)}\cdot\left[\frac{N}{j}-1\right]\cdot\ln(j) & \mathrm{if}\;\; \beta^{\mathrm{in}} = 1,\, \beta^{\mathrm{out}} = 1, \\
         
         m^{\mathrm{in}}+L\cdot\frac{1-\beta^{\mathrm{in}}}{\left[1-N^{-(1-\beta^{\mathrm{out}})}\right]\cdot\left[1-N^{-(1-\beta^{\mathrm{in}})}\right]\cdot\left(\beta^{\mathrm{in}}+\beta^{\mathrm{out}}-1\right)}\left[\left(\frac{N}{j}\right)^{\beta^{\mathrm{in}}+\beta^{\mathrm{out}}-1}-1\right]\cdot\left[1-j^{-(1-\beta^{\mathrm{out}})}\right] & \mathrm{otherwise}. \end{array} \right.
        \label{eq:expNumOfLinksATi_dirEPSO_in}
    \end{equation}
    
    As in the undirected model, the emergence of isolated sources and targets can be avoided by continuously repeating the connection trials until exactly $m^{\mathrm{out}}_j=\lfloor \bar{m}^{\mathrm{out}}_j \rfloor$ or $m^{\mathrm{out}}_j=\lceil \bar{m}^{\mathrm{out}}_j \rceil$, and $m^{\mathrm{in}}_j=\lfloor \bar{m}^{\mathrm{in}}_j \rfloor$ or $m^{\mathrm{in}}_j=\lceil \bar{m}^{\mathrm{in}}_j \rceil$ number of links are created.
    
    If the number of target nodes that appeared before node $s$ is not larger than $m_j^{\mathrm{out}}$, then all of them connects to node $s$. Otherwise (i.e., for $m_j^{\mathrm{out}}+1<j$), 
    \begin{itemize}
        \item[a)] if the temperature $T$ is $0$, then the source representation of node $s$ connects to the $m_j^{\mathrm{out}}$ hyperbolically closest target nodes, while
        \item[b)] at temperatures $0<T$, the source node $s$ gets connected to any older target node $q$ (appeared at time $i=1,2,...,j-1$) with probability
        \begin{equation}
            p(x_{s\rightarrow q})=\frac{1}{1+\mathrm{e}^{\frac{\zeta}{2T}(x_{s\rightarrow q}-R_j^{\mathrm{out}})}},
            \label{eq:PSO_link_prob_newSourceNode}
        \end{equation}
        where the hyperbolic distance $x_{s\rightarrow q}$ between the two nodes can be expressed from Eq.~(\ref{eq:hypDist}) and the cutoff distance $R_j^{\mathrm{out}}$ can be calculated as 
        \begin{equation}
            R_j^{\mathrm{out}} = \left\lbrace \begin{array}{ll} 
            r_{sj}^{\mathrm{source}}-\frac{2}{\zeta}\ln\left(\frac{2T}{\sin(T\pi)}\cdot
             \frac{1-\mathrm{e}^{-\frac{\zeta}{2}(1-\beta^{\mathrm{in}})r_{sj}^{\mathrm{source}}}}{m_j^{\mathrm{out}}(1-\beta^{\mathrm{in}})}\right) & \mathrm{if}\;\; \beta^{\mathrm{in}} < 1\,\left(\mathrm{i.e.,\,if}\,\,\gamma^{\mathrm{in}}=1+\frac{1}{\beta^{\mathrm{in}}}>2\right), \\
             r_{sj}^{\mathrm{source}} - \frac{2}{\zeta}\ln\left(\frac{T}{\sin(T\pi)}\cdot \frac{\zeta r_{sj}^{\mathrm{source}}}{m_j^{\mathrm{out}}}\right) & \mathrm{if} \;\; \beta^{\mathrm{in}}=1\,\left(\mathrm{i.e.,\, if}\,\,\gamma^{\mathrm{in}}=1+\frac{1}{\beta^{\mathrm{in}}}=2\right). \end{array} \right.
             \label{eq:cutoff_dirEPSO_out}
        \end{equation}
    \end{itemize}
    If the number of source nodes that appeared before node $t$ is not larger than $m_j^{\mathrm{in}}$, then node $t$ connects to all of them. Otherwise (i.e., for $m_j^{\mathrm{in}}+1<j$), 
    \begin{itemize}
        \item[a)] if the temperature $T$ is $0$, then the $m_j^{\mathrm{in}}$ hyperbolically closest source nodes connect the target representation of node $t$, while
        \item[b)] at temperatures $0<T$, any older source node $p$ (appeared at time $i=1,2,...,j-1$) gets connected to the target node $t$ with probability
        \begin{equation}
            p(x_{p\rightarrow t})=\frac{1}{1+\mathrm{e}^{\frac{\zeta}{2T}(x_{p\rightarrow t}-R_j^{\mathrm{in}})}},
            \label{eq:PSO_link_prob_newTargetNode}
        \end{equation}
        where the hyperbolic distance $x_{p\rightarrow t}$ between the two nodes can be expressed from Eq.~(\ref{eq:hypDist}) and the cutoff distance $R_j^{\mathrm{in}}$ can be calculated as
       \begin{equation}
            R_j^{\mathrm{in}} = \left\lbrace \begin{array}{ll} 
            r_{tj}^{\mathrm{target}}-\frac{2}{\zeta}\ln\left(\frac{2T}{\sin(T\pi)}\cdot
             \frac{1-\mathrm{e}^{-\frac{\zeta}{2}(1-\beta^{\mathrm{out}})r_{tj}^{\mathrm{target}}}}{m_j^{\mathrm{in}}(1-\beta^{\mathrm{out}})}\right) & \mathrm{if}\;\; \beta^{\mathrm{out}} < 1\,\left(\mathrm{i.e.,\,if}\,\,\gamma^{\mathrm{out}}=1+\frac{1}{\beta^{\mathrm{out}}}>2\right), \\
             r_{tj}^{\mathrm{target}} - \frac{2}{\zeta}\ln\left(\frac{T}{\sin(T\pi)}\cdot \frac{\zeta r_{tj}^{\mathrm{target}}}{m_j^{\mathrm{in}}}\right) & \mathrm{if} \;\; \beta^{\mathrm{out}}=1\,\left(\mathrm{i.e.,\, if}\,\,\gamma^{\mathrm{out}}=1+\frac{1}{\beta^{\mathrm{out}}}=2\right). \end{array} \right.
             \label{eq:cutoff_dirEPSO_in}
        \end{equation}
    \end{itemize}
    The derivation of Eqs.~(\ref{eq:cutoff_dirEPSO_out}) and (\ref{eq:cutoff_dirEPSO_in}) are analogous to that of Eq.~(\ref{eq:cutoff_EPSO}) in the undirected case~\cite{PSO,HyperMap}, founded on the condition that the expected number of older nodes to which the new source and target nodes connect at their appearance is indeed $m_j^{\mathrm{out}}$ and $m_j^{\mathrm{in}}$, respectively.
\end{enumerate}

\subsection{Comparison of the PSO-based and our new, model-independent Euclidean-hyperbolic conversion MIC on undirected E-PSO networks} 
\label{sect:PSObasedVsOurConversion}
\setcounter{figure}{0}
\setcounter{table}{0}
\setcounter{equation}{0}
\renewcommand{\thefigure}{S2.3.\arabic{figure}}
\renewcommand{\thetable}{S2.3.\arabic{table}}
\renewcommand{\theequation}{S2.3.\arabic{equation}}

In this section, we compare the performance of our new Euclidean-hyperbolic conversion MIC to the well-known PSO-based method regarding the reconstruction of hyperbolic networks of size $N=500$ generated by the two-dimensional E-PSO model at various parameter settings. We generated $3$ networks with each set of model parameters and carried out the embedding of each network once with each one of the parameter settings of the Euclidean methods. Note that because of the uncertainty of the radial order of the nodes having the same degree, the PSO-based conversion may lead to somewhat different results when repeated (while the examined Euclidean embedding algorithms and our new conversion method are fully deterministic); however, these differences can be smoothed out by averaging over the performances achieved on different network realisations obtained with the same parameter settings. Therefore, for a given network, we chose the radial order based on the node degrees only once and used this order for the PSO-based conversion in the case of all the Euclidean embeddings.

The Euclidean embeddings that we converted were created by HOPE-S, HOPE-R, TREXPEN-S and TREXPEN-R. We always performed both the above-described $d$-dimensional approaches of the PSO-based conversion: Eqs.~(\ref{eq:betaInPSOembedding1})--(\ref{eq:radCoordInPSOembedding1}), where the popularity fading parameter $\beta$ is independent of the number of dimensions $d$ but the achieved largest hyperbolic radial coordinate $r_N$ is not, and Eqs.~(\ref{eq:betaInPSOembedding2})--(\ref{eq:radCoordInPSOembedding2}), $r_N$ is independent of $d$ but $\beta$ is not. Thus, we executed the Euclidean-hyperbolic conversion with altogether $3$ methods. Besides, we allowed shifting the centre of mass (COM) of the Euclidean node arrangement before all the conversion processes.

The list of the tested $\alpha$ and $q$ parameters of the Euclidean embeddings were determined according to the descriptions in Sects.~\ref{sect:HOPE} and \ref{sect:TREXPEN}, respectively. As usual, the tested number of dimensions were restricted below $N/10$, yielding the values $d=2,3,4,8,16,32<N/10$, and in MIC the parameter $C$ of the largest hyperbolic radial coordinate $r_{\mathrm{hyp,max}}=\frac{C}{\zeta}\cdot\ln{N}$ was set to $2$, yielding the same values of $r_{\mathrm{hyp,max}}$ as the PSO-based conversions (see Eqs.~(\ref{eq:radCoordInPSOembedding1}) and (\ref{eq:radCoordInPSOembedding2})). We always set the curvature $K$ of the hyperbolic space to $-1$, using $\zeta=1$.

In all the measurements, the task was to reconstruct all the $E$ number of links of the embedded network. For the sake of simplicity, the only quality measure that we depict here is the proportion of the actual links among the first $E$ node pairs in the increasing order of the hyperbolic distances, denoted by $\mathrm{Prec}\in[0,1]$. Since dealing with undirected graphs, 
we considered each node pair once. Recall that for undirected graphs, all of our embedding methods create source node coordinates that are identical to the target node coordinates, meaning that they yield the same hyperbolic distance from node $i$ to node $j$ as from node $j$ to node $i$.

\begin{figure}[!h]
    \centering
    \includegraphics[width=0.84\textwidth]{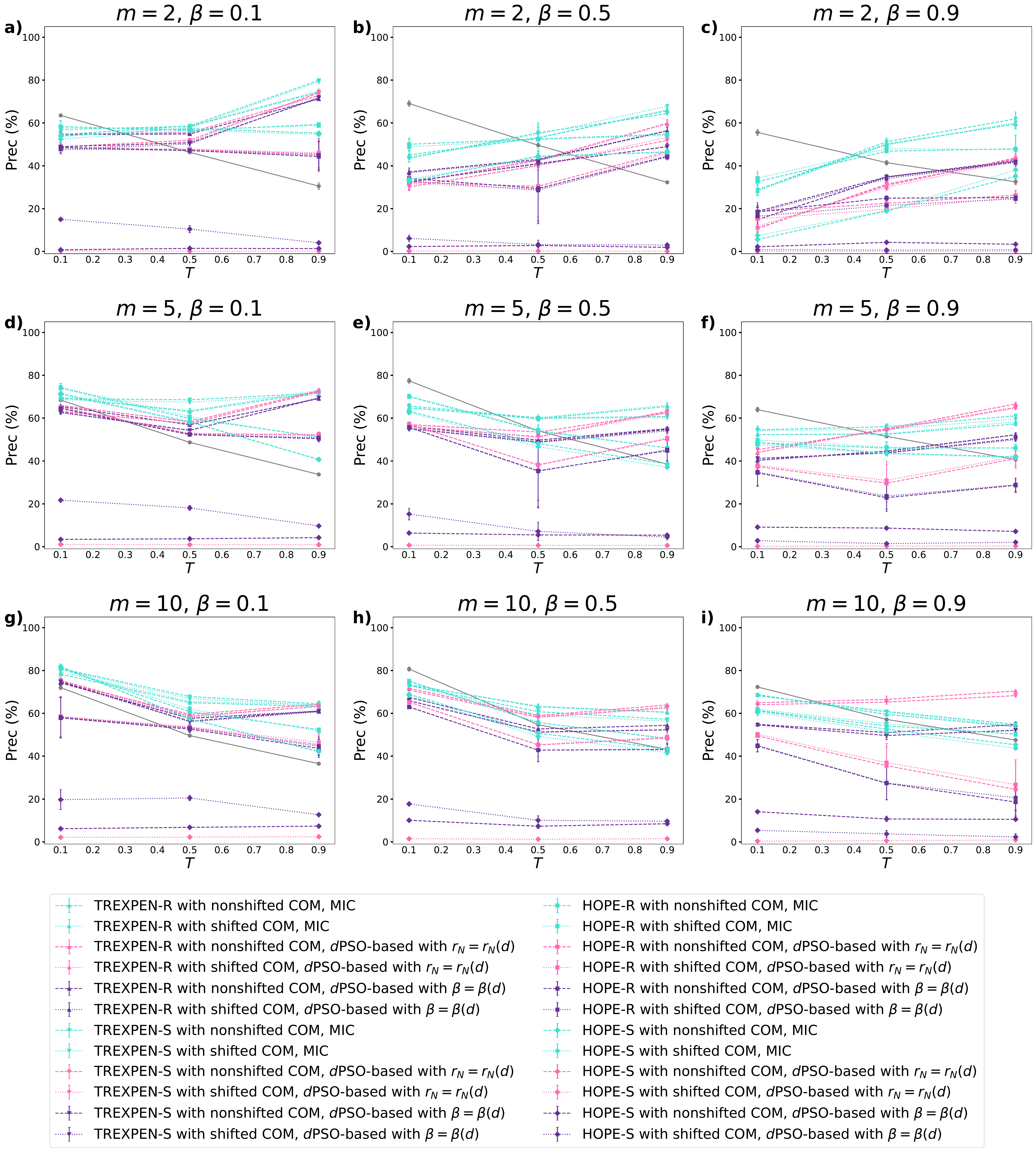}
    \caption{ {\bf Graph reconstruction performance of hyperbolic embeddings obtained with different Euclidean-hyperbolic conversion methods for two-dimensional, undirected PSO networks.} The different panels correspond to different parameter settings of the network generation: from top to bottom, the number of edges $E$ increases (the average degree $\bar{k}\approx2\cdot(m+L)$ is $4$, $10$ or $20$), while from left to right, the degree distribution becomes more fat-tailed (the degree decay exponent $\gamma=1+1/\beta$ is $11$, $3$ or $2.111$). All the networks consist of $N=500$ number of nodes and were generated using $\zeta=1$. The grey curve shows as a function of the network generation temperature $T$ the precision Prec obtained when reconstructing the first $E$ most probable links according to the hyperbolic distances measured in the original node arrangement that emerged during the network generation. The other curves depict the results of the embeddings, where the colouring denotes the type of the Euclidean-hyperbolic conversion, the line styles indicate the possible shift of the centre of mass of the Euclidean node arrangement before the conversion, and the markers show which dimension reduction method was used for creating the Euclidean embedding. For each network and embedding method, we considered only the best result achieved among all the tested settings of the embedding parameters and depicted at each set of the network generation parameters the average of these best results over $3$ network realisations, with the error bars displaying the standard deviations among the 3 networks of the same model parameters.}
    \label{fig:conversionComparison_undirPSO}
\end{figure}

\begin{figure}[!h]
    \centering
    \includegraphics[width=1.0\textwidth]{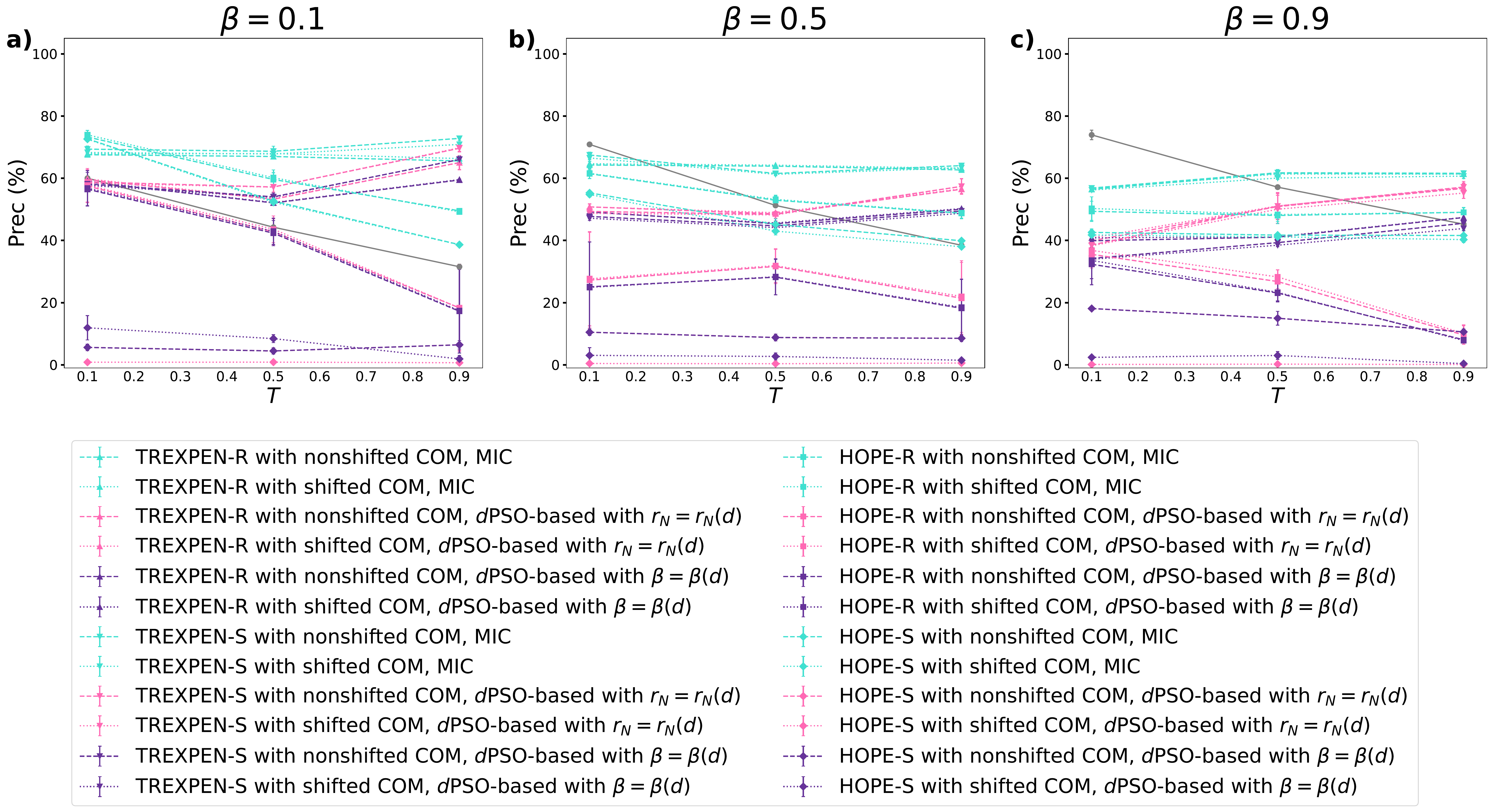}
    \caption{ {\bf Graph reconstruction performance of hyperbolic embeddings obtained with different Euclidean-hyperbolic conversion methods for two-dimensional, undirected E-PSO networks of $L>0$.} The different panels correspond to different parameter settings of the network generation: from left to right, the degree distribution becomes more fat-tailed (the degree decay exponent $\gamma=1+1/\beta$ is $11$, $3$ or $2.111$). All the networks consist of $N=500$ number of nodes and were generated using $\zeta=1$, $m=3$ and $L=2$ (yielding an average degree of $\bar{k}\approx2\cdot(m+L)=10$). The grey curve shows as a function of the network generation temperature $T$ the precision Prec obtained when reconstructing the first $E$ most probable links according to the hyperbolic distances measured in the original node arrangement that emerged during the network generation, where $E$ is the total number of links in the network. The other curves depict the results of the embeddings, where the colouring denotes the type of the Euclidean-hyperbolic conversion, the line styles indicate the possible shift of the centre of mass of the Euclidean node arrangement before the conversion, and the markers show which dimension reduction method was used for creating the Euclidean embedding. For each network and embedding method, we considered only the best result achieved among all the tested settings of the embedding parameters and depicted at each set of the network generation parameters the average of these best results over $3$ network realisations, with the error bars displaying the standard deviations among the 3 networks of the same model parameters.}
    \label{fig:conversionComparison_undirEPSO}
\end{figure}

In Fig.~\ref{fig:conversionComparison_undirPSO}, we compare the performance of the hyperbolic node arrangements obtained in different ways from the Euclidean embeddings of two-dimensional, undirected PSO networks, i.e., E-PSO networks of $L=0$. Then, Fig.~\ref{fig:conversionComparison_undirEPSO} shows the results for some two-dimensional, undirected E-PSO networks, where we increased the model parameter $L$ from $0$ to a positive value in order to ensure that the average internal degree of the subgraphs given by the nodes of larger degree than a certain threshold increases with the degree threshold, which may be considered to be a more realistic case~\cite{Mercator}. 
Based on these figures, the PSO-based conversion was able to outperform MIC only when embedding networks of a small degree decay exponent ($\beta=0.9$) and, simultaneously, a low average clustering coefficient ($T=0.9$). Accordingly, on locally random network structures having extremely large hubs, it may be beneficial to optimise the radial positions solely with respect to the global attractivities reflected by the node degrees, but otherwise, taking into account the interdependencies between the radial and the angular coordinates grasped by the dimension reduction, as in MIC, can more advantageous. Besides, especially in the case of real networks where the degree distribution does not necessarily follow a pure power law, it is important to remark that the fitting of a power law to a degree distribution is hampered by the uncertain identification of that range of the degree distribution where the power-law behaviour holds and the fact that the tail of the distribution usually falls into the regime of rare events, possibly significantly influenced by the occurrence of fluctuations. Therefore, the estimation of the popularity fading parameter $\beta$ based on the fitted degree decay exponent $\gamma$ can be problematic in many cases. Note that in our measurements, with the aim of minimising the inaccuracy of the PSO-based conversion that arises from the curve fitting, we implemented the method described in Ref.~\cite{degreeDistrFitting} for measuring $\gamma$.

\newpage
\null\newpage
\section{The automatic angular separation of communities in two-dimensional embeddings}
\label{sect:angSepOfComms}
\setcounter{figure}{0}
\setcounter{table}{0}
\setcounter{equation}{0}
\renewcommand{\thefigure}{S3.\arabic{figure}}
\renewcommand{\thetable}{S3.\arabic{table}}
\renewcommand{\theequation}{S3.\arabic{equation}}

Nodes of similar topological behaviour often form communities corresponding to apparent structural units in complex networks at the mesoscopic scale. In this study, we presented Euclidean and hyperbolic embeddings that aim at representing small distances measured along the graph between the network nodes as large inner products and small hyperbolic distances, respectively. In both cases, nodes that are topologically close to each other tend to become placed within small angular distances, since large angular distances are not favourable neither for the maximisation of the Euclidean inner product nor for the minimisation of the hyperbolic distance. Accordingly, these embeddings are able to reflect the community structure of a network through the angular organisation of its nodes, even in the absence of any direct input regarding the communities. In this section, we demonstrate this feature of the examined embedding methods on synthetic directed networks generated by the stochastic block model (SBM)~\cite{simplestSBMarticle,dirSBMgeneration,SBMcode} and on real networks.

\subsection{Layouts of SBM networks with assortative or disassortative block structure}
\label{sect:SBMlayouts}
\setcounter{figure}{0}
\setcounter{table}{0}
\setcounter{equation}{0}
\renewcommand{\thefigure}{S3.1.\arabic{figure}}
\renewcommand{\thetable}{S3.1.\arabic{table}}
\renewcommand{\theequation}{S3.1.\arabic{equation}}

\begin{figure}[!h]
    \centering
    \includegraphics[width=0.67\textwidth]{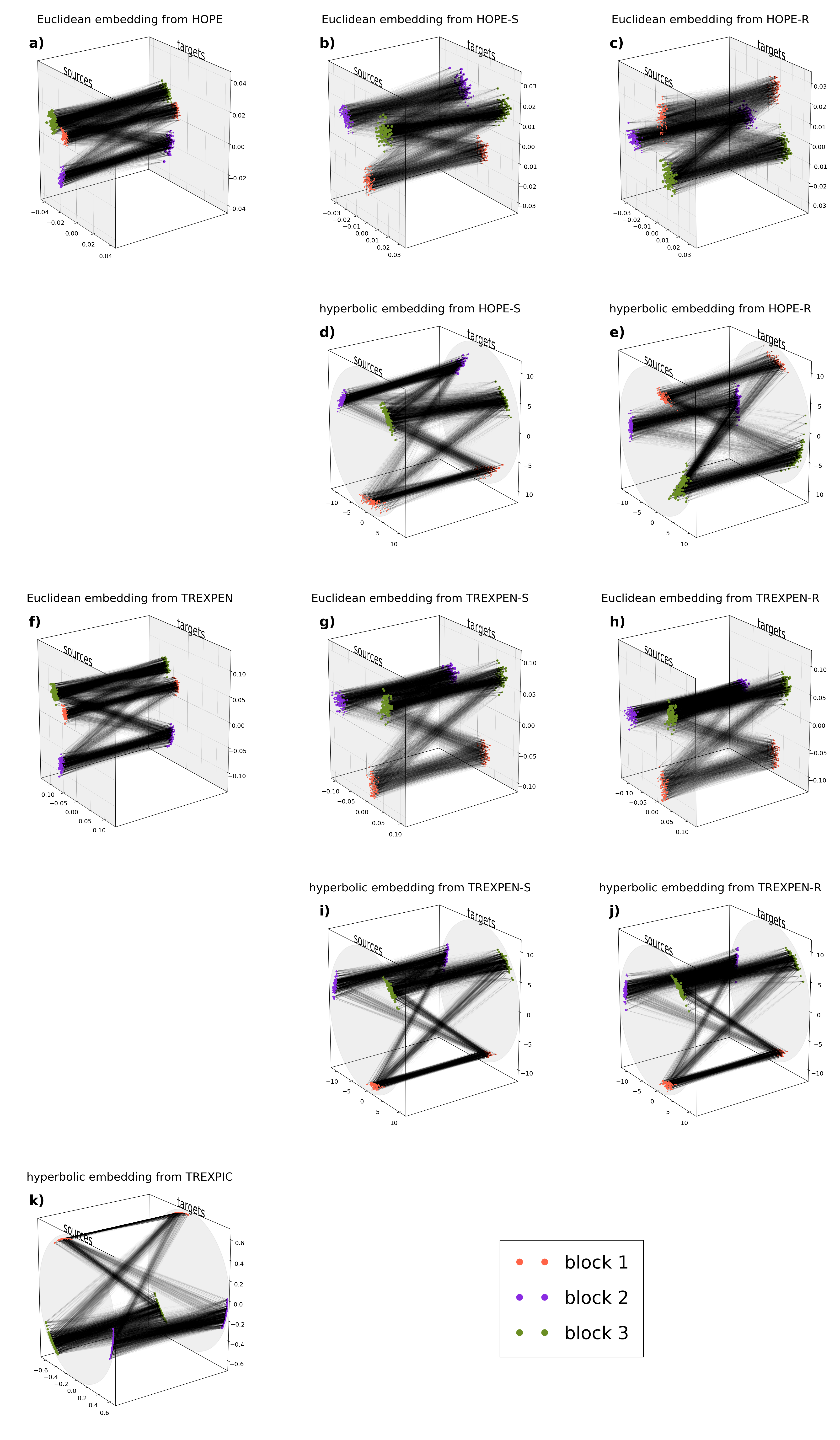}
    \caption{ {\bf Two-dimensional embeddings of an SBM network having an assortative block structure.} The different colours denote the different blocks in which the nodes were classed during the network generation. The node sizes are consistent with the node degrees. In the case of HOPE and its variants, we used $\alpha=2.01\cdot10^{-3}$. The embeddings with TREXPEN and its variants were obtained at $q=6.48$. The TREXPIC layout was created with $q=4.55\cdot10^{-2}$. We always used $C=2$ in MIC and $\zeta=1$ for the hyperbolic embeddings.}
    \label{fig:assortSBMlayouts}
\end{figure}

\begin{figure}[!h]
    \centering
    \includegraphics[width=0.67\textwidth]{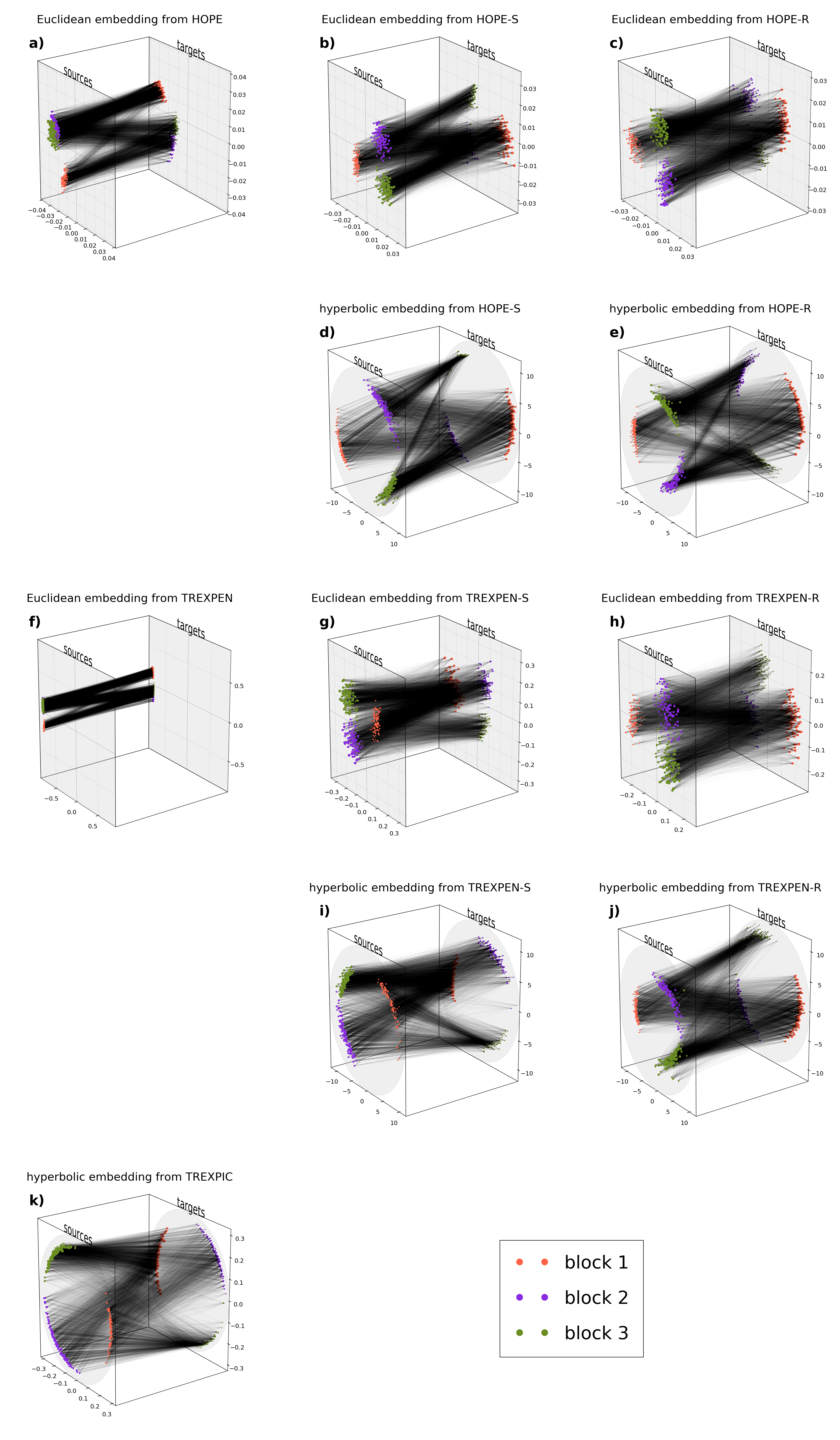}
    \caption{ {\bf Two-dimensional embeddings of an SBM network having a disassortative block structure.} The different colours denote the different blocks in which the nodes were classed during the network generation. The node sizes are consistent with the node degrees. In the case of HOPE and its variants, we used $\alpha=1.89\cdot10^{-3}$. The embeddings with TREXPEN and its variants were obtained at $q=0.26$. The TREXPIC layout was created with $q=1.65$. We always used $C=2$ in MIC and $\zeta=1$ for the hyperbolic embeddings.}
    \label{fig:dissortSBMlayouts}
\end{figure}

In Figs.~\ref{fig:assortSBMlayouts} and \ref{fig:dissortSBMlayouts}, we show two-dimensional embeddings of two directed SBM networks that were both created from 3 blocks of 100 nodes. In the case of Fig.~\ref{fig:assortSBMlayouts} (see also panels a)--f) in Fig.~\ref{fig:SBMLayout} of the main text), the edge densities between the different blocks were defined as
\begin{equation}
    \begin{bmatrix}
    0.25 & 0.05 & 0.1\\
    0.05 & 0.35 & 0.05\\
    0.1 & 0.15 & 0.4
    \end{bmatrix},
    \label{eq:assortSBMprobMatrix}
\end{equation}
while Fig.~\ref{fig:dissortSBMlayouts} (see also panels g)--l) in Fig.~\ref{fig:SBMLayout} of the main text) refers to a network where the edge densities were given by
\begin{equation}
    \begin{bmatrix}
    0.05 & 0.25 & 0.15\\
    0.35 & 0.05 & 0.2\\
    0.4 & 0.2 & 0.05
    \end{bmatrix}.
    \label{eq:dissortSBMprobMatrix}
\end{equation}
According to the edge densities, in the first network (defined by Eq.~(\ref{eq:assortSBMprobMatrix})) most of the links were formed within the blocks, yielding an assortative block structure that consists of such groups of the nodes that are more connected to each other than to the nodes of the other communities. On the contrary, in the second network (defined by Eq.~(\ref{eq:dissortSBMprobMatrix})) the nodes are connected mostly to nodes of other blocks, creating a disassortative block structure in which the members of a group are held together by their similar connection preference towards the other groups of the nodes. Despite the fundamental differences between the binding forces of the blocks in the above described two block structures, HOPE-S, HOPE-R, TREXPEN-S, TREXPEN-R and TREXPIC managed to angularly separate the 3 equally sized blocks in both SBM networks. Meanwhile, as expected, HOPE and TREXPEN placed both networks in a restricted angular range, resulting in a less clear separation of the blocks of the nodes.

\subsection{Arrangement of communities in the embeddings of real networks}
\label{sect:realLayouts}
\setcounter{figure}{0}
\setcounter{table}{0}
\setcounter{equation}{0}
\renewcommand{\thefigure}{S3.2.\arabic{figure}}
\renewcommand{\thetable}{S3.2.\arabic{table}}
\renewcommand{\theequation}{S3.2.\arabic{equation}}

\begin{figure}[!h]
    \centering
    \includegraphics[width=0.67\textwidth]{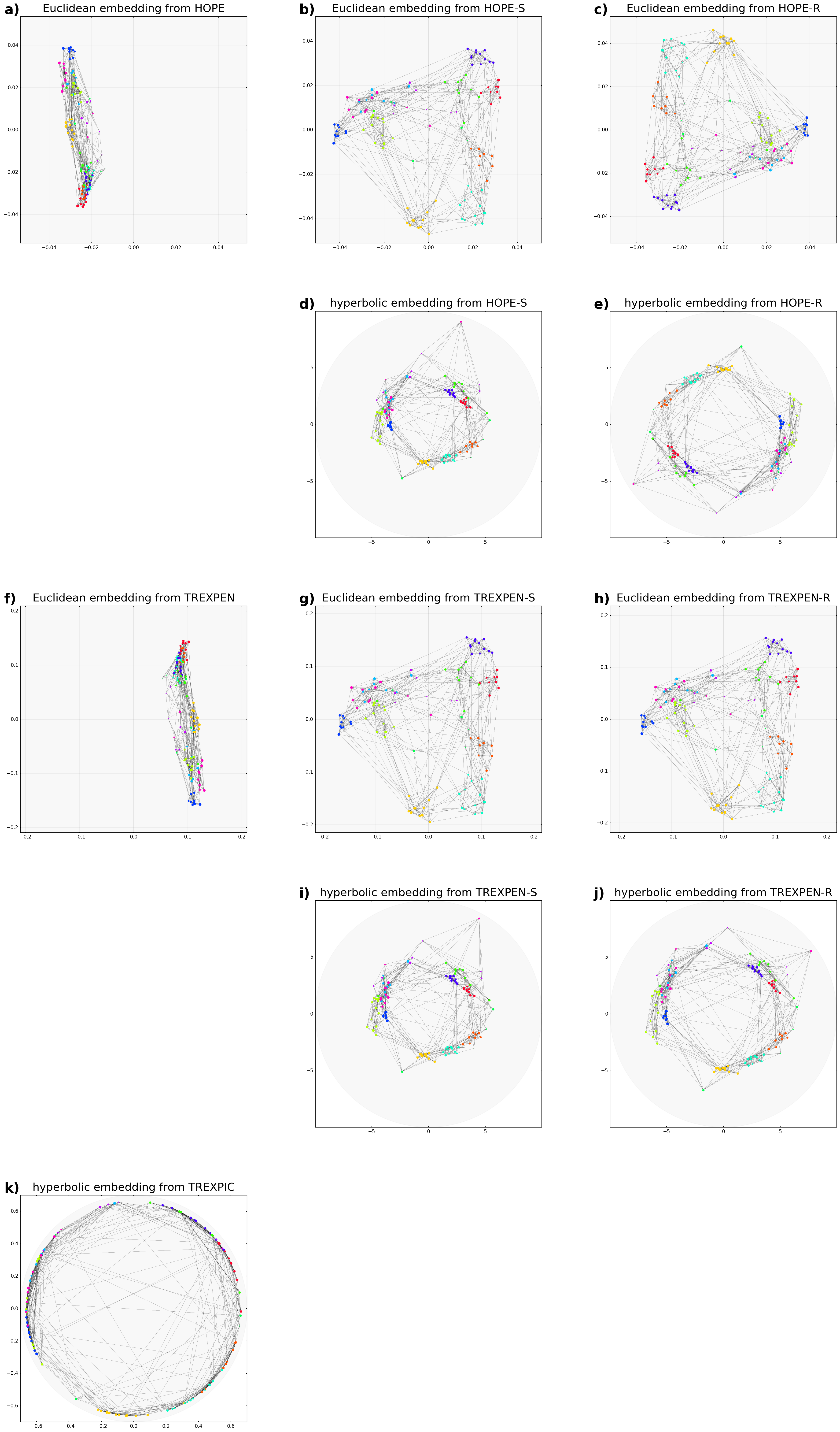}
    \caption{ {\bf Two-dimensional embeddings of the undirected American College Football network described in Section~\ref{sect:undirEmb}.} The colour of each node indicates the conference to which it belongs out of the 12 conferences. The node sizes are consistent with the node degrees. In the case of HOPE and its variants, we used $\alpha=6.56\cdot10^{-3}$. The embeddings with TREXPEN and its variants were obtained at $q=4.86$. The TREXPIC layout was created with $q=6.07\cdot10^{-2}$. We always used $C=2$ in MIC and $\zeta=1$ for the hyperbolic embeddings.} 
    \label{fig:footballLayouts}
\end{figure}

\begin{figure}[!h]
    \centering
    \includegraphics[width=0.67\textwidth]{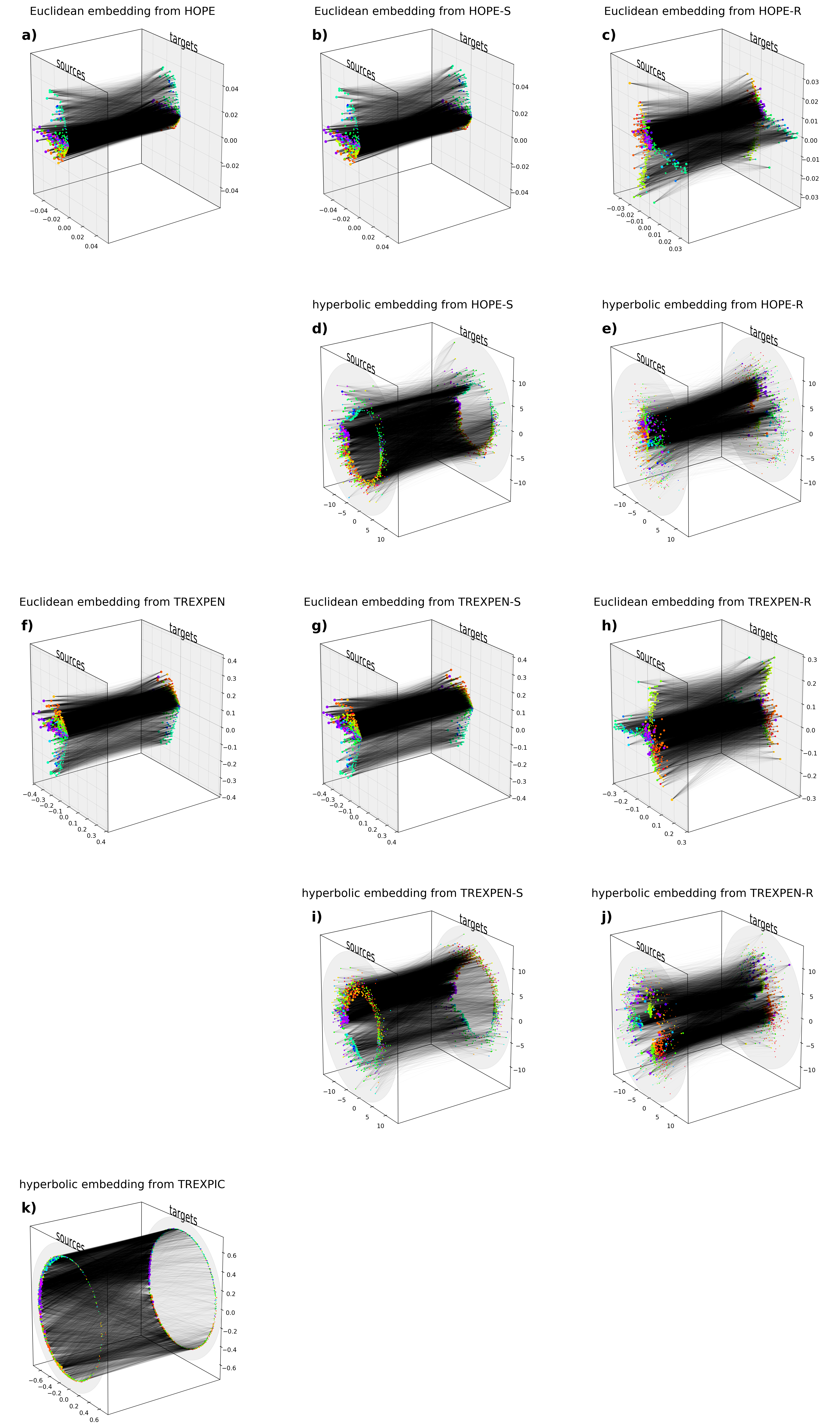}
    \caption{ {\bf Two-dimensional embeddings of the directed email network described in Sect.~\ref{sect:extraRealDirEmbeddings}.} The colour of each node indicates the department to which the corresponding person belongs out of the 42 departments. The node sizes are consistent with the node degrees. In the case of HOPE and its variants we used $\alpha=1.15\cdot10^{-3}$. The embeddings with TREXPEN and its variants were obtained at $q=2.78$. The TREXPIC layout was created with $q=0.22$. We always used $C=2$ in MIC and $\zeta=1$ for the hyperbolic embeddings.}
    \label{fig:emailLayouts}
\end{figure}

Fig.~\ref{fig:footballLayouts} depicts some two-dimensional layouts of the undirected American College Football network~\cite{football_net_data}, which is investigated in detail with respect to mapping accuracy, graph reconstruction, link prediction and greedy routing in Section~\ref{sect:undirEmb}. Besides, Fig.~\ref{fig:emailLayouts} shows two-dimensional embeddings of the email network~\cite{emailRef} that is studied in Sect.~\ref{sect:extraRealDirEmbeddings}. Based on these figures, all the examined embedding methods automatically provide some sort of separation between the known groups of the nodes given by the conferences of the football teams in the first case, and the departments of the research institute in the second.

It is important to note that although all the radial coordinates obtained from TREXPIC are very close to each other, even such small differences have a substantial impact on the distance relations in the system and can yield a completely reasonable distance-based ordering of the node pairs since in the hyperbolic distance formula the radial coordinates are inputted into rapidly changing functions like $\mathrm{sinh}$ and $\mathrm{cosh}$. Thus, as exemplified by Table~\ref{table:football_2DhypPerformances}, the depicted embedding provided by TREXPIC for the football network is of a similar, or even better quality from the point of view of mapping accuracy, graph reconstruction and greedy routing than those hyperbolic layouts in Fig.~\ref{fig:footballLayouts} that were generated from Euclidean embeddings created by HOPE-S, HOPE-R, TREXPEN-S and TREXPEN-R with our Euclidean-hyperbolic conversion method MIC.

\begin{table}[!ht]
\centering
\caption{{\bf The performance in mapping accuracy, graph reconstruction and greedy routing of the two-dimensional hyperbolic embeddings that are presented for the football network in Fig.~\ref{fig:footballLayouts}.} Despite its visually less pleasing radial arrangement, the embedding created by TREXPIC achieved the best scores in mapping accuracy and greedy routing. The best results are written in bold for each measure.}
\begin{tabular}{+l+c|c|c|c|c+}
\thickhline
\rowcolor{white}   & HOPE-S & HOPE-R & TREXPEN-S & TREXPEN-R & TREXPIC \\ \thickhline
\rowcolor{lightgray} $\mathrm{ACC}_{\mathrm{m}}$ & $0.350$ & $0.347$ & $0.357$ & $0.352$ & $\pmb{0.566}$ \\ \thickhline
\rowcolor{white} Prec (\%) in graph reconstruction & $\pmb{50.08}$ & $49.92$ & $49.76$ & $49.59$ & $40.95$ \\ \hline
\rowcolor{white} AUPR in graph reconstruction & $\pmb{0.473}$ & $0.441$ & $0.453$ & $0.439$ & $0.376$ \\ \hline
\rowcolor{white} AUROC in graph reconstruction & $0.815$ & $0.809$ & $0.816$ & $0.812$ & $\pmb{0.868}$ \\ \thickhline
\rowcolor{lightgray} GR-score & $0.566$ & $0.555$ & $0.561$ & $0.557$ & $\pmb{0.623}$ \\ \hline
\rowcolor{lightgray} percentage of successful greedy paths & $67.00$ & $65.63$ & $66.17$ & $66.01$ & $\pmb{73.38}$ \\ \hline
\rowcolor{lightgray} average hop-length of the successful greedy paths & $3.132$ & $\pmb{3.114}$ & $3.119$ & $3.128$ & $3.127$ \\ \thickhline
\end{tabular}
\label{table:football_2DhypPerformances}
\end{table}

\newpage
\null\newpage
\section{The dependence of the embedding performance on the parameters of the embedding methods}
\label{sect:embParams}
\setcounter{figure}{0}
\setcounter{table}{0}
\setcounter{equation}{0}
\renewcommand{\thefigure}{S4.\arabic{figure}}
\renewcommand{\thetable}{S4.\arabic{table}}
\renewcommand{\theequation}{S4.\arabic{equation}}

In the main text, we depicted only the best results achieved among all the tested parameter settings. Here, as an example, we present in detail how the performance of the examined embedding methods depends on their settings for the network of $505$ number of Wikipedia pages~\cite{wikipediaRef}. 

\begin{figure}[!h]
    \centering
    \includegraphics[width=1.0\textwidth]{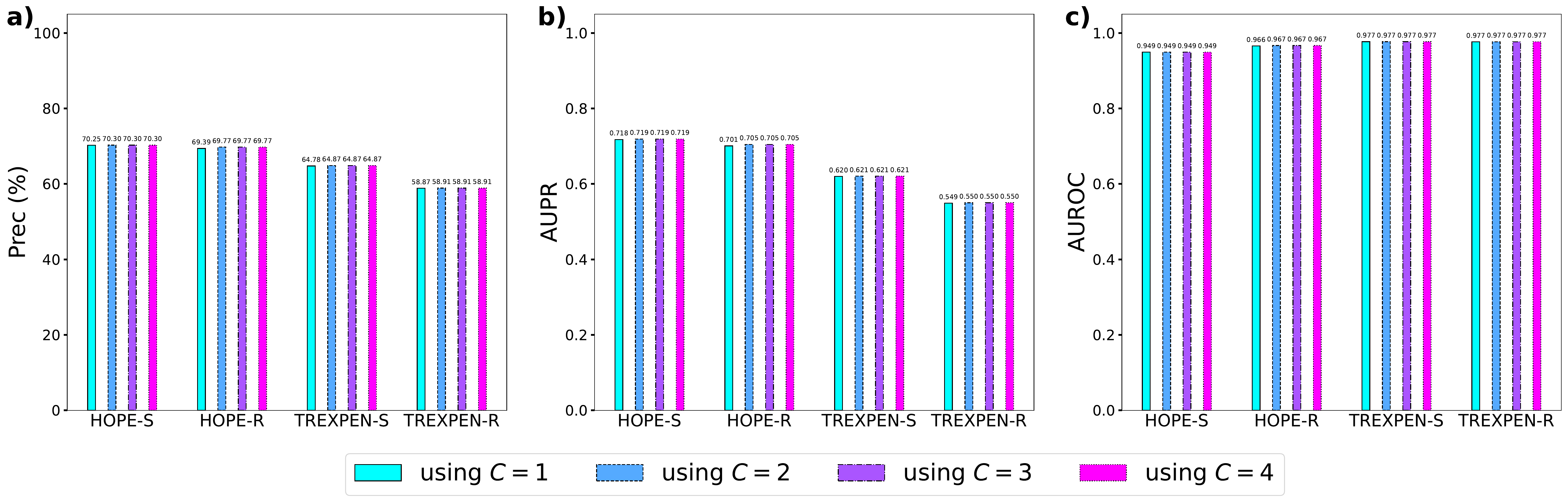}
    \caption{ {\bf Graph reconstruction performance on the network of $N=505$ number of Wikipedia pages at different settings of the $C$ parameter of our Euclidean-hyperbolic conversion method MIC.} The task was to reconstruct each one of the $E=2081$ number of links in the network based on the hyperbolic distances obtained from the hyperbolic conversion of the different Euclidean embeddings. The curvature $K$ of the hyperbolic space was set to $-1$ in each case. The tested number of dimensions $d\leq N/10$ were $d=2,3,4,8,16$ and $32$ for all the methods. We always performed the transformation to the hyperbolic space both with and without shifting the COM of the given Euclidean node arrangement. We plotted in each panel only the results of those settings that turned out to be the best regarding that given performance measure: the precision obtained when reconstructing the first $E$ most probable links in panel~a), the area under the precision-recall (PR) curve in panel~b) and the area under the receiver operating characteristic (ROC) curve in panel~c). The colours indicate the setting of the parameter $C$, as listed in the common legend at the bottom of the figure. The values depicted with dark blue for $C=2$ are the same as those given by the brown bars in panels~a)--c) of Fig.~\ref{fig:graRecMain} in the main text.}
    \label{fig:Cdependence_wikipedia_graRec}
\end{figure}

First, Fig.~\ref{fig:Cdependence_wikipedia_graRec} exemplifies via the graph reconstruction task that the parameter $C$ of the largest radial coordinate ${r_{\mathrm{hyp,max}} = \frac{C}{\zeta}\cdot\ln(N)}$ of the hyperbolic embeddings (see Eq.~(\ref{eq:fixedHypMaxR})) usually does not have a severe impact on the embedding quality. However, the performance obtained at $C=1$ (i.e. when using smaller radial coordinates) seems to be slightly weaker compared to that of the tested larger values of $C$, which may result from the reduced validity of the approximating formula of the hyperbolic distance given by Eq.~(\ref{eq:hypDistApprox}), which served as a basis for separating the radial conversion from the angular one and associating larger radial coordinates of the Euclidean space with smaller ones in the hyperbolic space that is an essential step of our Euclidean-hyperbolic conversion method MIC. Thus, our choice of setting $C$ to $2$ in MIC seems to be reasonable.

Using $C=2$, Figs.~\ref{fig:detailedMapAcc_wikipedia}--\ref{fig:detailedGR_wikipedia} demonstrate the impact of the number of dimensions of the embedding space and the multiplying factors $\alpha$ and $q$ that parametrise the reduced proximity and distance matrixes on the mapping accuracy, the graph reconstruction performance and the greedy routing score. To obtain a more complete view of the embeddings' behaviour as a function of the number of dimensions $d$, in Figs.~\ref{fig:detailedMapAcc_wikipedia}--\ref{fig:detailedGR_wikipedia} we increased our usual upper limit on $d$ from $N/10$ to $N/2$. Naturally, the embedding spaces of larger number of dimensions usually perform better. Besides, these figures validate our choice of the explored parameter ranges defined as
\begin{itemize}
    \item $\alpha\in\left[\frac{1}{200\cdot\rho_{\mathrm{spectral}}(\bm{A})},\frac{1}{\rho_{\mathrm{spectral}}(\bm{A})}\right]$ for HOPE and its variants (see Sect.~\ref{sect:HOPE}), where $\rho_{\mathrm{spectral}}(\bm{A})$ stands for the spectral radius of the adjacency matrix,
    \item $q\in[-\ln(0.9)/\mathrm{SPL}_{\mathrm{max}},-\ln(10^{-50})/\mathrm{SPL}_{\mathrm{max}}]$ for TREXPEN and its variants (see Sect.~\ref{sect:TREXPEN}) with $\mathrm{SPL}_{\mathrm{max}}$ denoting the largest finite shortest path length measured along the graph, and
    \item $q\in[\ln(1.0/0.9999)\cdot\mathrm{SPL}_{\mathrm{max}},\ln(10)\cdot\mathrm{SPL}_{\mathrm{max}}]$ for TREXPIC (see Sect.~\ref{sect:TREXPIC}),
\end{itemize}
since in most of the cases the achieved best results fall in the interior of these intervals and not at their boundaries. In addition, Figs.~\ref{fig:detailedMapAcc_wikipedia}-\ref{fig:detailedGR_wikipedia} show that shifting the centre of mass to the origin is rather disadvantageous for the Euclidean embeddings themselves from the point of view of the inner product, but can yield better inputs for the Euclidean-hyperbolic conversion, improving the quality of the resulted hyperbolic embeddings. Note that a uniform shifting of all the nodes in a Euclidean (or a hyperbolic) embedding does not affect the distances between the nodes, and thus, the distance-based performances of the given embedding.

\begin{figure}[!h]
    \centering
    \includegraphics[width=0.61\textwidth]{detailedMapAcc_wikipedia.pdf}
    \caption{ {\bf Mapping accuracy on the network of $N=505$ number of Wikipedia pages at different settings of the examined embedding methods.} $\mathrm{ACC}_{\mathrm{m}}$ was measured considering each node pair connected by at least one directed path. One row of panels depicts all the examined variations of one embedding method, differing in the geometric measure that was used for the evaluation and whether or not the centre of mass (COM) of the angularly not restricted Euclidean patterns was shifted to the origin. The parameter $C$ was set to $2$ in MIC. The curvature $K$ of the hyperbolic space was always set to $-1$.}
    \label{fig:detailedMapAcc_wikipedia}
\end{figure}

\begin{figure}[!h]
    \centering
    \includegraphics[width=0.58\textwidth]{detailedPrec_wikipedia.pdf}
    \caption{ {\bf Graph reconstruction on the network of $N=505$ number of Wikipedia pages at different settings of the examined embedding methods.} The task was to reconstruct each one of the $E=2081$ number of links in the network based on the node coordinates. The embedding performance is measured here by the precision obtained when reconstructing the first $E$ links that are the most probable according to the given geometric measure. One row of panels depicts all the examined variations of one embedding method, differing in the applied geometric measure and whether or not the centre of mass (COM) of the angularly not restricted Euclidean patterns was shifted to the origin. The parameter $C$ was set to $2$ in MIC. The curvature $K$ of the hyperbolic space was always set to $-1$.}
    \label{fig:detailedPrec_wikipedia}
\end{figure}

\begin{figure}[!h]
    \centering
    \includegraphics[width=0.6\textwidth]{detailedGR_wikipedia.pdf}
    \caption{ {\bf Greedy routing on the network of $N=505$ number of Wikipedia pages at different settings of the examined embedding methods.} The task was to perform greedy routing between each node pair connected by at least one directed path. The embedding performance is measured here by the GR-score. One row of panels depicts all the examined variations of one embedding method, differing in the geometric measure that was used for the routing and whether or not the centre of mass (COM) of the angularly not restricted Euclidean patterns was shifted to the origin. The parameter $C$ was set to $2$ in MIC. The curvature $K$ of the hyperbolic space was always set to $-1$.}
    \label{fig:detailedGR_wikipedia}
\end{figure}

\subsection{The role of the curvature of the hyperbolic space}
\label{sect:embParams_curvature}
\setcounter{figure}{0}
\setcounter{table}{0}
\setcounter{equation}{0}
\renewcommand{\thefigure}{S4.1.\arabic{figure}}
\renewcommand{\thetable}{S4.1.\arabic{table}}
\renewcommand{\theequation}{S4.1.\arabic{equation}}

In the case of our model-independent Euclidean-hyperbolic conversion MIC, the setting of the curvature $K=-\zeta^2<0$ of the hyperbolic space does not have an impact on the distance-based ordering of the different node pairs. This is due to the fact that the curvature appears only in a $1/\zeta$ multiplier in the radial coordinate formula given by Eq.~(\ref{eq:rHyp(rEuc)}), which becomes eliminated by the $\zeta$ multipliers on the right-hand side of the hyperbolic law of cosines written in Eq.~(\ref{eq:hypDist}). Based on these two equations, varying the curvature of the hyperbolic space corresponds to a simple rescaling of all the hyperbolic distances and does not change the distance-based ordering of the node pairs. Consequently, the value of the curvature $K$ does not affect the embedding performance achieved by MIC in mapping accuracy, graph reconstruction or greedy routing. Thus, according to the usual custom ~\cite{HyperMap,HyperMap-CN,LPCS,coalescentEmbedding,Mercator,descriptionOfMeasuresOfGraRecAndLinkPred,ourEmbedding}, we always used $K=-1$, i.e. $\zeta=1$. 

\begin{figure}[H]
    \centering
    \includegraphics[width=1.0\textwidth]{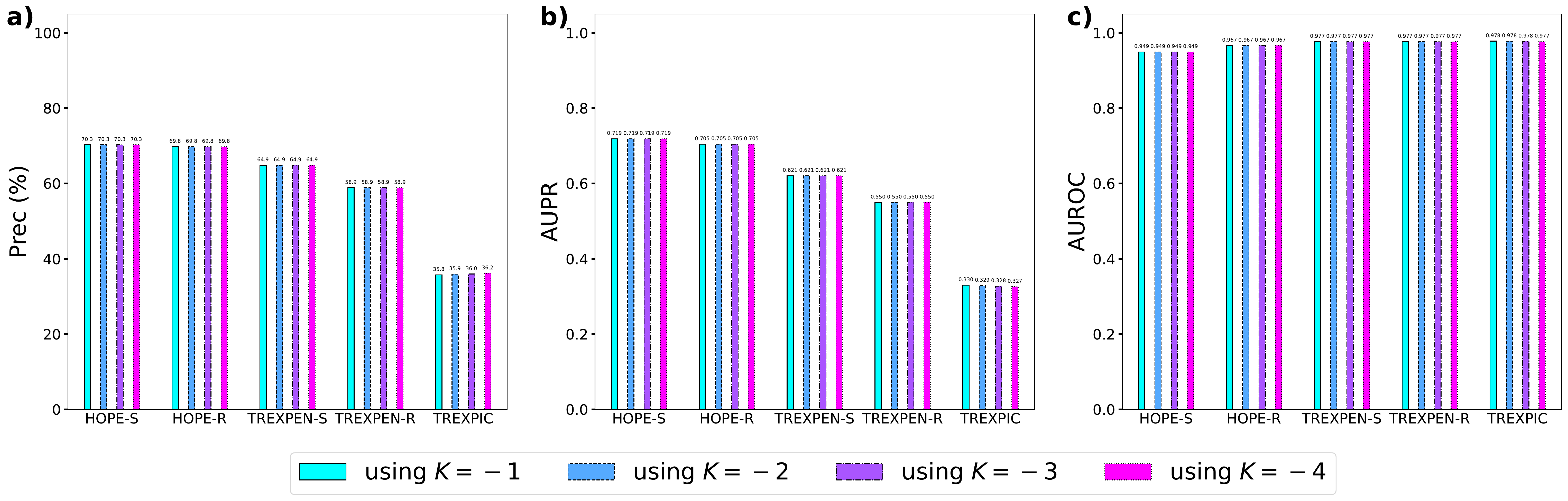}
    \caption{ {\bf Graph reconstruction performance on the network of $N=505$ number of Wikipedia pages at different settings of the $K$ curvature of the hyperbolic space in our hyperbolic embedding methods.} The task was to reconstruct each one of the $E=2081$ number of links in the network based on the hyperbolic distances measured in our hyperbolic embeddings. The parameter $C$ was set to $2$ in each case of the Euclidean-hyperbolic conversion MIC. The tested number of dimensions $d\leq N/10$ were $d=2,3,4,8,16$ and $32$ for all the methods. We always performed the transformation to the hyperbolic space both with and without shifting the COM of the given Euclidean node arrangement. We plotted in each panel only the results of those settings that turned out to be the best regarding that given performance measure: the precision obtained when reconstructing the first $E$ most probable links in panel~a), the area under the precision-recall (PR) curve in panel~b) and the area under the receiver operating characteristic (ROC) curve in panel~c). The colours indicate the setting of the curvature $K$, as listed in the common legend at the bottom of the figure. The values depicted with light blue for $K=-1$ are the same as those given by the brown bars in panels~a)--c) of Fig.~\ref{fig:graRecMain} in the main text.}
    \label{fig:Kdependence_wikipedia_graRec}
\end{figure}

In contrast, nothing precludes that changing the hyperbolic curvature $K$ has an effect on the performance of TREXPIC. However, as exemplified by Fig.~\ref{fig:Kdependence_wikipedia_graRec}, the decrease in the curvature $K$ may be disadvantageous in TREXPIC, albeit all the changes presented in the figure are very small. Nevertheless, e.g. in Ref.~\cite{curvatureInHypEmb} it was shown that the decrease in the curvature $K$ may be advantageous in hyperbolic embeddings. Yet, the detailed exploration of the role of the curvature in TREXPIC is out of the scope of this study, and we used in our measurements the $K=-1$ setting for TREXPIC -- just like it was used in the case of hydra~\cite{Hydra}, the method which inspired TREXPIC.

\newpage
\section{Additional examples for the embedding performance on directed real networks}
\label{sect:extraRealDirEmbeddings}
\setcounter{figure}{0}
\setcounter{table}{0}
\setcounter{equation}{0}
\renewcommand{\thefigure}{S5.\arabic{figure}}
\renewcommand{\thetable}{S5.\arabic{table}}
\renewcommand{\theequation}{S5.\arabic{equation}}

In the main text, we studied the embedding performance on two smaller and two larger directed real networks with respect to the mapping accuracy, the graph reconstruction and the greedy routing. Here, to broaden the scope of investigation, 
we present the results of the same measurements that are described in the main text for the following four additional directed real networks:
\begin{itemize}
    \item A food web of $N=99$ number of nodes and $E=901$ number of links, named \textit{Dutch Microfauna food web PlotC}, downloaded from Ref.~\cite{foodWebRef}. 
    \item A neural network~\cite{CelegansRef} of C. elegans, consisting of $N=297$ number of nodes and $E=2345$ number of directed links.
    \item A network~\cite{emailRef} connecting to each other $N=986$ coworkers of a European research institute according to email sendings. A link $s\rightarrow t$ means that person $s$ sent at least one email to person $t$. The total number of edges is $E=24929$. Each node is labelled by the department to which the corresponding person belongs out of the 42 departments.
    \item A network~\cite{airportRef} of $N=2905$ number of airports connected by $E=30442$ number of routes.
\end{itemize}
During the measurements, we took into consideration all the possible node pairs in each task for the two smaller additional graphs (namely the food web and the neural network), but -- because of the high computational intensity -- accomplished the evaluation of the embedding performance only on sampled sets of node pairs in the case of the two larger graphs above (i.e. the email network and the airport network). The details of the applied sampling procedures are given in the Methods section of the main text.

\begin{figure}[!h]
    \centering
    \includegraphics[width=0.81\textwidth]{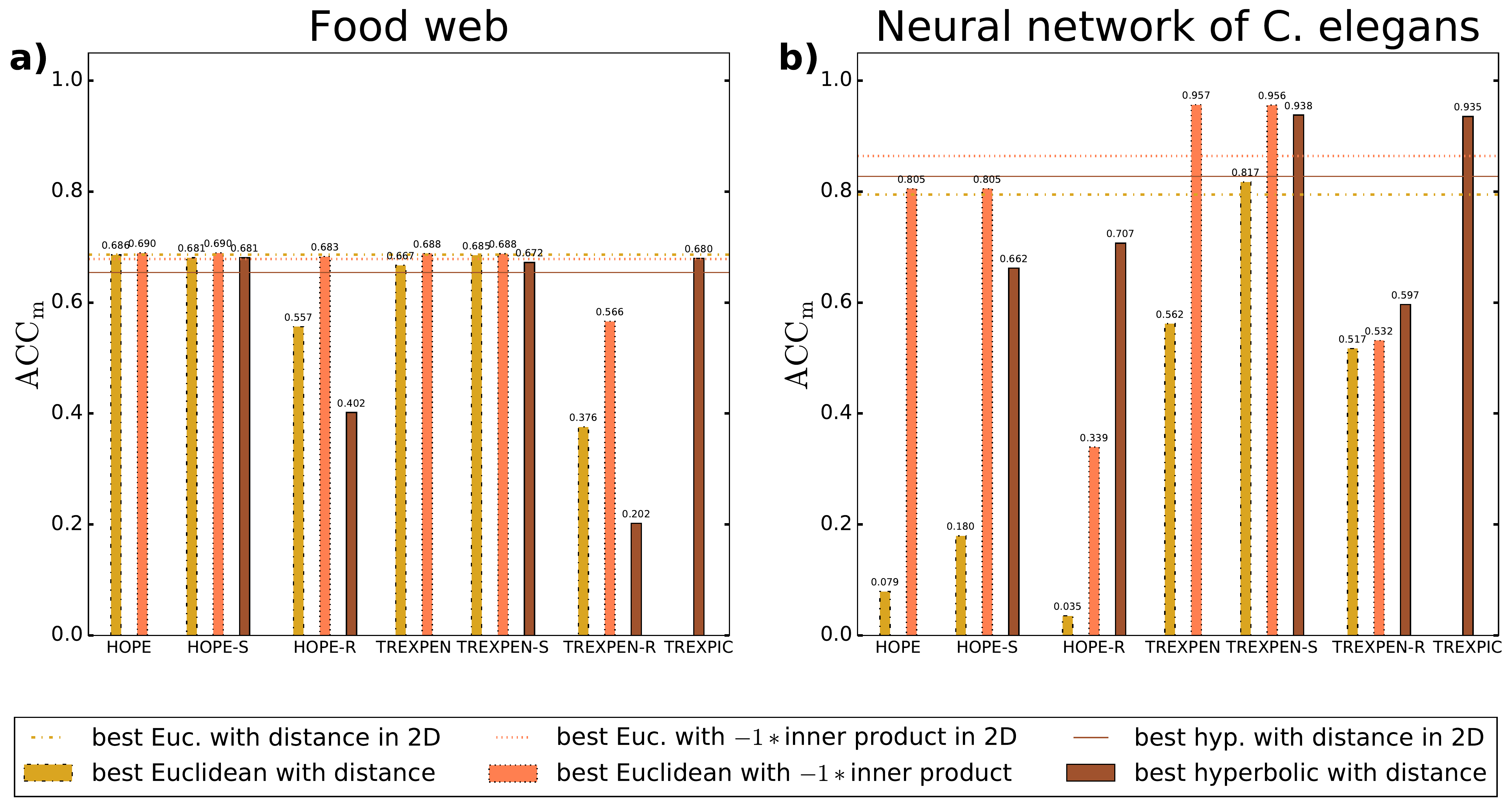}
    \caption{ {\bf Mapping accuracy on the additional directed real networks of smaller sizes.} For each network, we measured the mapping accuracy considering each node pair connected by at least one directed path. The colours indicate what geometric measure was used, as listed in the common legend at the bottom of the figure. We plotted in each panel only the results of those parameter settings that turned out to be the best, i.e. which yielded the highest values of the mapping accuracy. 
    The bars were created considering all the tested number of dimensions, whereas the horizontal lines show the best two-dimensional performances achieved among all the embedding methods. Each panel refers to a real network named in the title of the row.}
    \label{fig:mapAcc_smallExtraDirNetworks}
\end{figure}

\begin{figure}[!h]
    \centering
    \includegraphics[width=0.81\textwidth]{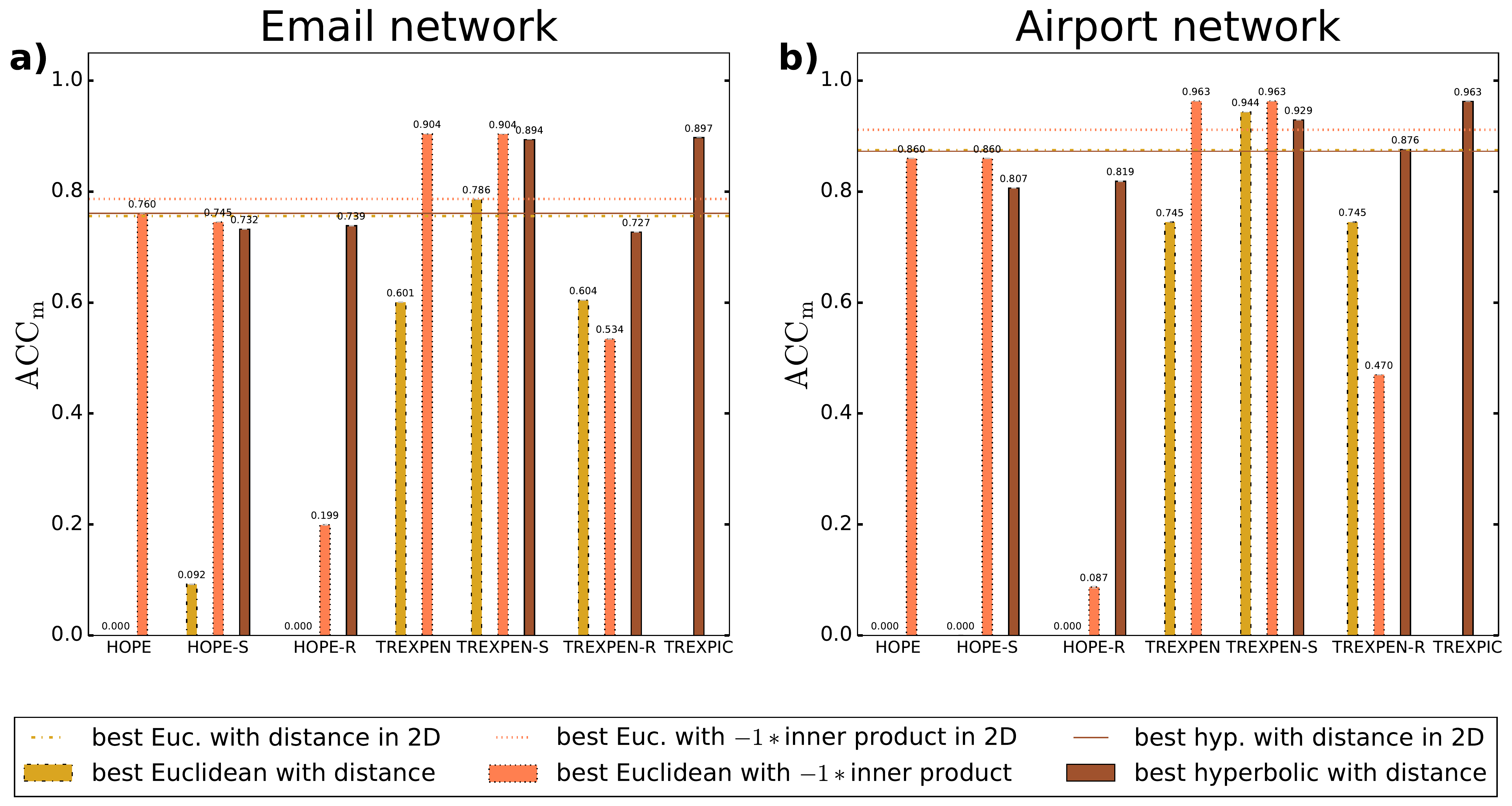}
    \caption{ {\bf Mapping accuracy on the additional directed real networks of larger sizes.} For each network, we measured the mapping accuracy on 3 samples of $500000$ node pairs connected by at least one directed path. The colours indicate what geometric measure was used, as listed in the common legend at the bottom of the figure. The plotted values were obtained by averaging the results of the 3 samples, and the error bars show the standard deviations among the different samples. We depicted in each panel only the results of those parameter settings that turned out to be the best, i.e. which yielded the highest values of the average mapping accuracy over the 3 samples. Note that the 0 values denote that the given methods have not achieved any positive average value. The bars were created considering all the tested number of dimensions, whereas the horizontal lines show the best two-dimensional average performances achieved among all the embedding methods. Each panel refers to a real network named in the title of the row.}
    \label{fig:mapAcc_largeExtraDirNetworks}
\end{figure}

\begin{figure}[!h]
    \centering
    \includegraphics[width=1.0\textwidth]{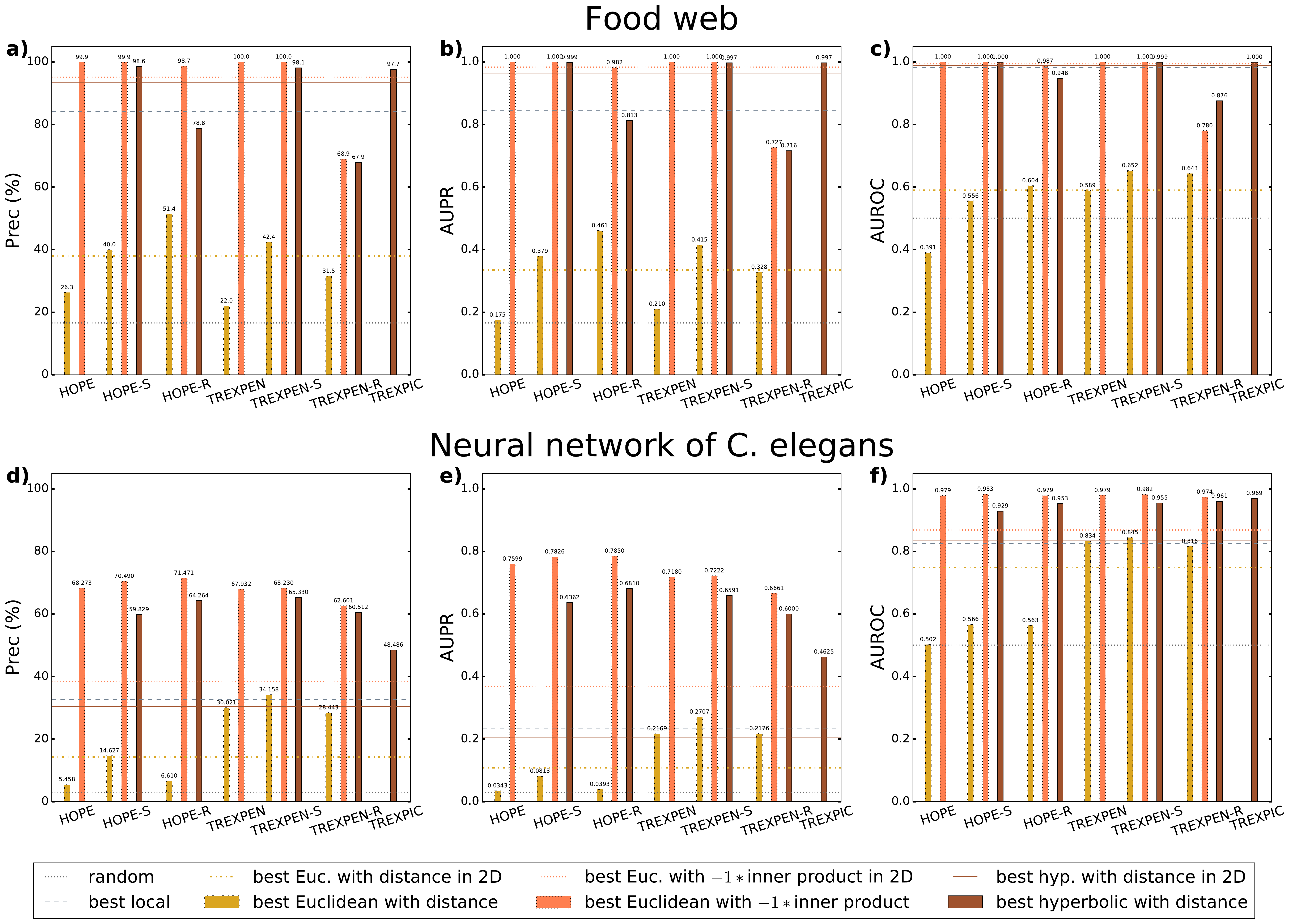}
    \caption{ {\bf Graph reconstruction performance on the additional directed real networks of smaller sizes.} For each network, the task was to reconstruct each one of the $E$ number of links. The colours indicate what geometric measure was used, as listed in the common legend at the bottom of the figure. We plotted in each panel only the results of those parameter settings that turned out to be the best regarding that given performance measure. The bars were created considering all the tested number of dimensions, whereas the colourful horizontal lines show the best two-dimensional performances achieved among all the embedding methods. Each row of panels refers to a real network named in the title of the row. The different columns of panels correspond to different measures: panels~a) and d) to the precision obtained when reconstructing the first $E$ most probable links, panels~b) and e) to the area under the precision-recall (PR) curve, while panels~c) and f) to the area under the receiver operating characteristic (ROC) curve. The grey horizontal lines show the baselines: the performance of the random predictor and the best local method.}
    \label{fig:graRec_smallExtraDirNetworks}
\end{figure}

\begin{figure}[!h]
    \centering
    \includegraphics[width=1.0\textwidth]{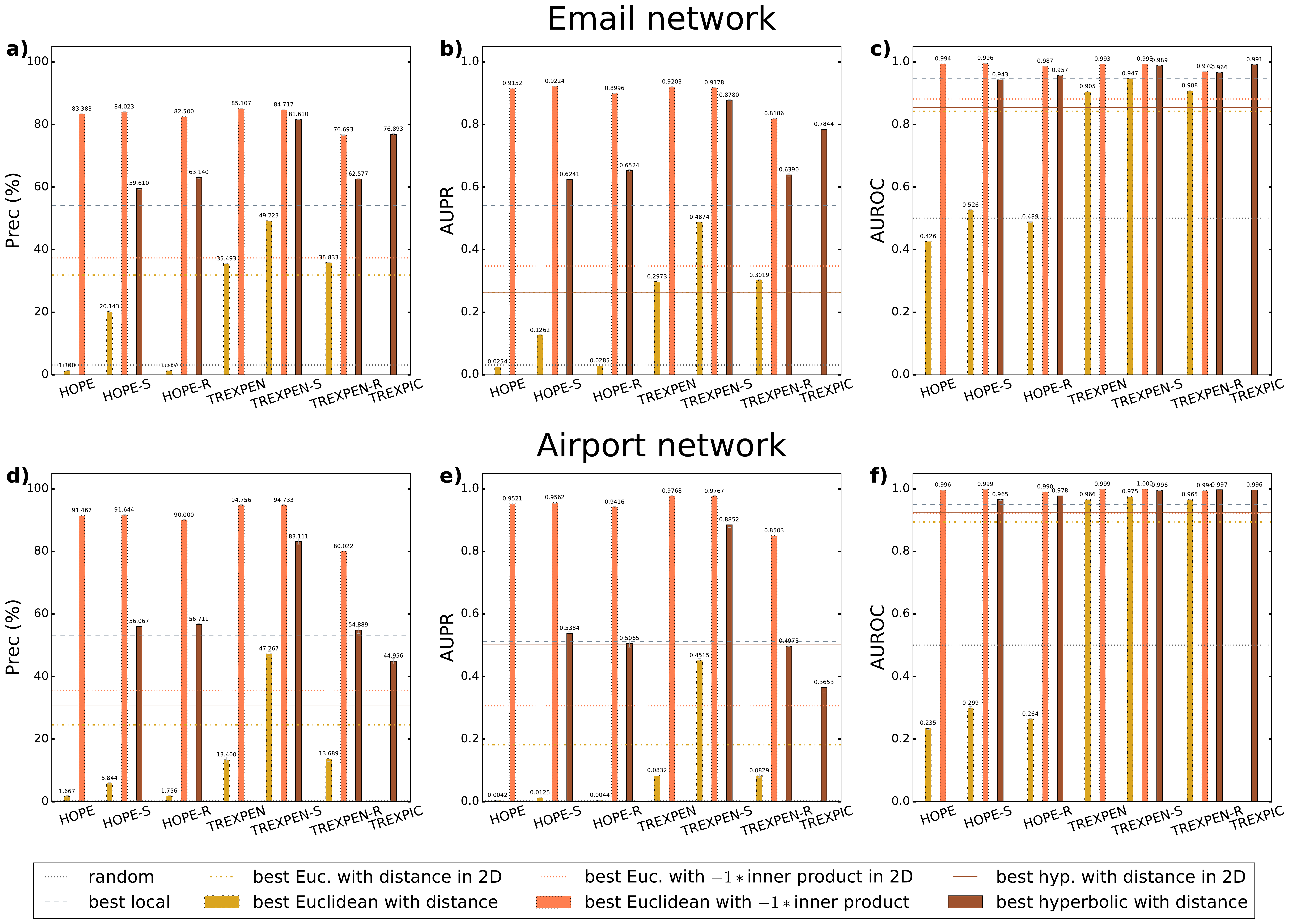}
    \caption{ {\bf Graph reconstruction performance on the additional directed real networks of larger sizes.} For each network, the task was to reconstruct $3$ samples of the links. The number of links per sample was $E_{\mathrm{sampled}}=10000$ for the email network and $E_{\mathrm{sampled}}=1500$ for the airport network. The colours indicate what geometric measure was used, as listed in the common legend at the bottom of the figure. The plotted values were obtained by averaging the results of the 3 samples, and the error bars show the standard deviations among the different samples. We depicted in each panel only the results of those parameter settings that turned out to be the best regarding the average of that given performance measure over the 3 samples. The bars were created considering all the tested number of dimensions, whereas the colourful horizontal lines show the best two-dimensional average performances achieved among all the embedding methods. Each row of panels refers to a real network named in the title of the row. The different columns of panels correspond to different measures: panels~a) and d) to the precision obtained when reconstructing the first $E_{\mathrm{sampled}}$ most probable links, panels~b) and e) to the area under the precision-recall (PR) curve, while panels~c) and f) to the area under the receiver operating characteristic (ROC) curve. The grey horizontal lines show the baselines: the average performance of the random predictor and the best average performance achieved among the local methods.}
    \label{fig:graRec_largeExtraDirNetworks}
\end{figure}

\begin{figure}[!h]
    \centering
    \includegraphics[width=1.0\textwidth]{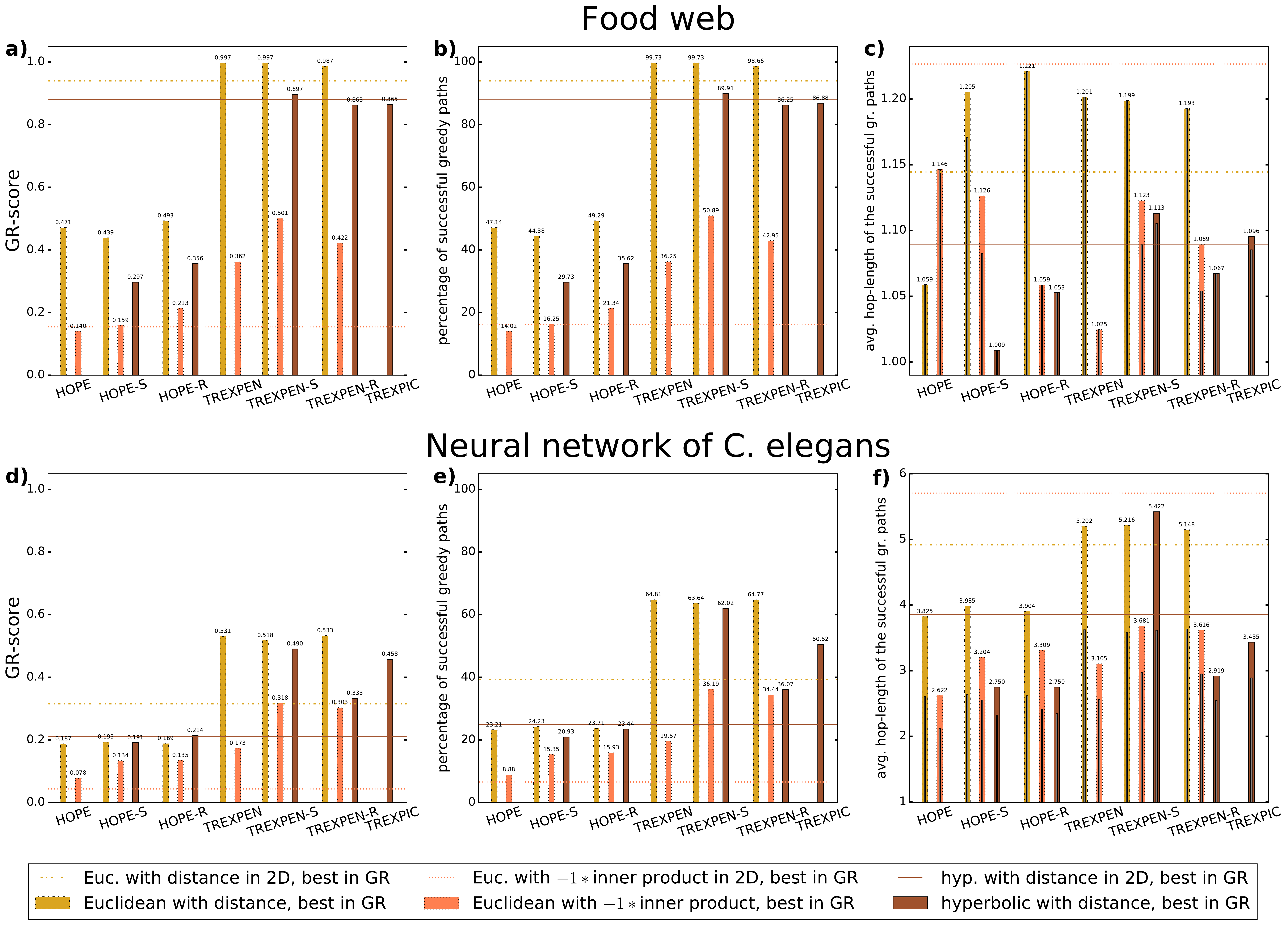}
    \caption{ {\bf Greedy routing performance on the additional directed real networks of smaller sizes.} For each network, the task was to perform greedy routing between each node pair connected by at least one directed path. The colours indicate what geometric measure was used, as listed in the common legend at the bottom of the figure. We plotted in each panel only the results of those parameter settings that turned out to be the best regarding the $\mathrm{GR\text{-}score}$. The bars were created considering all the tested number of dimensions, whereas the horizontal lines show the best two-dimensional performances achieved among all the embedding methods. Each row of panels refers to a real network named in the title of the row. The different columns of panels correspond to different measures: panels~a) and d) to the greedy routing score (the higher the better), panels~b) and e) to the success rate of greedy routing (the higher the better), while panels~c) and f) to the average hop-length of the successful greedy paths (the smaller the better), depicting with grey bars also the average of the hop-length of the shortest paths connecting those node pairs for which the greedy routing was successful.}
    \label{fig:GR_smallExtraDirNetworks}
\end{figure}

\begin{figure}[!h]
    \centering
    \includegraphics[width=1.0\textwidth]{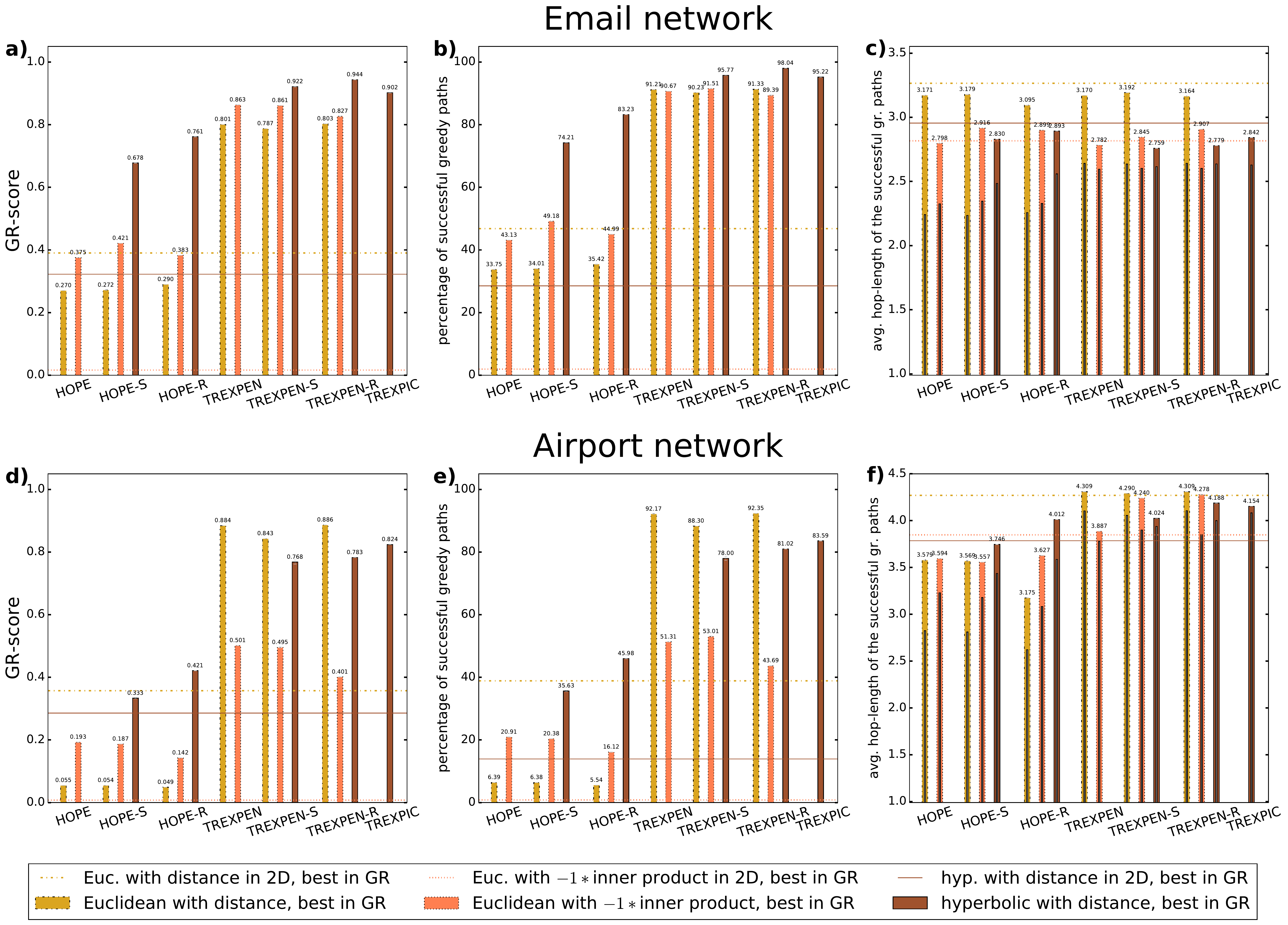}
    \caption{ {\bf Greedy routing performance on the additional directed real networks of larger sizes.} For each network, the task was to perform greedy routing in 3 samples of $500000$ node pairs connected by at least one directed path. The colours indicate what geometric measure was used, as listed in the common legend at the bottom of the figure. The plotted values were obtained by averaging the results of the 3 samples, and the error bars show the standard deviations among the different samples. We depicted in each panel only the results of those parameter settings that turned out to be the best regarding the average of the $\mathrm{GR\text{-}score}$ over the 3 samples. The bars were created considering all the tested number of dimensions, whereas the horizontal lines show the best two-dimensional average performances achieved among all the embedding methods. Each row of panels refers to a real network named in the title of the row. The different columns of panels correspond to different measures: panels~a) and d) to the greedy routing score (the higher the better), panels~b) and e) to the success rate of greedy routing (the higher the better), while panels~c) and f) to the average hop-length of the successful greedy paths (the smaller the better), depicting with grey bars also the average of the hop-length of the shortest paths connecting those node pairs for which the greedy routing was successful.}
    \label{fig:GR_largeExtraDirNetworks}
\end{figure}

First, in Fig.~\ref{fig:mapAcc_smallExtraDirNetworks} we show the mapping accuracy on the two smaller additional test networks, namely the food web and the neural network, while Fig.~\ref{fig:mapAcc_largeExtraDirNetworks} depicts the results obtained for the two larger additional networks, i.e. the email network and the airport network. Then, Figs.~\ref{fig:graRec_smallExtraDirNetworks} and \ref{fig:graRec_largeExtraDirNetworks} present the embedding quality with respect to the graph reconstruction task of the examined two smaller and two larger additional networks, respectively. Finally, Figs.~\ref{fig:GR_smallExtraDirNetworks} and \ref{fig:GR_largeExtraDirNetworks} depict the achieved greedy routing scores with the corresponding success rates and average hop-lengths for the examined starting node-destination node pairs in the studied two smaller and two larger additional test networks, respectively. Regarding these figures, basically the same conclusions can be drawn that were described in the main text: first, HOPE and its variants may fall behind our new methods in mapping accuracy and especially in greedy routing, but can be similarly well applied for graph reconstruction, and second, the Euclidean inner product and the hyperbolic distance seem to be a better indicator of both the shortest path lengths and the connection probabilities than the Euclidean distance. However, for certain networks (Fig.~\ref{fig:GR_smallExtraDirNetworks}a, d and Fig.~\ref{fig:GR_largeExtraDirNetworks}d), instead of the hyperbolic, the Euclidean distance was proven to be the most suitable geometric measure for the navigation task -- nevertheless, the hyperbolic distance also performed well in these cases too.

\newpage
\null\newpage
\null\newpage
\null\newpage
\null\newpage
\null\newpage
\section{The impact of the links' directedness on the embeddings}
\label{sect:directednessOfLinks}
\setcounter{figure}{0}
\setcounter{table}{0}
\setcounter{equation}{0}
\renewcommand{\thefigure}{S6.\arabic{figure}}
\renewcommand{\thetable}{S6.\arabic{table}}
\renewcommand{\theequation}{S6.\arabic{equation}}

In this section, we quantify how important it is to consider the links' directedness in the embedding process by comparing the results presented in the main text and Sect.~\ref{sect:extraRealDirEmbeddings} to the performances achieved in mapping accuracy, graph reconstruction 
and greedy routing on the embeddings of the undirected version of the actually directed networks that we studied, where only one position vector is assigned to each node, which can be interpreted both as its source and as its target position. To provide a fair comparison, we carried out exactly the same tasks with the undirected embeddings as in the directed case, meaning that during the evaluation we always took into account the directedness of the links -- i.e., we examined the same set of potentially connected node pairs in graph reconstruction, 
and considered the directed paths on the graphs in the case of mapping accuracy and greedy routing. The list of the tested embedding parameters $\alpha$ and $q$ were determined the same way in the undirected cases as for the directed networks (see Sects.~\ref{sect:HOPE}, \ref{sect:TREXPEN} and \ref{sect:TREXPIC}), but based on the adjacency matrix and the shortest path lengths of the undirected graphs.

We present two figures concerning each task, of which the first one always refers to the four smaller directed real networks that we studied in the main text or in Sect.~\ref{sect:extraRealDirEmbeddings}, while the second one deals with the altogether four networks of size close to or above $N=1000$. In the latter case, we always evaluated the undirected embeddings on exactly the same 3 samples of the node pairs that we used for evaluating the corresponding directed embeddings, meaning that the differences in the measured performances arose solely from the differences between the directed and the undirected embeddings and not from the evaluation process, not only for the smaller networks but also for the larger networks. 

\begin{figure}[!h]
    \centering
    \includegraphics[width=1.0\textwidth]{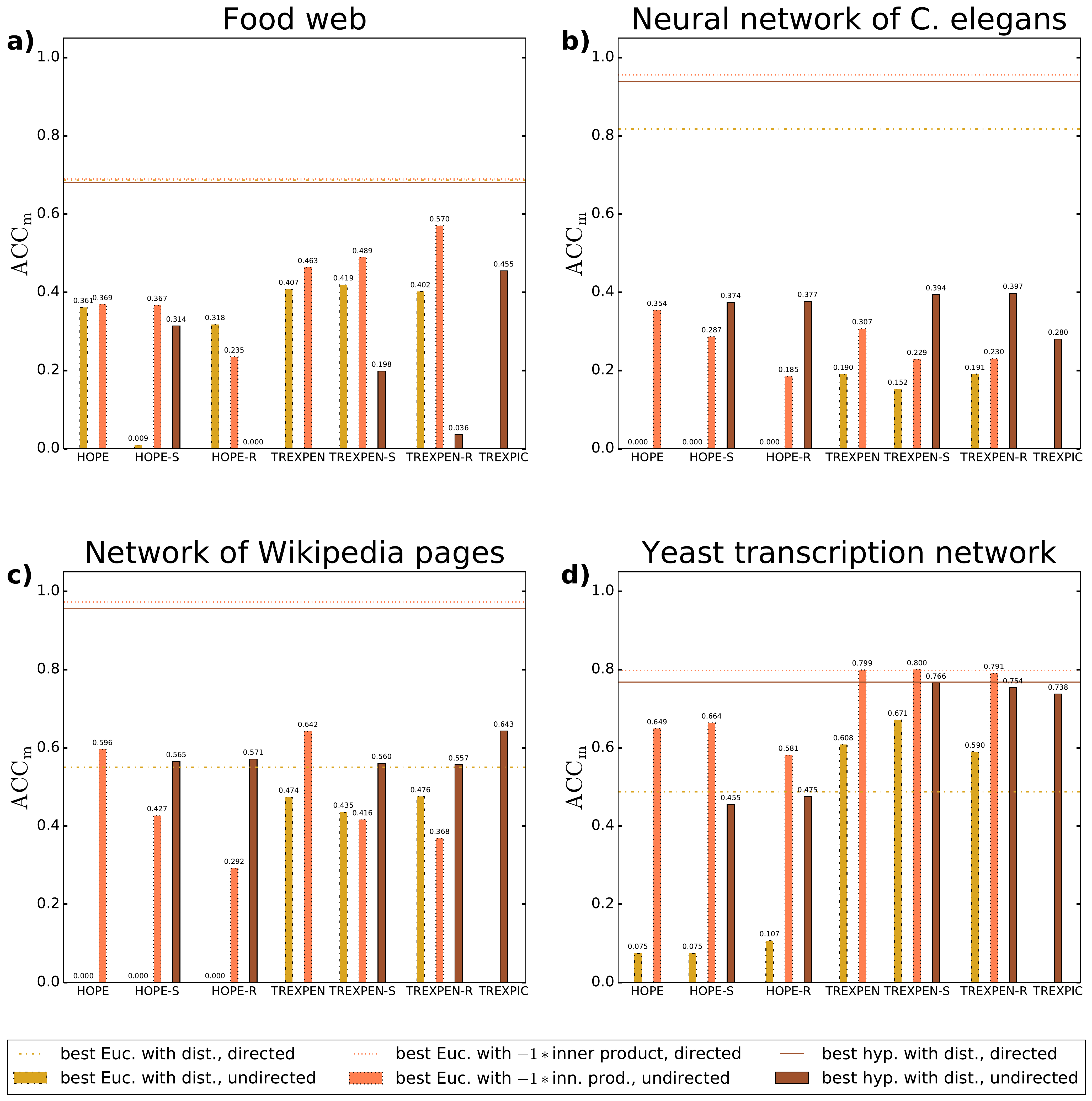}
    \caption{ {\bf Mapping accuracy on the undirected version of the smaller real networks that were examined in panels~a) and b) in Fig.~\ref{fig:mapAccMain} of the main text and in Fig.~\ref{fig:mapAcc_smallExtraDirNetworks} of the previous section.} The horizontal lines depict the best performance that was achieved with a given geometric measure among all the examined embedding methods in the directed case. The bars show the performance of the different embedding methods at their best parameter setting when inputting in the undirected version of the network. Note that the 0 values denote that the given methods have not achieved any positive value.}
    \label{fig:mapAcc_symm_small}
\end{figure}

\begin{figure}[!h]
    \centering
    \includegraphics[width=1.0\textwidth]{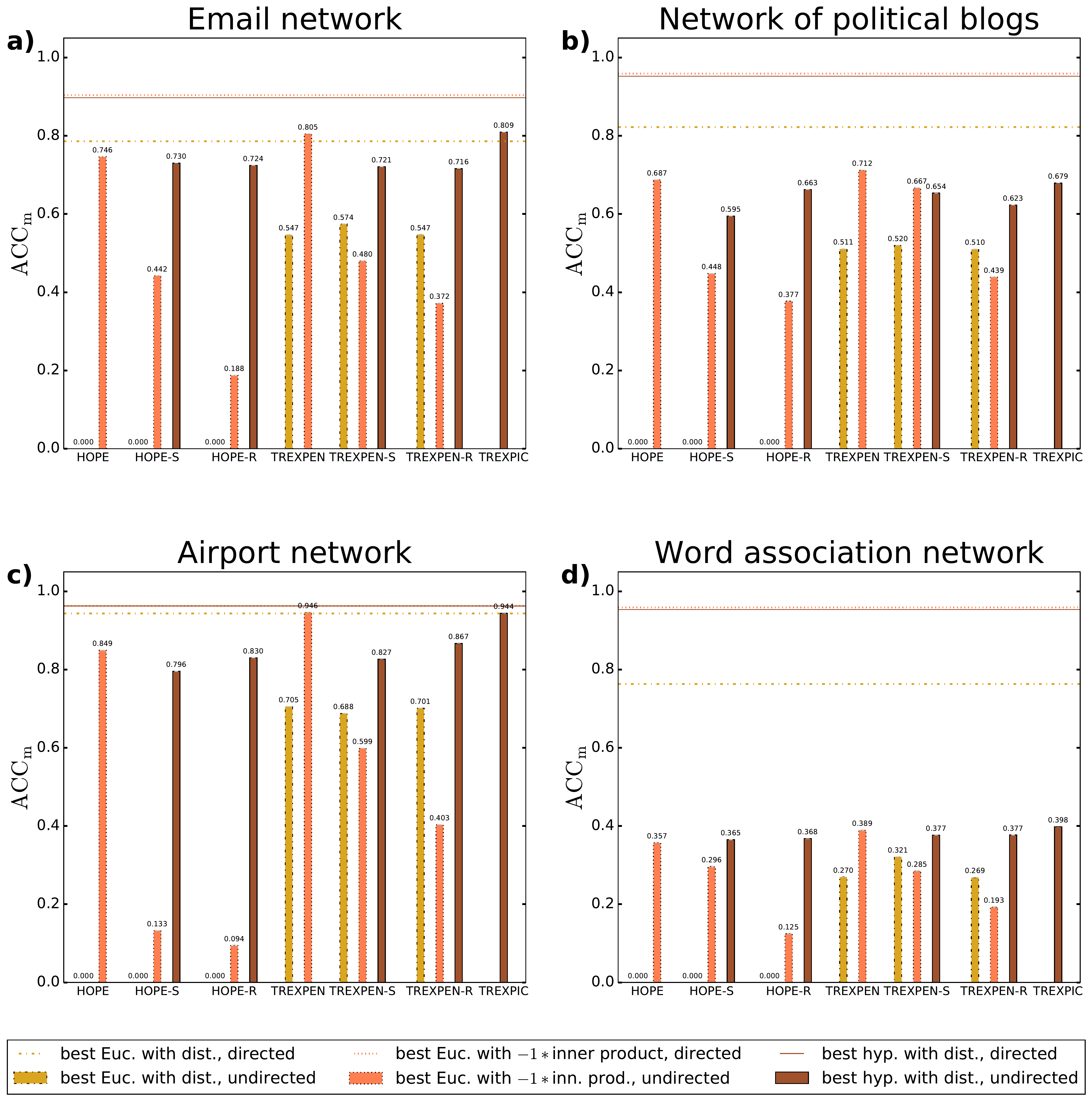}
    \caption{ {\bf Mapping accuracy on the undirected version of the larger real networks that were examined in panels~c) and d) in Fig.~\ref{fig:mapAccMain} of the main text and in Fig.~\ref{fig:mapAcc_largeExtraDirNetworks} of the previous section.} The horizontal lines depict the best average performance that was achieved with a given geometric measure among all the examined directed embedding methods. The bars show the average performance for the different embedding methods at the parameter setting that was the best on average when inputting in the undirected version of the network. Note that the 0 values denote that the given methods have not achieved any positive average value. The error bars show the corresponding standard deviations between the 3 tests.}
    \label{fig:mapAcc_symm_large}
\end{figure}

\begin{figure}[!h]
    \centering
    \includegraphics[width=0.88\textwidth]{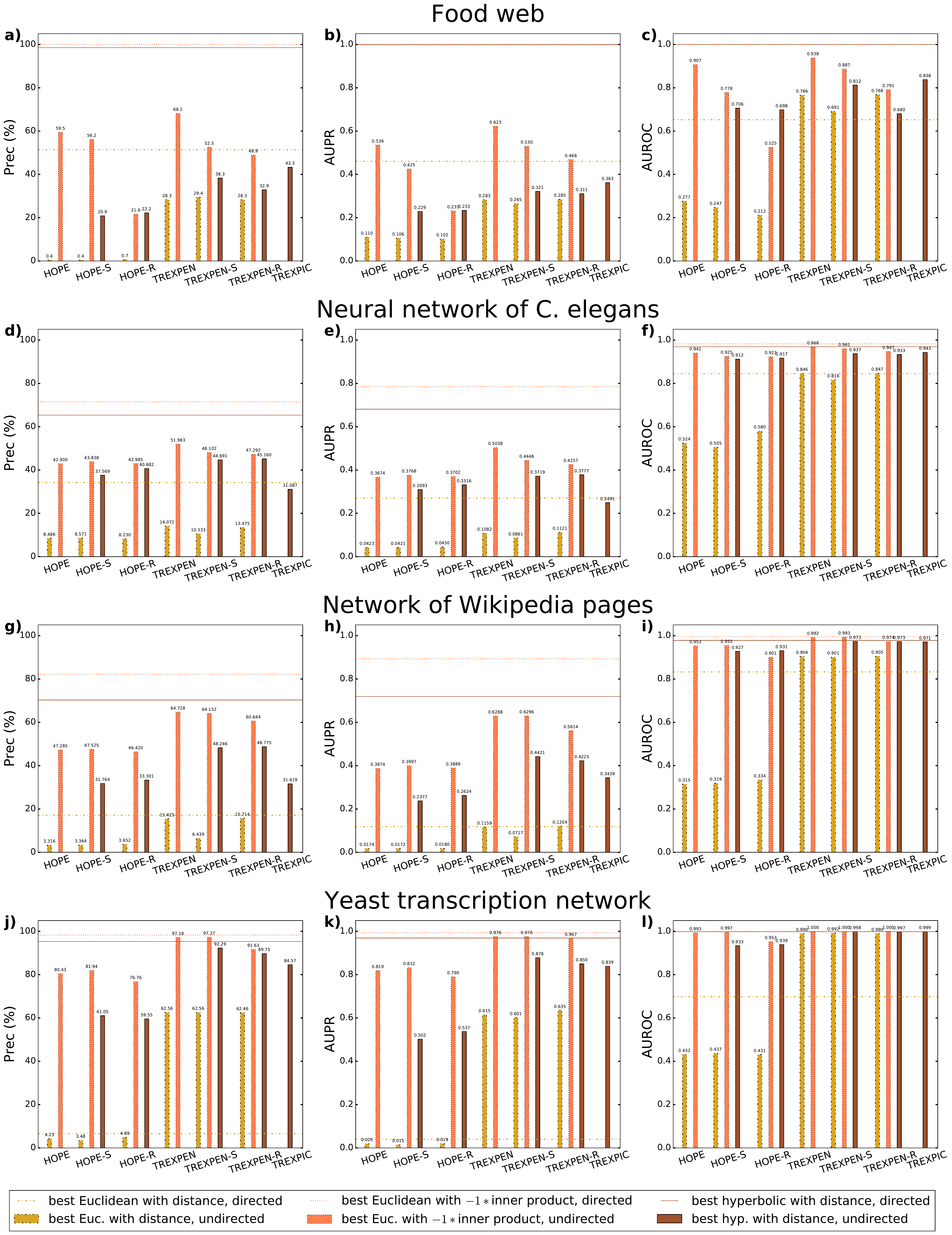}
    \caption{ {\bf Graph reconstruction on the undirected version of the smaller real networks that were examined in panels~a)--f) in Fig.~\ref{fig:graRecMain} of the main text and in Fig.~\ref{fig:graRec_smallExtraDirNetworks} of the previous section.} The horizontal lines depict for each measure the best performance that was achieved with a given geometric measure among all the examined embedding methods in the directed case. The bars show the performance of the different embedding methods at their best parameter setting when inputting in the undirected version of the network.}
    \label{fig:graRec_symm_small}
\end{figure}

\begin{figure}[!h]
    \centering
    \includegraphics[width=0.87\textwidth]{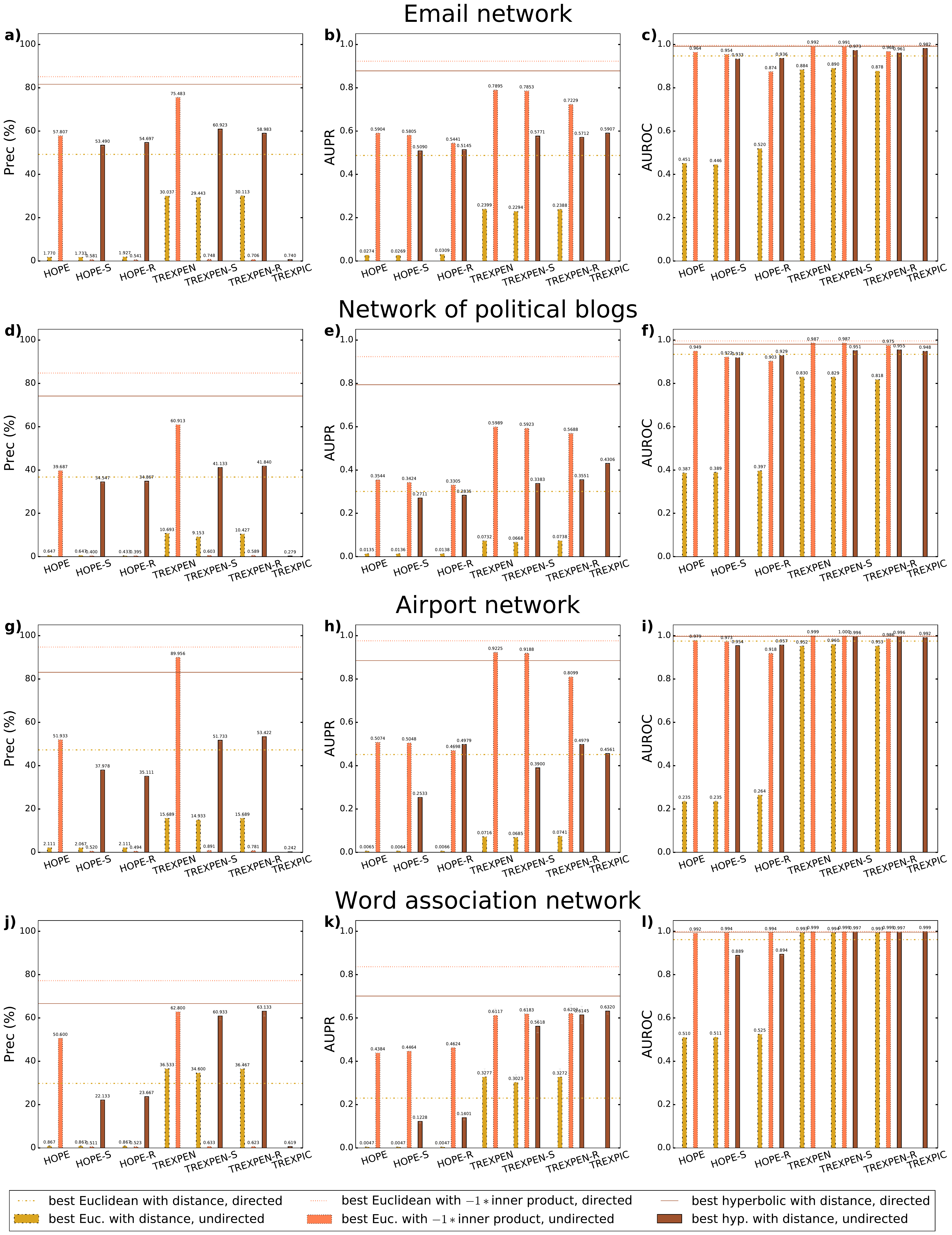}
    \caption{ {\bf Graph reconstruction on the undirected version of the larger real networks that were examined in panels~g)--l) in Fig.~\ref{fig:graRecMain} of the main text and in Fig.~\ref{fig:graRec_largeExtraDirNetworks} of the previous section.} The horizontal lines depict for each measure the best average performance that was achieved with a given geometric measure among all the examined directed embedding methods. The bars show the average performance for the different embedding methods at the parameter setting that was the best on average when inputting in the undirected version of the network. The error bars show the corresponding standard deviations between the 3 tests.}
    \label{fig:graRec_symm_large}
\end{figure}

\begin{figure}[!h]
    \centering
    \includegraphics[width=0.88\textwidth]{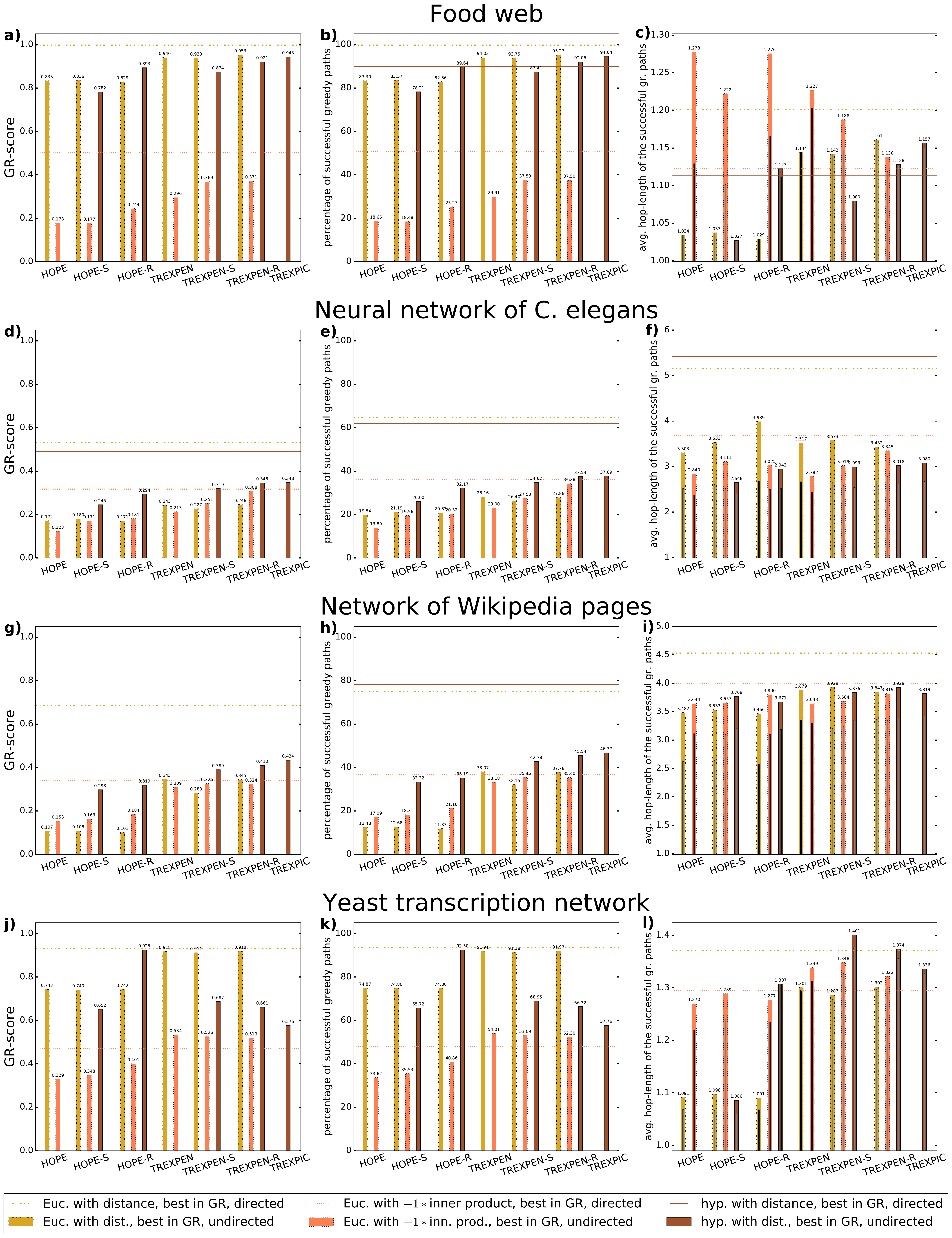}
    \caption{ {\bf Greedy routing performance on the undirected version of the smaller real networks that were examined in panels~a)--f) in Fig.~\ref{fig:GRmain} of the main text and in Fig.~\ref{fig:GR_smallExtraDirNetworks} of the previous section.} The horizontal lines depict the results belonging to the best GR-score that was achieved with a given geometric measure among all the examined embedding methods in the directed case. The bars show the performance of the different embedding methods at that parameter setting that was their best with regard to the GR-score when inputting in the undirected version of the network.}
    \label{fig:GR_symm_small}
\end{figure}

\begin{figure}[!h]
    \centering
    \includegraphics[width=0.86\textwidth]{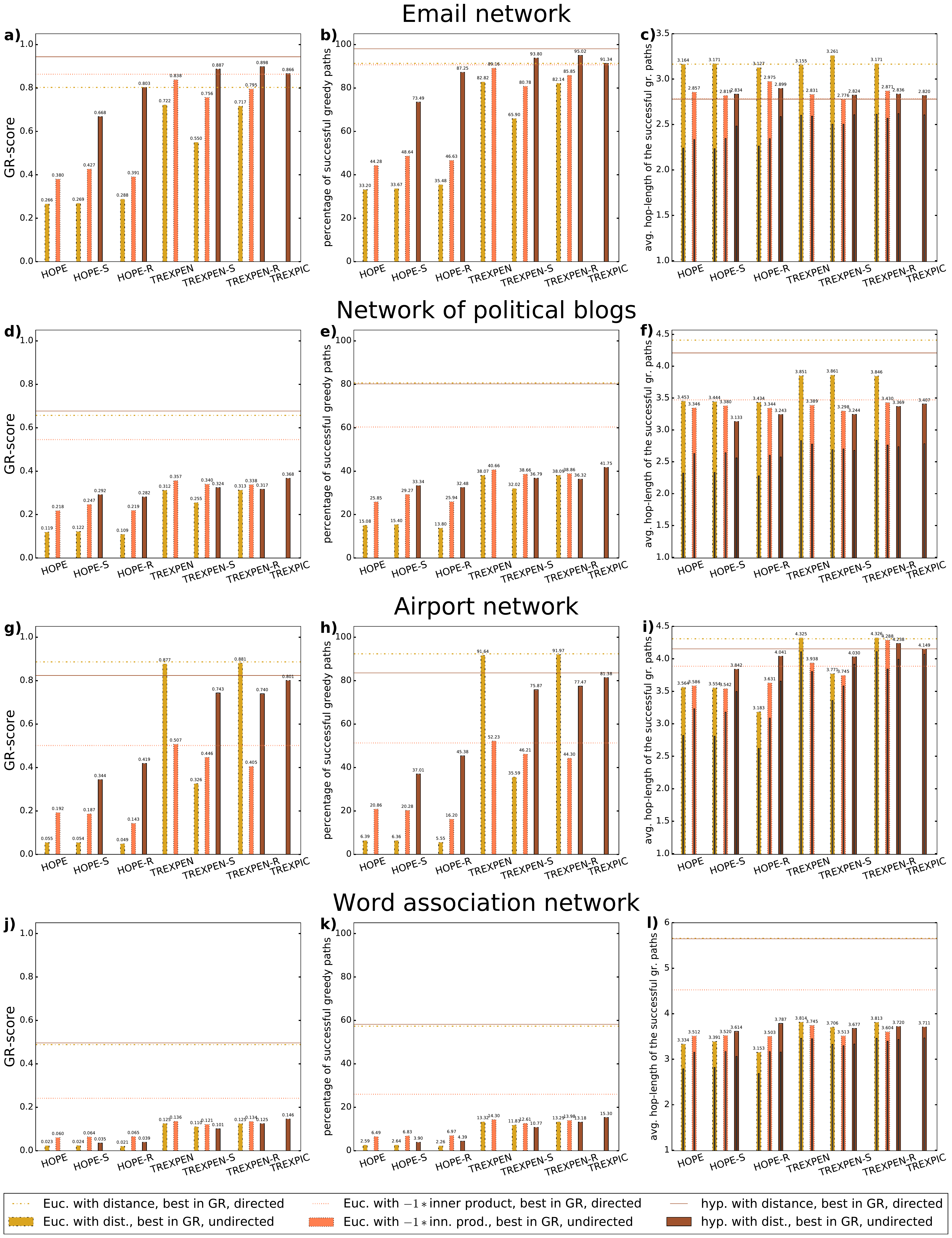}
    \caption{ {\bf Greedy routing performance on the undirected version of the larger real networks that were examined in panels~g)--l) in Fig.~\ref{fig:GRmain} of the main text and in Fig.~\ref{fig:GR_largeExtraDirNetworks} of the previous section.} The horizontal lines depict the average results belonging to the best average GR-score that was achieved with a given geometric measure among all the examined directed embedding methods. The bars show the average performance for the different embedding methods at that parameter setting that was their best with regard to the average of the GR-score when inputting in the undirected version of the network. The error bars show the corresponding standard deviations between the 3 tests.}
    \label{fig:GR_symm_large}
\end{figure}

First, Figs.~\ref{fig:mapAcc_symm_small} and \ref{fig:mapAcc_symm_large} deal with the mapping accuracy, as Fig.~\ref{fig:mapAccMain} in the main text and Figs.~\ref{fig:mapAcc_smallExtraDirNetworks}--\ref{fig:mapAcc_largeExtraDirNetworks} in the previous section. Then, Figs.~\ref{fig:graRec_symm_small} and \ref{fig:graRec_symm_large} can be paired to Fig.~\ref{fig:graRecMain} in the main text and Figs.~\ref{fig:graRec_smallExtraDirNetworks}--\ref{fig:graRec_largeExtraDirNetworks}, presenting how the embeddings performed in graph reconstruction. 
Finally, Figs.~\ref{fig:GR_symm_small} and \ref{fig:GR_symm_large} depict the examined measures concerning the greedy routing, just like Fig.~\ref{fig:GRmain} in the main text and Figs.~\ref{fig:GR_smallExtraDirNetworks}--\ref{fig:GR_largeExtraDirNetworks}. All of these figures clearly confirm that disregarding the link directions in directed networks can significantly hinder the performance of the examined embedding methods, showing that these methods are actually able to grasp the directedness of the inputted networks and providing a strong motivation for the application of such embedding techniques that are capable of utilizing the information of link directions.

\newpage
\null\newpage
\section{Embedding performance on undirected real networks}
\label{sect:undirEmb}
\setcounter{figure}{0}
\setcounter{table}{0}
\setcounter{equation}{0}
\renewcommand{\thefigure}{S7.\arabic{figure}}
\renewcommand{\thetable}{S7.\arabic{table}}
\renewcommand{\theequation}{S7.\arabic{equation}}

This section demonstrates that HOPE, TREXPEN, TREXPIC and their variants can provide good performance even when used on undirected networks (placing each node as a source of links at its target position and vice versa) and are able to cope with well-known dimension reduction techniques that have already been proven to be outstanding in the creation of undirected embeddings. The undirected methods that we used as references were the following:
\begin{itemize}
    \item hydra (hyperbolic distance recovery and approximation)~\cite{Hydra,Hydra_code}, which places network nodes directly in the $d$-dimensional hyperbolic space of curvature $K=-\zeta^2=-1$ by performing the eigendecomposition of a matrix that estimates the pairwise Lorentz products in the hyperboloid representation based on the shortest path lengths of the network to be embedded,
    \item Laplacian eigenmaps (LE)~\cite{LE,LaplEigmap_code} that uses the eigendecomposition of the network’s Laplacian matrix to place its nodes in the $d$-dimensional Euclidean space,
    \item Isomap (ISO)~\cite{Isomap} and noncentred Isomap (ncISO)~\cite{coalescentEmbedding}, which embed the network nodes in the $d$-dimensional Euclidean space via the singular value decomposition (SVD) on the centred or the noncentred version of the matrix of pairwise shortest path lengths, respectively,
    \item and the hyperbolic versions~\cite{coalescentEmbedding} of LE, ISO and ncISO, in which the Euclidean radial node arrangement obtained from the dimension reduction is replaced according to the PSO model of hyperbolic network growth~\cite{PSO} while keeping the angular coordinates unaltered.
\end{itemize}
Note that while hydra is similar to TREXPIC, the latter uses exponential distances and SVD instead of the plain shortest path lengths themselves and eigendecomposition. Furthermore, to obtain well-comparable hyperbolic embeddings, we converted the results yielded by the hydra method in the Poincaré ball representation of the hyperbolic space to the equivalent native ball representation, where all the other examined methods place the nodes. In addition, while the Euclidean ISO and ncISO methods are similar to the Euclidean HOPE-S, HOPE-R, TREXPEN-S and TREXPEN-R in the sense that all of these methods are fundamentally based on the path lengths measured along the links of a network and create angularly not restricted patterns based on SVD, there is an important difference between them: ISO and ncISO reduce a matrix that estimates pairwise distances in the Euclidean space, while our new methods search for such a Euclidean node arrangement that reproduces a matrix of pairwise proximities.

\begin{figure}[!h]
    \centering
    \includegraphics[width=1.0\textwidth]{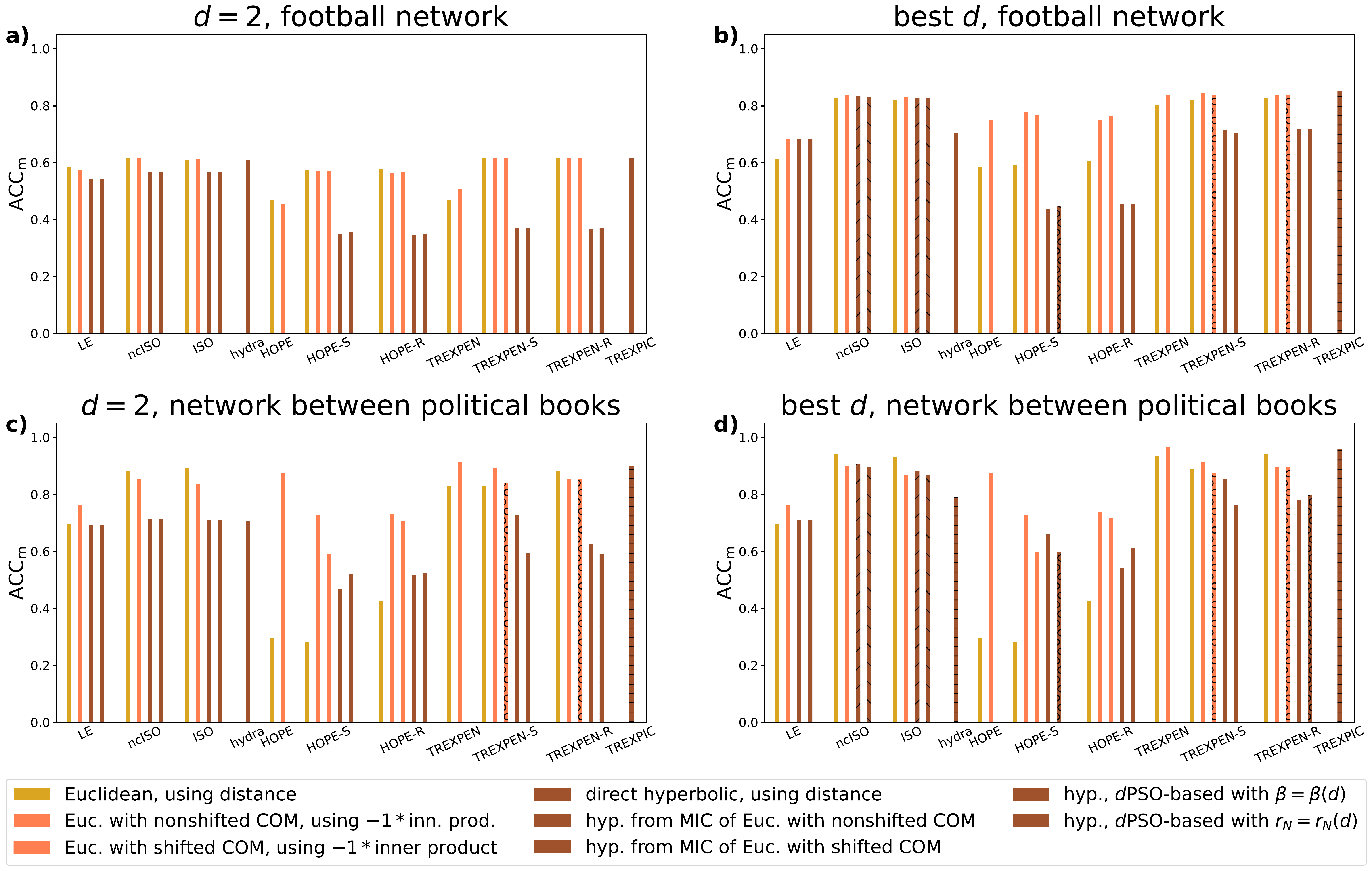}
    \caption{ {\bf Mapping accuracy on the examined two undirected real networks.} Each row of panels refers to a network named in the title of the corresponding two panels. For both networks, we measured the mapping accuracy considering all the (unordered) node pairs. The colours indicate what geometric measure was used, while the different patterns denote different characteristics regarding the way the node arrangements were created in the given geometry, as listed in the common legend at the bottom of the figure. We plotted for each method only the result of the parameter setting or trial that turned out to be the best, i.e. which yielded the highest mapping accuracy. In panels~a) and c), we embedded the networks only in the Euclidean or hyperbolic plane, while the bars in panels~b) and d) were created considering all the tested number of dimensions $d$.}
    \label{fig:undirMapAcc}
\end{figure}

\begin{figure}[!h]
    \centering
    \includegraphics[width=1.0\textwidth]{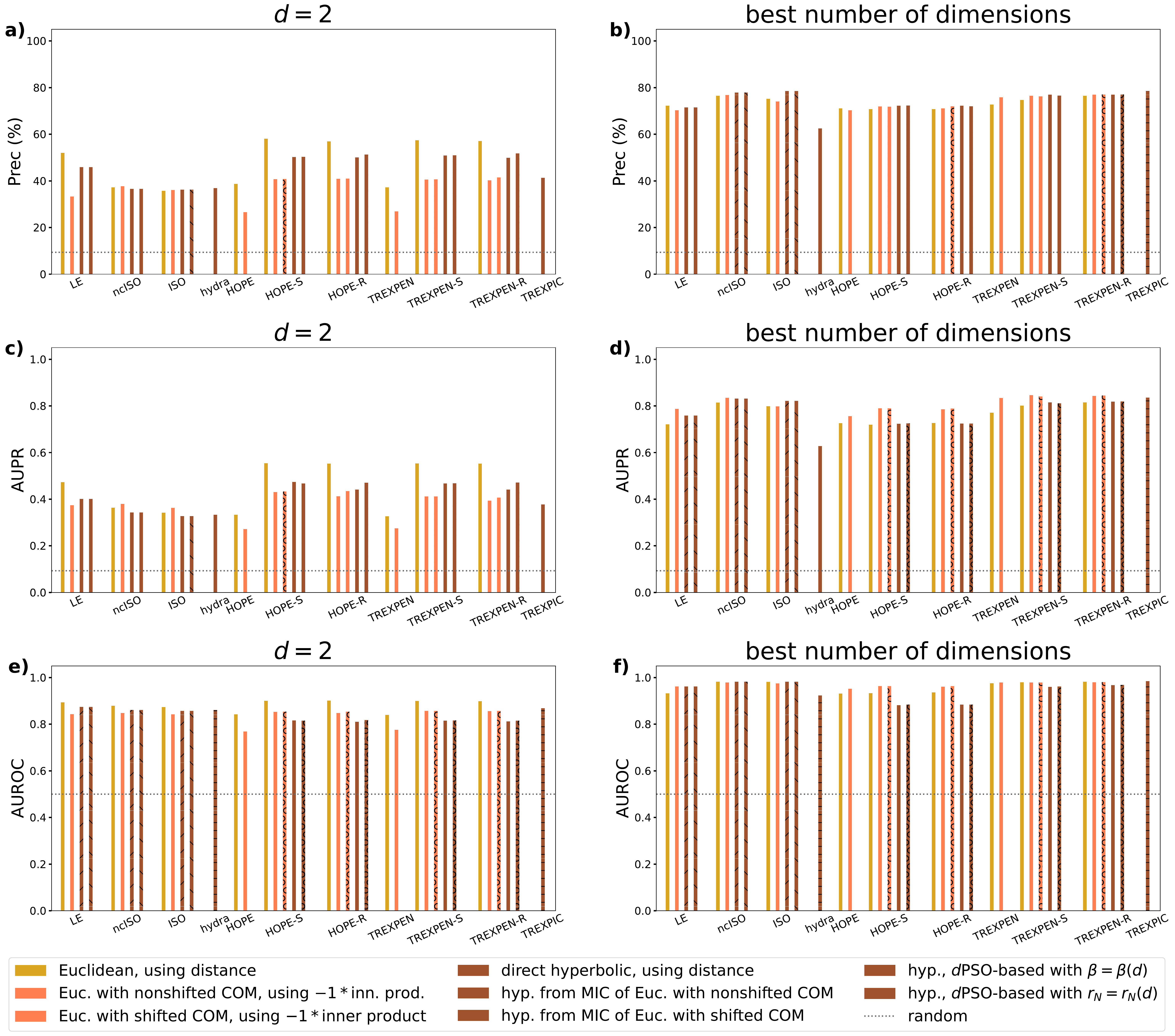}
    \caption{ {\bf Graph reconstruction performance on the football network.} The task was to reconstruct each one of the $E$ number of links in the network by ranking all the (unordered) node pairs using the node positions obtained when embedding the network knowing all of its links. The colours indicate what geometric measure was used, while the different patterns denote different characteristics regarding the way the node arrangements were created in the given geometry, as listed in the common legend at the bottom of the figure. Each row of panels refers to a given measure of embedding quality: panels~a) and b) to the precision obtained when reconstructing the first $E$ most probable links, panels~c) and d) to the area under the precision-recall (PR) curve, whilst panels~e) and f) to the area under the receiver operating characteristic (ROC) curve. We plotted in each panel for each method only the result of the parameter setting or trial that turned out to be the best regarding the given performance measure. In panels~a), c) and e), we embedded the networks only in the Euclidean or hyperbolic plane, while the bars in panels~b), d) and f) were created considering all the tested number of dimensions $d$. The grey horizontal lines show the performance of the random predictor.}
    \label{fig:undirGraRec_football}
\end{figure}

\begin{figure}[!h]
    \centering
    \includegraphics[width=1.0\textwidth]{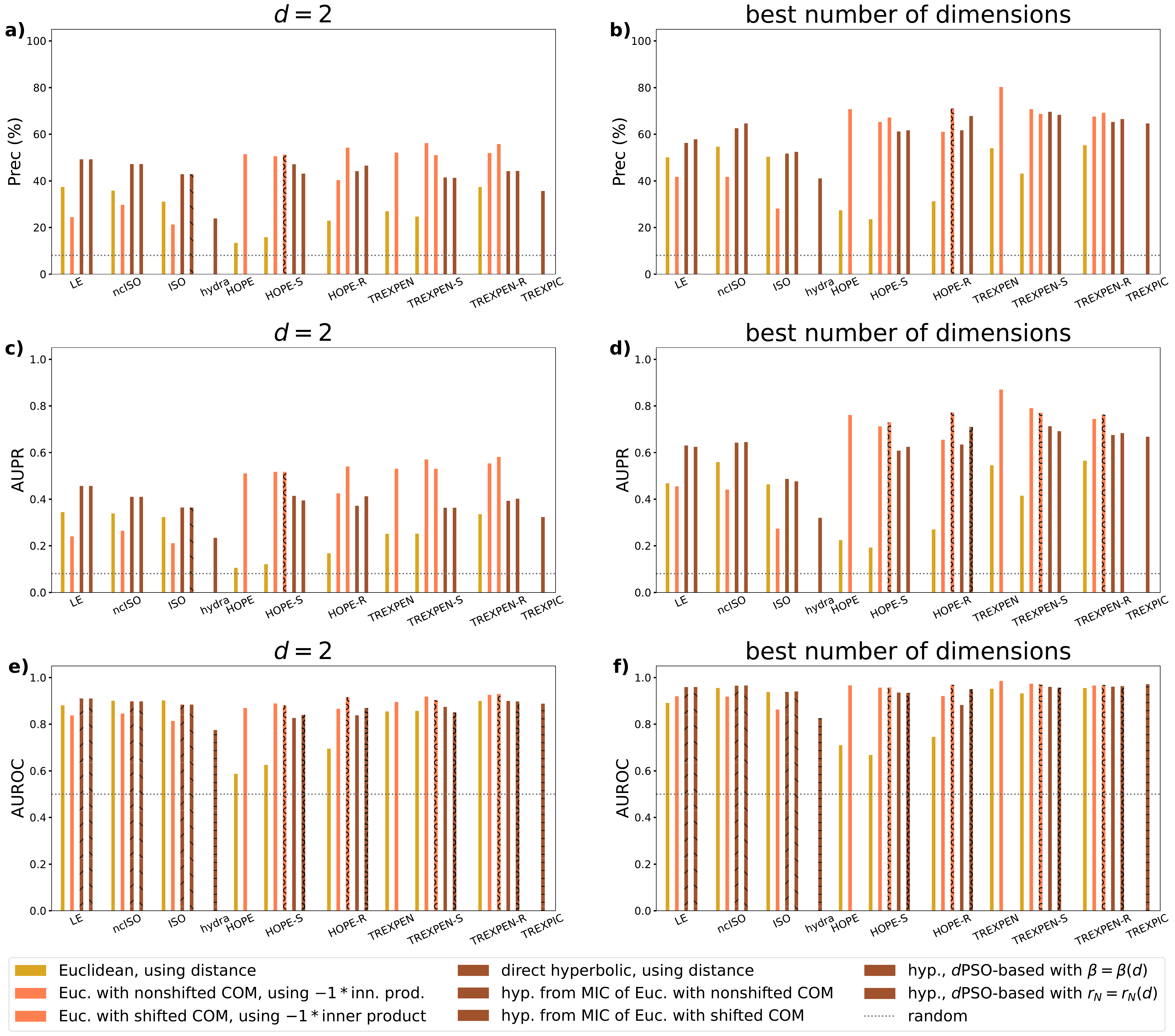}
    \caption{ {\bf Graph reconstruction performance on the network of political books.} The task was to reconstruct each one of the $E$ number of links in the network by ranking all the (unordered) node pairs using the node positions obtained when embedding the network knowing all of its links. The colours indicate what geometric measure was used, while the different patterns denote different characteristics regarding the way the node arrangements were created in the given geometry, as listed in the common legend at the bottom of the figure. Each row of panels refers to a given measure of embedding quality: panels~a) and b) to the precision obtained when reconstructing the first $E$ most probable links, panels~c) and d) to the area under the precision-recall (PR) curve, whilst panels~e) and f) to the area under the receiver operating characteristic (ROC) curve. We plotted in each panel for each method only the result of the parameter setting or trial that turned out to be the best regarding the given performance measure. In panels~a), c) and e), we embedded the networks only in the Euclidean or hyperbolic plane, while the bars in panels~b), d) and f) were created considering all the tested number of dimensions $d$. The grey horizontal lines show the performance of the random predictor.}
    \label{fig:undirGraRec_polBooks}
\end{figure}

\begin{figure}[!h]
    \centering
    \includegraphics[width=1.0\textwidth]{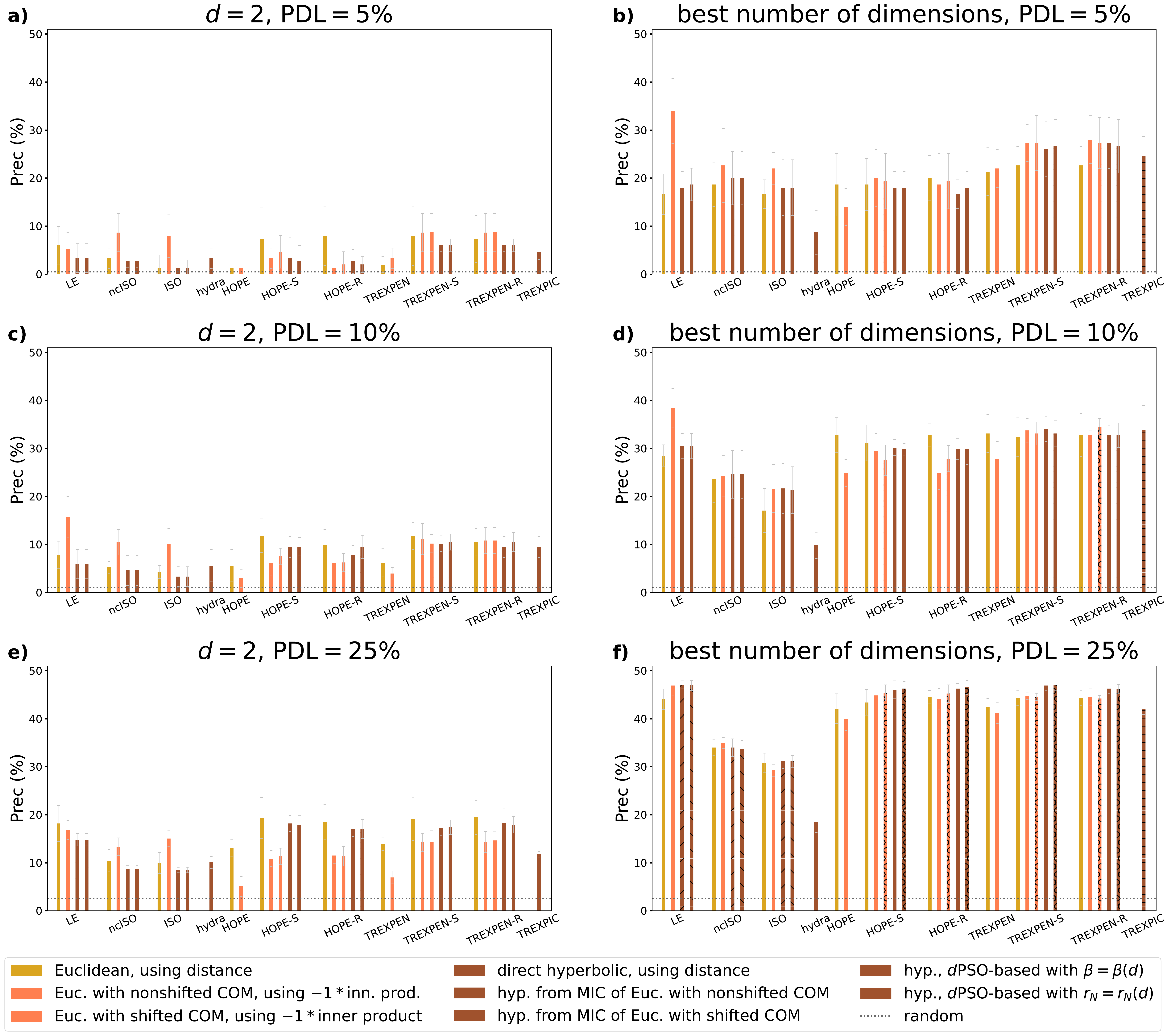}
    \caption{ {\bf Link prediction performance on the football network.} In this task, we removed a given number of links from the original network, embedded the LCC of the resulted pruned graph and by ranking all the unconnected (unordered) node pairs that were embedded according to a given geometric measure associated with connection probability, tried to reconstruct all the $E_{\mathrm{missing}}$ number of the deleted links that connected in the original graph such nodes that both were embedded eventually. Each row of panels refers to a given proportion of deleted links (PDL) among all the connections. The link removal and prediction was repeated $5$ times with each PDL. We considered in each case only the result of the parameter setting or trial that turned out to be the best regarding the precision obtained when reconstructing the first $E_{\mathrm{missing}}$ most probable links. The plotted values were obtained by averaging the results of the $5$ link prediction tasks and the error bars show the standard deviations among the quality scores achieved on the $5$ different sets of missing links. The colours indicate what geometric measure was used, while the different patterns denote different characteristics regarding the way the node arrangements were created in the given geometry, as listed in the common legend at the bottom of the figure. In panels~a), c) and e), we embedded the networks only in the Euclidean or hyperbolic plane, while the bars in panels~b), d) and f) were created considering all the tested number of dimensions $d$. The grey horizontal lines show the average performance of the random predictor.}
    \label{fig:undirLinkPred_football}
\end{figure}

\begin{figure}[!h]
    \centering
    \includegraphics[width=1.0\textwidth]{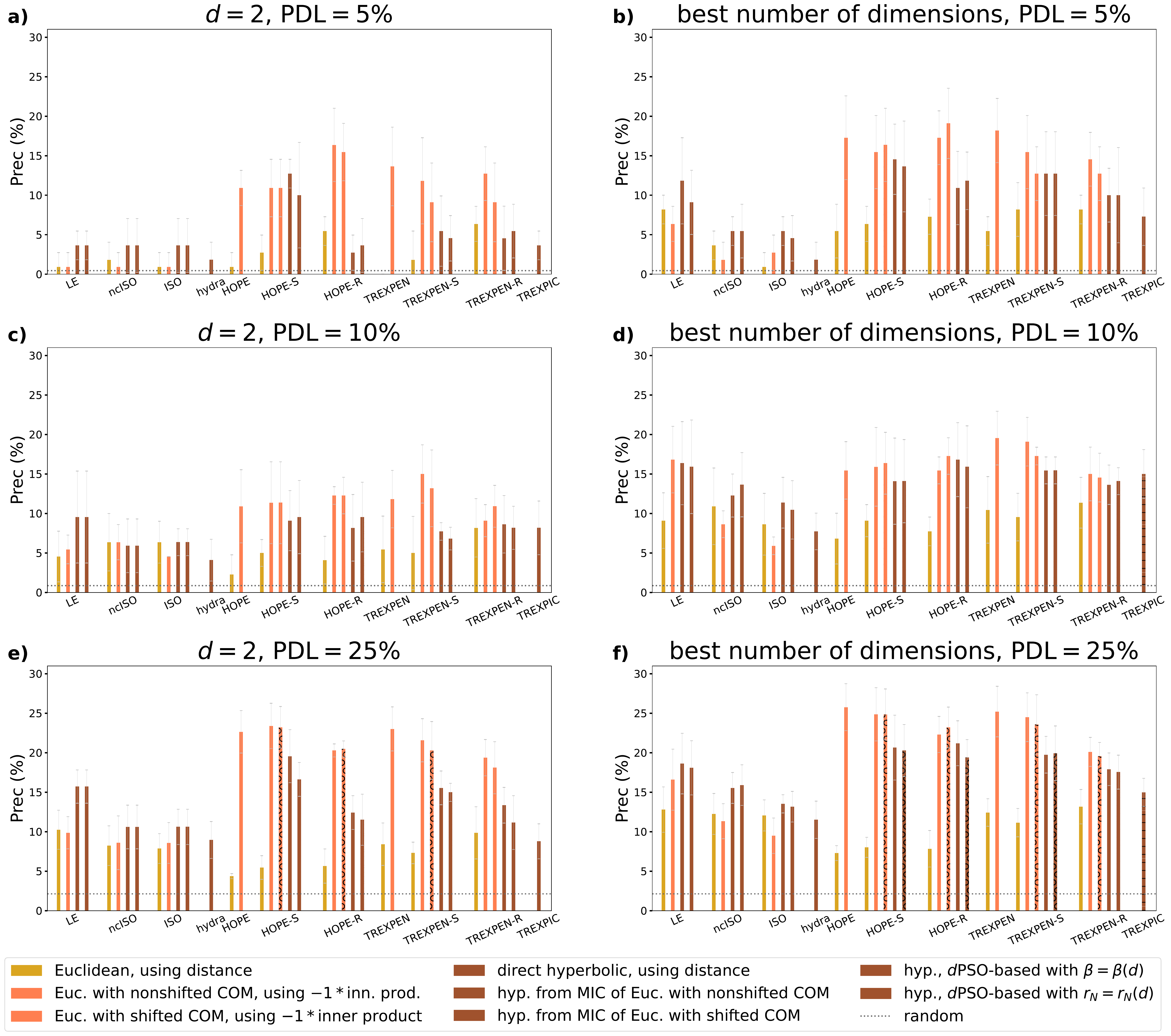}
    \caption{ {\bf Link prediction performance on the network of political books.} In this task, we removed a given number of links from the original network, embedded the LCC of the resulted pruned graph and by ranking all the unconnected (unordered) node pairs that were embedded according to a given geometric measure associated with connection probability, tried to reconstruct all the $E_{\mathrm{missing}}$ number of the deleted links that connected in the original graph such nodes that both were embedded eventually. Each row of panels refers to a given proportion of deleted links (PDL) among all the connections. The link removal and prediction was repeated $5$ times with each PDL. We considered in each case only the result of the parameter setting or trial that turned out to be the best regarding the precision obtained when reconstructing the first $E_{\mathrm{missing}}$ most probable links. The plotted values were obtained by averaging the results of the $5$ link prediction tasks and the error bars show the standard deviations among the quality scores achieved on the $5$ different sets of missing links. The colours indicate what geometric measure was used, while the different patterns denote different characteristics regarding the way the node arrangements were created in the given geometry, as listed in the common legend at the bottom of the figure. In panels~a), c) and e), we embedded the networks only in the Euclidean or hyperbolic plane, while the bars in panels~b), d) and f) were created considering all the tested number of dimensions $d$. The grey horizontal lines show the average performance of the random predictor.}
    \label{fig:undirLinkPred_polBooks}
\end{figure}

\begin{figure}[!h]
    \centering
    \includegraphics[width=1.0\textwidth]{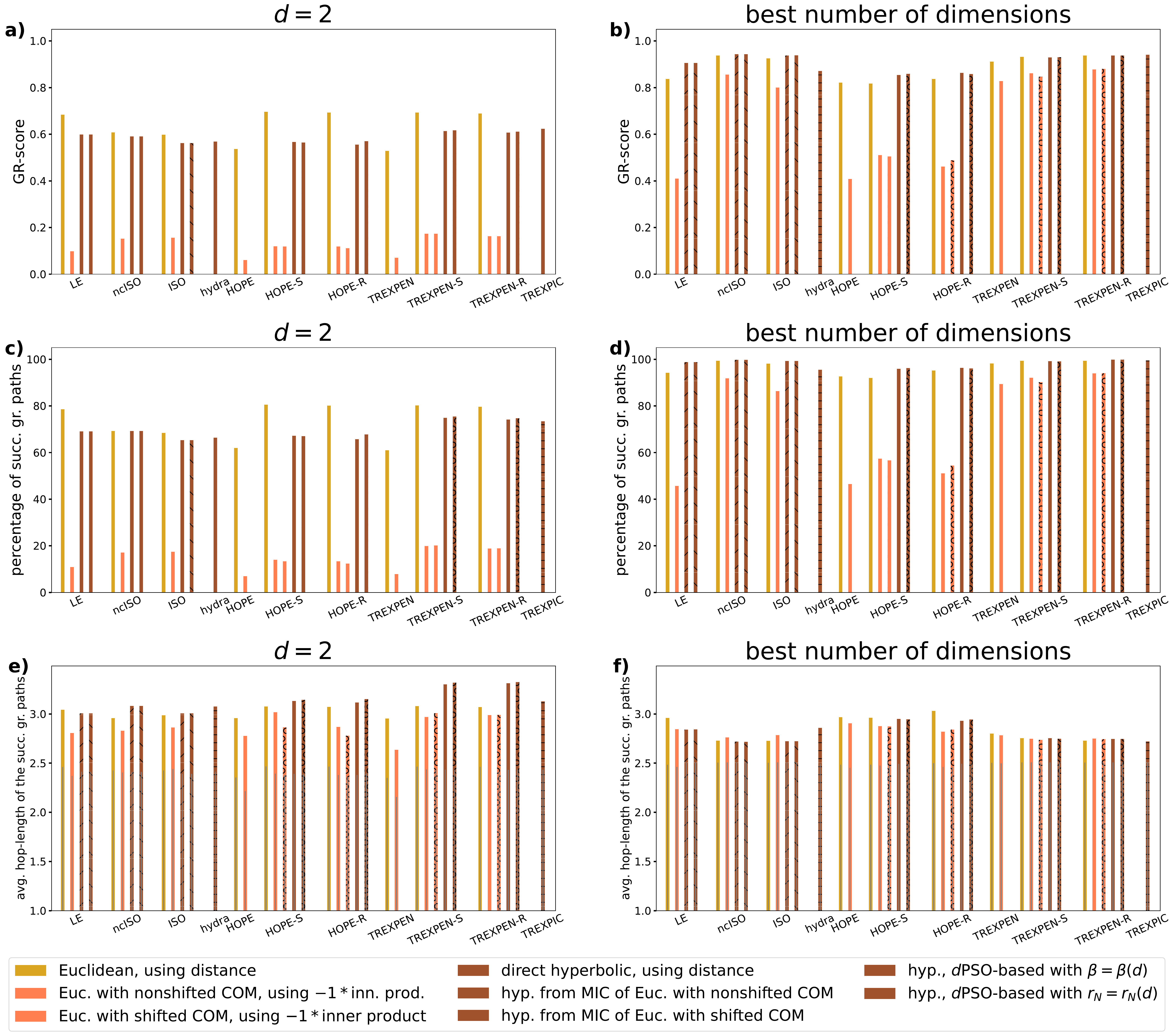}
    \caption{ {\bf Greedy routing performance on the football network.} The task was to perform greedy routing between each one of the (unordered) node pairs. The colours indicate what geometric measure was used, while the different patterns denote different characteristics regarding the way the node arrangements were created in the given geometry, as listed in the common legend at the bottom of the figure. Each row of panels refers to a given measure of embedding quality: panels~a) and b) to the greedy routing score (the higher the better), panels~c) and d) to the success rate of greedy routing (the higher the better), whilst panels~e) and f) to the average hop-length of the successful greedy paths (the smaller the better), depicting with grey bars also the average of the hop-length of the shortest paths connecting those node pairs for which the greedy routing was successful. We plotted in each panel for each method only the result of the parameter setting or trial that turned out to be the best regarding the GR-score. In panels~a), c) and e), we embedded the networks only in the Euclidean or hyperbolic plane, while the bars in panels~b), d) and f) were created considering all the tested number of dimensions $d$.}
    \label{fig:undirGR_football}
\end{figure}

\begin{figure}[!h]
    \centering
    \includegraphics[width=1.0\textwidth]{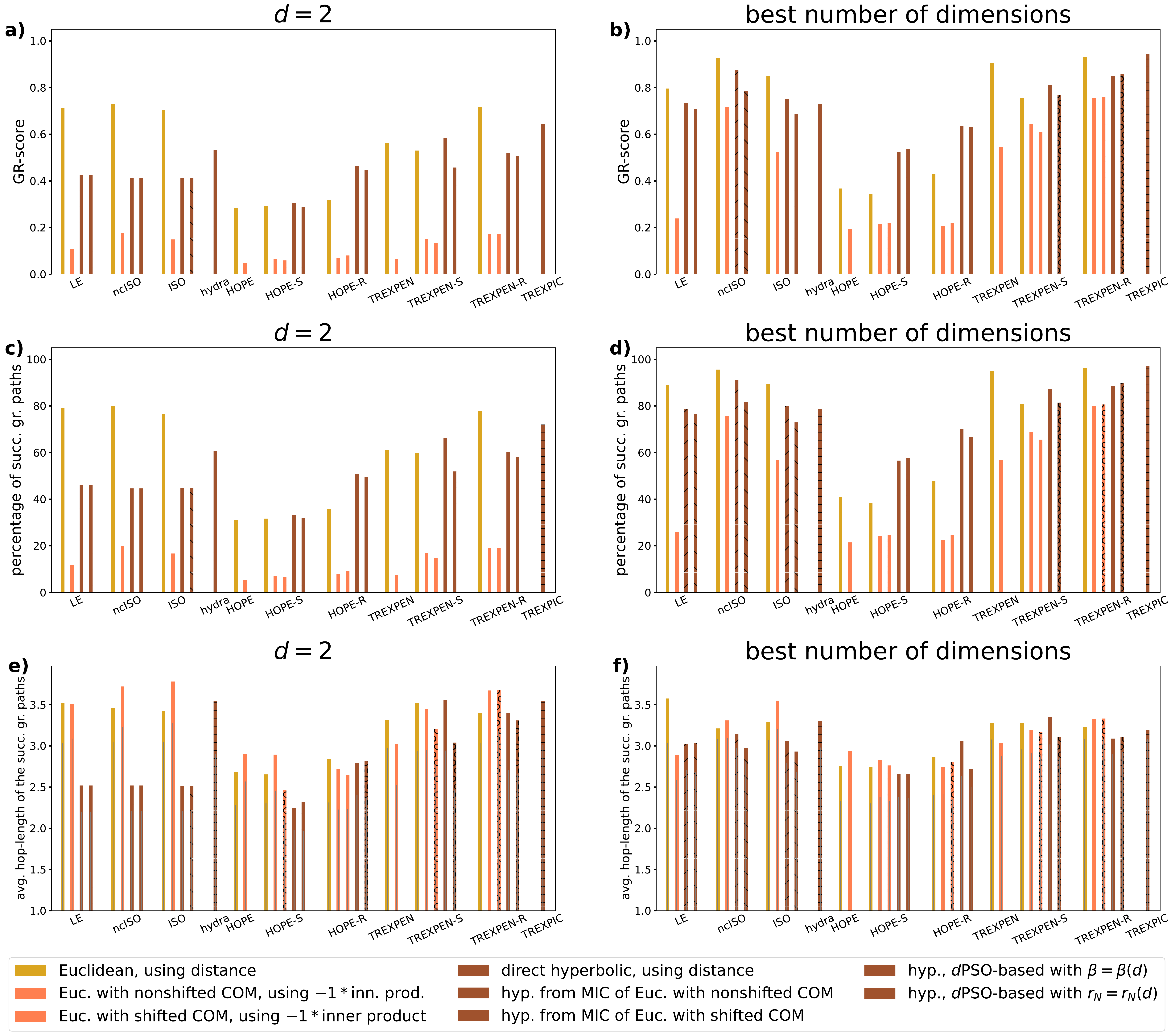}
    \caption{ {\bf Greedy routing performance on the network of political books.} The task was to perform greedy routing between each one of the (unordered) node pairs. The colours indicate what geometric measure was used, while the different patterns denote different characteristics regarding the way the node arrangements were created in the given geometry, as listed in the common legend at the bottom of the figure. Each row of panels refers to a given measure of embedding quality: panels~a) and b) to the greedy routing score (the higher the better), panels~c) and d) to the success rate of greedy routing (the higher the better), whilst panels~e) and f) to the average hop-length of the successful greedy paths (the smaller the better), depicting with grey bars also the average of the hop-length of the shortest paths connecting those node pairs for which the greedy routing was successful. We plotted in each panel for each method only the result of the parameter setting or trial that turned out to be the best regarding the GR-score. In panels~a), c) and e), we embedded the networks only in the Euclidean or hyperbolic plane, while the bars in panels~b), d) and f) were created considering all the tested number of dimensions $d$.}
    \label{fig:undirGR_polBooks}
\end{figure}

Figures~\ref{fig:undirMapAcc}--\ref{fig:undirGR_polBooks} compare the performance of the different embedding methods on the following two undirected networks:
\begin{itemize}
    \item The American College Football network~\cite{football_net_data}, connecting $N=115$ number of Division IA colleges via $E=613$ number of games played during regular season Fall 2000. Each node has an attribute denoting to which of the 12 conferences it belongs.
    \item A network~\cite{pol_blogs_data} between $N=105$ number of books about U.S. politics sold by Amazon.com, where the $E=441$ number of links represent frequent co-purchasing by the same buyers. Each node has an attribute indicating whether they are "liberal", "neutral", or "conservative".
\end{itemize}
Note that when dealing with an undirected network, we always embedded only the largest connected component (LCC). The above-listed $N$ and $E$ values refer to the number of nodes and edges in the LCCs.

In the Euclidean version of LE, ncISO and ISO, the only tunable parameter was the number of dimensions $d$. To give room for the performance-improving effect of the increase in the number of dimensions but, besides, ensure a significant dimension reduction ($d\leq N/10$), we tested the $d=2$, $3$, $4$ and $8$ settings for all the embedding methods and both networks. Besides, we used the same setting of the curvature $K=-\zeta^2$ of the hyperbolic space, namely $K=-1$ in all of the examined embedding algorithms, including our new methods too.

For the radial conversion of the Euclidean embeddings provided by LE, ISO and ncISO we used the $d$-dimensional PSO model~\cite{dPSO} instead of the original, two-dimensional one~\cite{PSO}. For this task, we determined the degree decay exponent $\gamma$ by fitting a power law ($\mathrm{CCDF}(k)=P(k\leq K)\sim k^{-(\gamma-1)}$) to the complementary cumulative distribution function of the node degrees $k$ using the method described in Ref.~\cite{degreeDistrFitting}. We always tested both $d$PSO-based approaches that are described in Sect.~\ref{sect:embOfPSO} -- making either the largest hyperbolic radial coordinate $r_N$ (Eqs.~(\ref{eq:betaInPSOembedding1})--(\ref{eq:radCoordInPSOembedding1})) or the popularity fading parameter $\beta$ (Eqs.~(\ref{eq:betaInPSOembedding2})--(\ref{eq:radCoordInPSOembedding2})) dependent on the number of dimensions -- since according to our measurements, in $d$-dimensional spaces of $d>2$ the performance of these two approaches can differ from each other. Besides, since the radial order between the nodes having the same degree is arbitrary in the $d$PSO-based Euclidean-hyperbolic conversion, we re-ran it $15$ times in each case. Note that repeating any of the other procedures examined in this section (i.e., the Euclidean embeddings, the hydra method and our new Euclidean-hyperbolic conversion MIC) is not reasonable due to their fully deterministic nature. Although several link weighting formulas were proposed in Ref.~\cite{coalescentEmbedding} that were shown to be capable of improving the hyperbolic embeddings obtained from the LE, the ISO and the ncISO methods, in order to moderate the number of adjustable embedding parameters, we did not use any pre-weighting strategies and inputted the plain, unweighted networks to all the examined embedding methods. However, as it is shown in Sect.~\ref{sect:realLinkWeights}, weighted networks can be easily inputted to TREXPEN and TREXPIC, meaning that just like LE, ISO and ncISO, these new methods are also able to utilize pre-weights assigned to the links.

Regarding the embedding parameter $\alpha$ of HOPE, HOPE-S, HOPE-R and $q$ of TREXPEN, TREXPEN-S, TREXPEN-R and TREXPIC, we tested the same $15$ settings as always (see Sects.~\ref{sect:HOPE}, \ref{sect:TREXPEN} and \ref{sect:TREXPIC} for the details). Besides, as usual, we allowed shifting the centre of mass (COM) of the Euclidean node arrangements obtained from HOPE-S, HOPE-R, TREXPEN-S and TREXPEN-R to the origin. As in the other measurements, in MIC we set the parameter $C$ of the largest hyperbolic radial coordinate $r_{\mathrm{hyp,max}}=\frac{C}{\zeta}\cdot\ln{N}$ to $2$. Note that in the PSO-based conversion $r_{\mathrm{hyp,max}}\equiv r_{N}=\frac{2}{\zeta}\cdot\ln{N}$ for both of the above-described approaches (see Eqs.~(\ref{eq:radCoordInPSOembedding1}) and (\ref{eq:radCoordInPSOembedding2})), which coincides with the largest hyperbolic radial coordinate of our hyperbolic methods at $C=2$. 

In the figures of this section we always plotted only the best results achieved among the several trials for each embedding method. 
First, in Fig.~\ref{fig:undirMapAcc}, we show the mapping accuracies achieved by the different methods for both of the examined undirected real networks. Regarding the other tasks, we created one figure for each network separately: Figs.~\ref{fig:undirGraRec_football} and \ref{fig:undirGraRec_polBooks} deal with the graph reconstruction task examining how well the embeddings can learn to differentiate between the inputted connected and unconnected node pairs (in-sample prediction), Figs.~\ref{fig:undirLinkPred_football} and \ref{fig:undirLinkPred_polBooks} show the results of some link prediction tasks investigating how well the embeddings can learn to differentiate between the missing links and the actually unconnected node pairs without having any specific input of the correct labels (out-of-sample prediction), while Figs.~\ref{fig:undirGR_football} and \ref{fig:undirGR_polBooks} present the performances in greedy routing. Generally, it can be concluded that TREXPIC and the variants of TREXPEN performed on the examined undirected networks similarly to or even better than the earlier undirected methods in all the tasks, while HOPE and its variants seem to be less suitable than the other algorithms for the greedy routing task.

\newpage
\null\newpage
\null\newpage
\null\newpage
\null\newpage
\null\newpage
\null\newpage
\section{Representation of link weights in the embeddings}
\label{sect:realLinkWeights}
\setcounter{figure}{0}
\setcounter{table}{0}
\setcounter{equation}{0}
\renewcommand{\thefigure}{S8.\arabic{figure}}
\renewcommand{\thetable}{S8.\arabic{table}}
\renewcommand{\theequation}{S8.\arabic{equation}}

In this section, we demonstrate via the neural network of C. elegans~\cite{CelegansRef} that the variants of TREXPEN and TREXPIC are able to interpret the link weights of an inputted network. For this, given link weights that suggest distances or dissimilarities (and not proximities or similarities), we simply have to identify the shortest path length in the exponential proximity and distance formula (Eqs.~(\ref{eq:ourProxMatrixElements}) and (\ref{eq:ourFiniteDistMatrixElements})) with the sum of the weight of the involved links instead of the number of hops. Note that using the formula
\begin{equation}
    P_{st}=\sum_{0<\ell} \alpha^{\ell}\cdot n_{s\rightarrow t}^{\mathrm{paths}}(\ell)
    \label{eq:KatzMatrixElementsForWeightedGraph}
\end{equation}
instead of the original definition of the Katz index given for unweighted networks by Eq.~(\ref{eq:KatzDefinition}), HOPE and its variants could also be applied for embedding weighted graphs. Nevertheless, the calculation of the sum in Eq.~(\ref{eq:KatzMatrixElementsForWeightedGraph}) is tremendously compute-intensive due to the necessity of the exploration of all the possible paths between the nodes, while the matrix reformulation of the Katz index written in Eq.~(\ref{eq:KatzMatrixElements}) is not valid for weighted networks (since $n_{s\rightarrow t}^{\mathrm{paths}}(\ell)=\bm{A}^{\ell}$ is fulfilled only by unweighted graphs, i.e. when all the link weights are $1$). Therefore, here we study only our new methods -- that use exponential proximities and distances instead of the Katz indexes -- from the point of view of the representation of link weights in embeddings.

To quantify how accurately the inputted weights of the connections (interpreted as distances) are reflected by our embeddings, we calculate their Spearman's correlation coefficient~\cite{SpearmanCorrCode} $\mathrm{ACC}_{\mathrm{w}}$ with the different geometric measures that we considered to be an increasing function of the topological distances: the Euclidean distance and the additive inverse of the inner product in Euclidean node arrangements, and the hyperbolic distance in hyperbolic embeddings. In successful mappings, we expect the emergence of high positive values of $\mathrm{ACC}_{\mathrm{w}}$.

\begin{figure}[hbt]
    \centering
    \includegraphics[width=\textwidth]{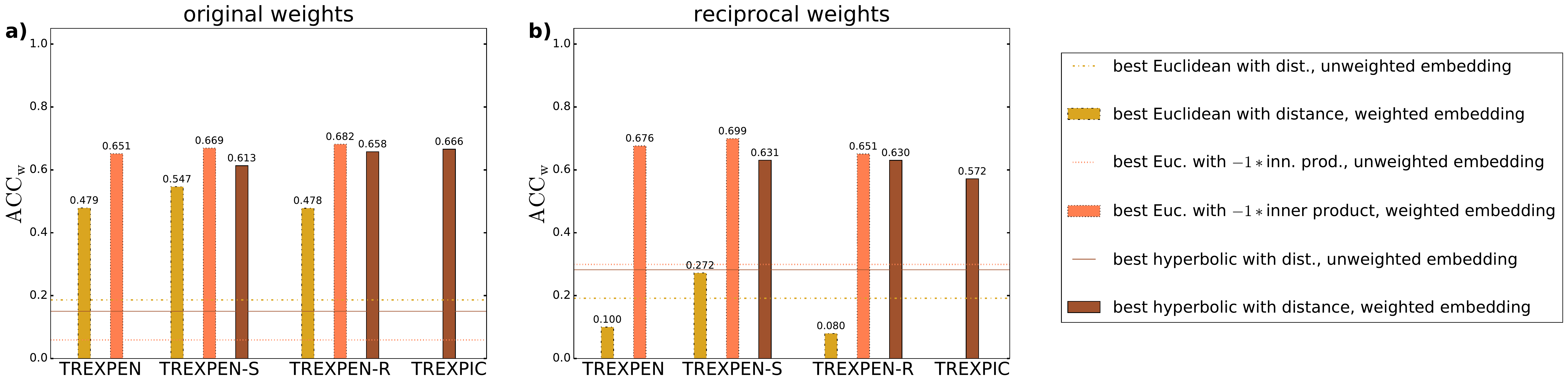}
    \caption{ {\bf Spearman's correlation coefficient between the non-zero link weights and the examined geometric measures in the embeddings of the neural network of C. elegans.} In panel~a), we identified the original link weights given in the data set with distances, while in panel~b) we re-weighted the network with the reciprocal of the original link weights, interpreting the original link weights as proximities. The colours indicate what geometric measure was used, as listed in the legend. We plotted only the results of those parameter settings that turned out to be the best, i.e. which yielded the highest values of the correlation. The bars were created by inputting the link weights to the embedding methods, whereas the horizontal lines show the highest correlations achieved among all the algorithms when the unweighted version of the network was embedded.}
    \label{fig:weightCorr}
\end{figure}

In Fig.~\ref{fig:weightCorr}a, we show the correlation coefficient $\mathrm{ACC}_{\mathrm{w}}$ for the non-zero link weights of the neural network of C. elegans given by Ref.~\cite{CelegansRef} when these are inputted in their original form to our embedding methods, i.e. when we interpret the link weights given in the data set as distance measures. Besides, in Fig.~\ref{fig:weightCorr}b we depict the $\mathrm{ACC}_{\mathrm{w}}$ values that were calculated for the reciprocal of the original weights of the connections when running our embedding algorithms using these inverted weight values, i.e. when we read the link weights in the data set as proximities. As a baseline, we plot in both cases for each geometric measure the highest correlation that was achieved when we embedded the network setting all the link weights to 1. According to the figure, our embedding methods were capable of the optimisation of the node arrangement according to the inputted link weights, suggesting that it might be worthwhile to test the embeddings also with the consideration of the link weights when such information is provided. Furthermore, just like it was proposed for undirected networks in Ref.~\cite{coalescentEmbedding}, one may even apply artificial link weights derived from the network topology itself to promote the embedding process.

\newpage
\bibliographystyle{vancouver}
\bibliography{references}

\begin{thebibliography}{10}
\bibitem{SBMcode}
\bibinfo{title}{For generating graphs with the stochastic block model, we used
  the {P}ython function `stochastic\_block\_model' available in the
  `{NetworkX}' package.}
\bibitem{SpearmanCorrCode}
\bibinfo{title}{We calculated the {S}pearman's correlation coefficients with
  the {P}ython function `spearmanr' available in the `scipy.stats' package.}
\bibitem{precRecCurveCode}
\bibinfo{title}{{We computed the precision-recall pairs for different
  probability thresholds with the Python function `precision\_recall\_curve'
  available in the `sklearn.metrics' package.}}
\bibitem{areaUnderCurveCode}
\bibinfo{title}{{We calculated the area under the PR and the ROC curves with
  the Python function `auc' available in the `sklearn.metrics' package.}}
\bibitem{ROCcurveCode}
\bibinfo{title}{{We computed the receiver operating characteristic curve with
  the Python function `roc\_curve' available in the `sklearn.metrics'
  package.}}
\end{thebibliography}

\end{document}